\documentclass[12pt]{article}

\usepackage{amssymb,amsmath,amsthm,thmtools,amsfonts,bbm,eurosym,geometry,ulem,graphicx,caption,color,setspace,comment,footmisc,caption,natbib,pdflscape,array,enumerate,nomencl,etoolbox,mathtools,booktabs,makecell,subcaption,algorithm,multirow,multicol}

\usepackage{tikz}          % core TikZ package
\usepackage{float}         % for [H] placement option in figure
\usepackage{scalefnt}      % for \scalebox (actually comes via graphicx)
\usepackage{graphicx}      % provides \scalebox
\usetikzlibrary{positioning,calc,arrows.meta}

\usepackage{algpseudocode,footmisc}
\makenomenclature
\usepackage[colorlinks=true]{hyperref}
\hypersetup{%
  citecolor=blue
}
\renewcommand\nomgroup[1]{%
  \item[\bfseries
  \ifstrequal{#1}{A}{General descriptive variables}{%
  \ifstrequal{#1}{B}{Model-specific variables}{}}%
]}

\mathtoolsset{showonlyrefs=true}
\theoremstyle{remark}

\renewcommand{\today}{\ifcase \month \or January \or February \or March \or %
April \or May \or June \or July \or August \or September \or October \or November \or %
December \fi \number \year} 

\title{Deep Learning for Art Market Valuation}

\author{Jianping Mei\thanks{CKGSB. Email: \href{mailto:jpmei@ckgsb.edu.cn}{jpmei@ckgsb.edu.cn}.} \and
Michael Moses\thanks{Art Market Consultancy. Email: \href{mailto:mam1@stern.nyu.edu}{mam1@stern.nyu.edu}.}  \and
Jan Wälty\thanks{University of Zurich. Email: \href{mailto:jan.waelty@uzh.ch}{jan.waelty@uzh.ch}.} \and
Yucheng Yang\thanks{University of Zurich and Swiss Finance Institute. Email: \href{mailto:yucheng.yang@uzh.ch}{yucheng.yang@uzh.ch}. }
}

\date{\today\\
}

\begin{document}
\maketitle

\begin{abstract}
We study how deep learning can improve valuation in the art market by incorporating the visual content of artworks into predictive models. Using a large repeated-sales dataset from major auction houses, we benchmark classical hedonic regressions and tree-based methods against modern deep architectures, including  multi-modal models that fuse tabular and image data. We find that while artist identity and prior transaction history dominate overall predictive power, visual embeddings provide a distinct and economically meaningful contribution for \textit{fresh-to-market works} where historical anchors are absent. Interpretability analyses using Grad-CAM and embedding visualizations show that models attend to compositional and stylistic cues. Our findings demonstrate that multi-modal deep learning delivers significant value in precisely the situations where valuation is hardest—first-time sales—and thus offers new insights for both academic research and practice in art market valuation.
\end{abstract}

\newpage

\section{Introduction}
The art market is a stringent setting for valuation. Each asset is unique, trading is infrequent, and the information set is uneven: condition and provenance are partly private, reputational signals evolve slowly, and auction houses publish presale estimates that guide participation while also reflecting strategy. In such illiquid, history-sparse environments, price anchors—when they exist—carry disproportionate weight. When they do not, valuation is difficult. We ask whether modern deep learning can extract price-relevant information directly from the image of the artwork and, crucially, when this visual information is incremental to what metadata and price history already provide. Bringing computer vision back to the most economically important image-native asset—fine art—lets us study this question where it matters.

We design the study around a simple economic contrast. If images act as a substitute for missing history, their marginal value should concentrate in first-time sales. If images mainly proxy for structured attributes or reputation, they should add little once history is available. To test this, we assemble a repeated-sales panel from major auction houses covering 1970–2024 and match each transaction to the underlying image. We impose a forward-looking temporal split and evaluate strictly out of sample. Throughout, we report results separately for fresh-to-market works (first sale in our panel) and repeated sales (with price anchors).

We benchmark standard valuation tools against multi-modal deep learning in a common framework. On the “tabular” side, we estimate hedonic regressions, a strong tree-based model (XGBoost), and a compact neural net using the same structured predictors (artist, medium, dimensions, venue, timing). On the “vision” side, we build a deliberately simple multi-modal architecture that fuses a pretrained ResNet image embedding with the tabular projection and trains end-to-end on log prices. To situate the exercise in practice, we benchmark predictions not only against realized prices but also against presale estimates, the operative yardstick for market participants. Finally, we complement forecasting with interpretation: we use light-touch tools—Grad-CAM saliency maps on representative lots and low-dimensional projections of the fused embeddings—to show what the model attends to; and we run ablations and feature-importance checks to quantify how visual and structured signals interact.

The forecasting results are organized by the information environment. When history exists, it dominates: in repeated sales, models that include prior prices and artist identity achieve high out-of-sample fit, and adding image features contributes little beyond strong tabular baselines. This confirms that structured signals—especially price history—soak up most of the predictable variation in seasoned states. By contrast, when history is absent, images matter. For fresh-to-market works, multi-modal models materially improve accuracy relative to tabular-only approaches, indicating that visual content supplies orthogonal information exactly where valuation is hardest. The magnitude is economically meaningful: in our holdout period, the multi-modal model lifts explanatory power for first-time sales relative to trees and tabular nets, whereas the lift is negligible for repeated sales. We also position these results against expert guidance. Presale estimates remain the single most accurate predictor overall, consistent with experts’ access to condition, provenance, and anchoring. Nonetheless, the incremental value of images is largest precisely where estimates are least anchored by history—first-time sales. A complementary exercise predicts deviations from estimates; performance is modest overall, but again the relative gains from images concentrate in fresh-to-market lots.

Heterogeneity checks are consistent with this state-dependent view. Image features improve predictions most in visually distinctive segments (e.g., certain media or contemporary categories) and least where valuation is primarily reputational. The gains are robust to alternative specifications and persist when we remove past prices to avoid imputation artifacts, but they attenuate when the image embedding is made excessively high-dimensional, consistent with a bias–variance trade-off in small fresh-to-market samples.

The interpretation and diagnostics reinforce the economic story. Saliency highlights that the vision branch attends to composition and texture and sometimes to signatures or focal objects; embedding projections show partial clustering by style or medium. At the same time, attention can be superficial, underscoring that visual cues complement rather than replace expertise. Feature-importance and ablation analyses quantify the division of informational labor: prior prices and artist identity dominate in repeated sales; object-level attributes and images have the largest marginal contribution in first-time sales; and images interact with those object-level features—consistent with visual content supplying the missing record when anchors are absent.

Taken together, the evidence yields a simple organizing principle with implications beyond art. When history is rich, it is hard to beat: tabular models with price history and reputation signals perform best, and images add little. When history is thin, computer vision can substitute for part of the missing record and meaningfully complement expert judgment. The paper proceeds by implementing this design, presenting the forecasting results by state, benchmarking against presale estimates, and then documenting what the models learn from images and how those signals interact with structured predictors.

\paragraph{Related Literature} We contribute to several literatures. First, we speak to art valuation without image data. Classic hedonic and repeated–sales approaches relate prices to observed characteristics or exploit resale pairs to separate common time components and idiosyncratic variation, emphasizing illiquidity, selection in observed transactions, and the central roles of artist identity and price history \citep[e.g.,][]{ginsburgh2006computation,MeiMoses,Renneboog,graddy2014anchoring,mei2023residual}. We adopt this perspective but make the information set explicit: we maintain a repeated–sales panel and split all evaluations into fresh–to–market versus repeated sales. This design allows us to characterize \emph{state dependence} in the marginal value of additional signals and to report strictly out–of–sample performance under a temporal holdout.

Second, we contribute to the growing literature that incorporates image data. Multi–modal models that fuse images with metadata typically find modest average gains from images once rich tabular features are included; pure image models generally underperform metadata–only baselines in pricing tasks \citep[e.g.,][]{aubry,Bailey}. We refine these results by showing that the informational content of images is strongly state–dependent: images add economically meaningful information when history is missing (fresh–to–market) and are largely redundant when history is rich (repeated). Methodologically, we keep the architecture deliberately simple (pretrained ResNet plus a compact tabular projection) and enforce a forward–looking temporal split so that improvements are attributable to \emph{information content} rather than heavy tuning.

Third, we contribute to understanding valuation \emph{relative to} expert benchmarks rather than only to realized outcomes. Presale estimate ranges guide participation and reserves, anchor bidding, and are known to exhibit systematic forecast errors and strategic or behavioral components \citep[e.g.,][]{Beggs,Mei2005,reserve}. We benchmark directly against these estimates and analyze residuals to map where machine predictions complement expert guidance. The key result is that the marginal value of images is largest when estimates are least anchored by history—first–time sales—clarifying how expert judgment and machine learning divide informational roles.

Finally, we contribute to the emerging literature on machine learning for asset pricing and prediction. On structured data, tree–based methods frequently rival or outperform generic deep nets at medium scales; in finance, recent work demonstrates the gains and limits of ML for prediction and representation learning \citep[e.g.,][]{gu2020empirical,leippold2022machine,cong2025textual,yang2020knowledge,mauer2024tabular,CARUGNO2025126468,liu2022,jiao2025interpretable}. We confirm that strong tabular baselines are hard to beat in seasoned states (repeated sales with prior prices and artist identity) and show that multi–modal learning delivers incremental value exactly where the structured information set is thin. Conceptually, our results provide a simple organizing principle relevant beyond art: when history is rich, it is hard to beat; when history is thin, unstructured signals (images here) can substitute for part of the missing record.

The remainder of this paper is structured as follows. In Section \ref{sec:method}, we introduce our dataset and methodological approaches for price predictions. Section \ref{sec:results} presents empirical results with a particular emphasis on predictive performance and the interpretation of visual features.  Finally, we conclude.

\section{Data and Methodology}\label{sec:method}
\subsection{Data} \label{sec:data}
We use proprietary data from Art Market Consultancy, which includes auction transaction data and the corresponding image data from various auction houses. The raw data span 1970–2024 and are organized as repeated-sales pairs at the object level. We convert these pairs into a transaction-level panel. It is worth noting that our data include paintings that were for sale but was not sold (i.e. bought-in). Thus, some transactions only 
had one sale price.

\paragraph{Variables.}
For each transaction, we observe a standard set of structured attributes used in hedonic valuation and in practice. These include: (i) artist identity; (ii) auction house and location; (iii) sale timing (year and month); (iv) market segment (category) and medium; (v) physical characteristics (height, width, and shape); (vi) indicators and counts capturing catalog and provenance-related information (e.g., signed, dated, number of exhibitions); and (vii) the auction house presale estimate range (low and high). We also observe a catalog image for each lot, which we use as input to the computer-vision models. Appendix~\ref{app:abbr} lists abbreviations used in figures and tables.

\paragraph{Fresh-to-market versus repeated sales.}
A central feature of our design is the information environment faced by the valuer. We therefore classify each transaction as either \emph{fresh-to-market} (first appearance in our auction panel) or \emph{previously auctioned} (a repeat appearance). For previously auctioned works, we construct a lagged-price feature equal to the most recent prior transaction price observed in the panel. For fresh-to-market works, the lagged price is missing by construction. We handle this using a standard missing-indicator specification: we set the missing lagged price to zero and include an indicator for whether a prior price is observed. This preserves the economic meaning of the state variable (availability of a price anchor) while allowing us to estimate a single model on the full sample. \cite{imputer} showed that this method can provide an effective approach to deal with missing values.  In robustness checks, we also consider specifications that exclude the lagged-price level and retain only the indicator.

\begin{figure*}[h]
	\centering
	\begin{subfigure}[b]{0.48\textwidth}
		\centering
		\includegraphics[width=0.95\textwidth]{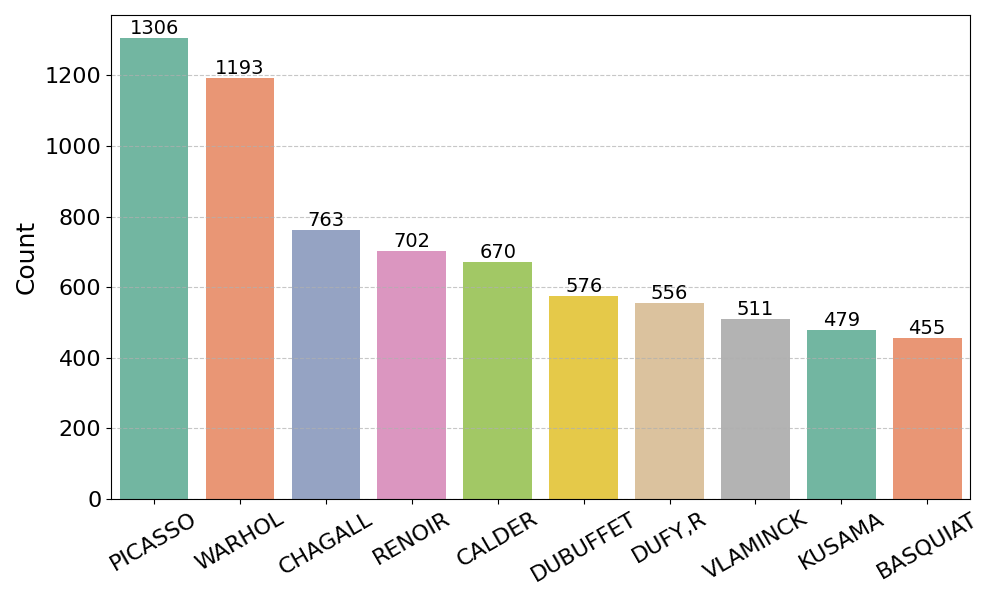}
		\caption[]%
		{{\small Artist}}    
		
	\end{subfigure}
	%\hfill
	\begin{subfigure}[b]{0.48\textwidth}  
		\centering 
		\includegraphics[width=0.95\textwidth]{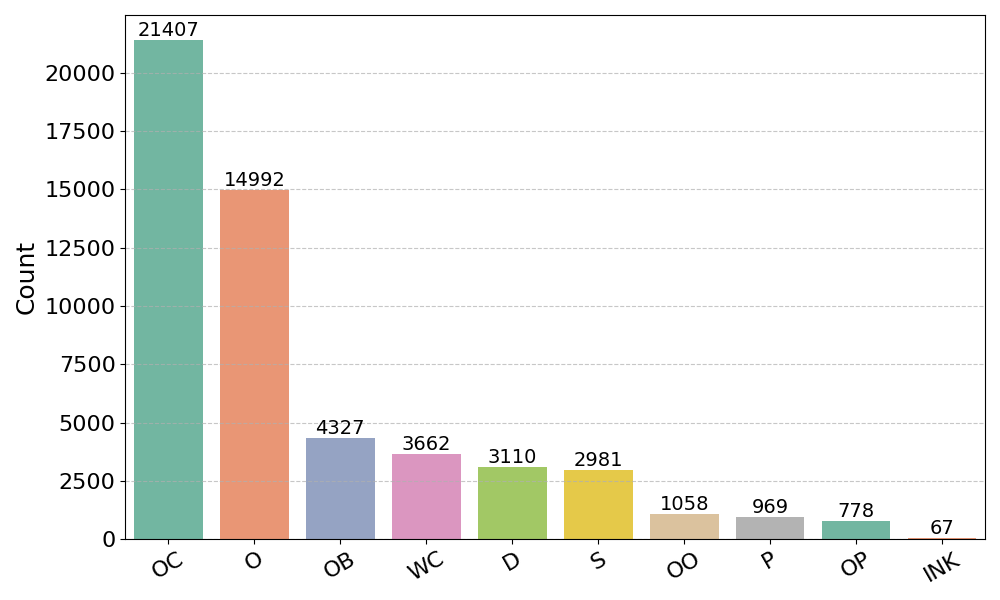}
		\caption[]%
		{{\small Medium}}    
		
	\end{subfigure}
	\vskip\baselineskip
	\begin{subfigure}[b]{0.48\textwidth}   
		\centering
		\includegraphics[width=0.95\textwidth]{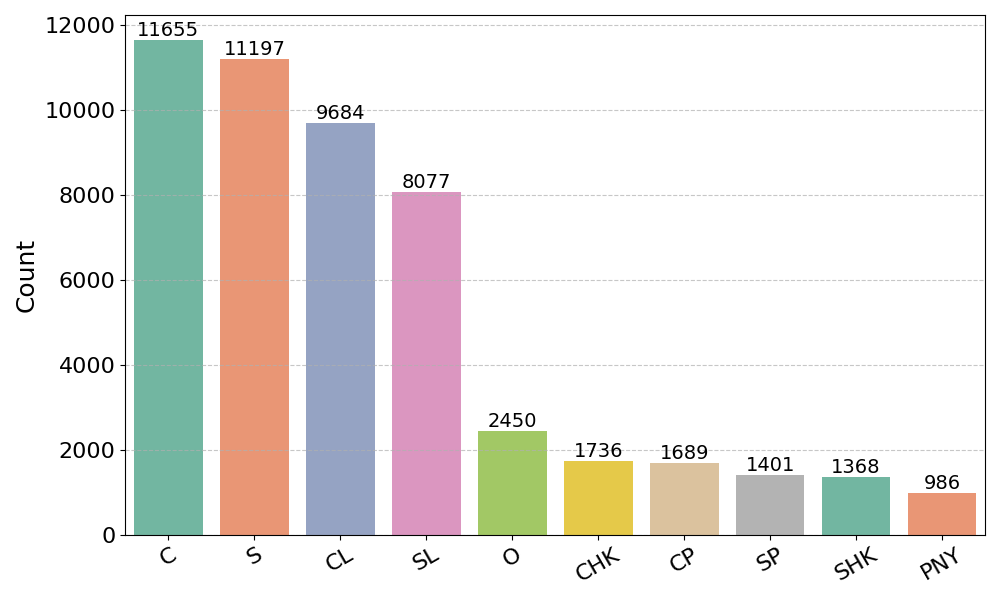}
		\caption[]%
		{{\small Transaction house}}    
		
	\end{subfigure}
	%\hfill
	\begin{subfigure}[b]{0.48\textwidth}   
		\centering 
		\includegraphics[width=0.95\textwidth]{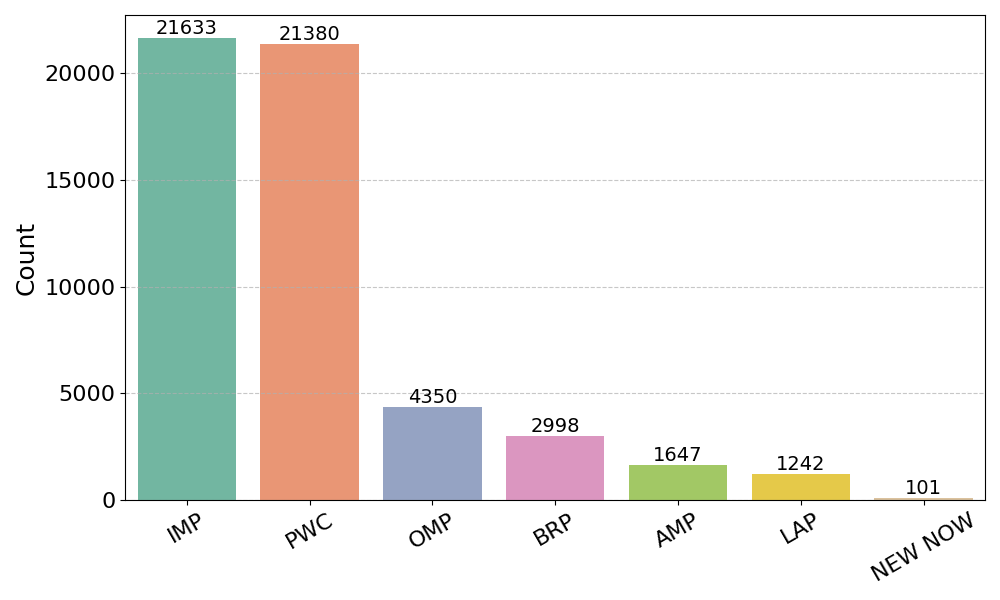}
		\caption[]%
		{{\small Category}}    
	
	\end{subfigure}
	\caption[Distribution of selected categorical variables]
	{\small Distribution of selected categorical variables} 
	\label{fig:sdist_cat}
\end{figure*}

\paragraph{Data filtering and preprocessing.}
% The data distributions suggest careful data preprocessing.  shows distributions of relevant categorical features. 
We apply three preprocessing steps to ensure a stable out-of-sample evaluation. First, we restrict attention to transactions with non-missing images and presale estimates, since these are required for our multi-modal and benchmarking exercises. Second, we impose a minimum transaction price of \$10,000 to focus on the segment where catalog information and auction-house processes are comparable and where pricing is economically meaningful. Third, for high-cardinality categorical variables (e.g., artist, auction house, medium), we keep categories that appear at least 20 times in the training sample and apply the same mapping to the test period so that evaluation does not depend on categories never seen during training. Figure \ref{fig:sdist_cat} summarizes the distribution of key categorical variables after filtering.

% Given the unbalanced and highly concentrated categorical variables, we filter the dataset by retaining only those category instances that occur more than 20 times in the training set. The same filter is then applied to the test data, which ensures that no previously unseen categories are included during evaluation. This helps avoid overfitting to infrequent artists, auction houses, and so forth. Additionally, we impose a minimum transaction price of \$ 10'000 to remove irrelevant artworks whose prices might not be adequately described by patterns in their characteristics. We also remove transactions for which we do not get the image data and/or price estimates. 

Table \ref{tab:desc_top5} presents descriptive statistics for the five most frequent artists and auction houses in the filtered dataset. All prices are converted to U.S. dollars for comparability. The pronounced deviation of the mean from the median indicates the presence of a heavy right-tailed distribution, which is consistent with the empirical characteristics of art markets, where a relatively small number of extremely high-value transactions exert a disproportionate influence on average price levels. This skewness is further reflected in the wide interquartile ranges and the considerable distance between upper percentiles (P95, P99) and the median, suggesting high price dispersion and the presence of outliers. Notably, leading artists such as Picasso and Warhol exhibit substantially higher mean transaction prices compared to others, reflecting their market prominence, while artists like Calder or Chagall occupy a lower, though still elevated, price segment. Similarly, New York auction houses dominate in terms of average transaction prices and transaction volume. 
\begin{table}[H]
\centering

\scalebox{0.80}{
\begin{tabular}{lrrrrrrr}
%\toprule
\toprule
 &  & \multicolumn{6}{c}{Transaction price (in \$)} \\
\cmidrule(lr){3-8}
 & N & Mean & P25 & P50 & P75 & P95 & P99\\
\toprule
 All & 53'351 & 630'576 & 40'453 & 105'664 & 367'111 & 2'688'000& 9'199'798 \\
 \midrule
 
 Picasso & 1'306 & 1'822'175 & 97'124 & 312'750 & 1'284'712 & 8'618'437 & 21'683'802 \\
Warhol  & 1'193 & 1'321'387 & 82'249 & 236'756 & 854'999 & 6'252'702 & 18'391'419 \\
  Chagall & 763 & 767'356 & 141'734 & 357'999 & 771'481 & 2'767'912 & 6'010'362 \\
  Renoir  & 702 & 765'400 & 126'877 & 297'500 & 673'315 & 3'164'868 & 8'720'552 \\
 Calder & 670 & 661'913 & 56'250 & 124'999 & 532'374 & 3'213'474& 5'742'794 \\

 \midrule

 Christie's New York  & 11'655& 780'244 & 44'650 & 118'749 & 425'000 & 3'375'999 & 12'204'579 \\
  Sotheby's New York & 11'197 & 737'306 & 43'750 & 118'749 & 418'000 & 3'299'999 & 11'091'599 \\
 Christie's London & 9'684 & 555'724 & 38'279 & 98'699 & 346'719 & 2'370'473 & 7'613'926 \\
 Sotheby's London & 8'077 & 667'572 & 42'836 & 110'491 & 394'623 & 2'953'964 & 9'246'671 \\
 
 Others & 2'450& 145'336 & 24'345& 49'635 & 116'275 & 502'197 & 1'495'285 \\
\bottomrule
\end{tabular}}
\caption{Transaction price summary of top five artist and auction houses (after filtering)}
\label{tab:desc_top5}
\end{table}

As we use one-hot encodings for our numerous categorical variables, we standardize the other numerical features to account for range differences.
The image data is preprocessed with a standard ResNet or ViT transformation, which rescales and standardizes the images such that they can be fed into the corresponding image model.

\begin{figure}[H]
    \centering
    \includegraphics[width=0.7\textwidth]{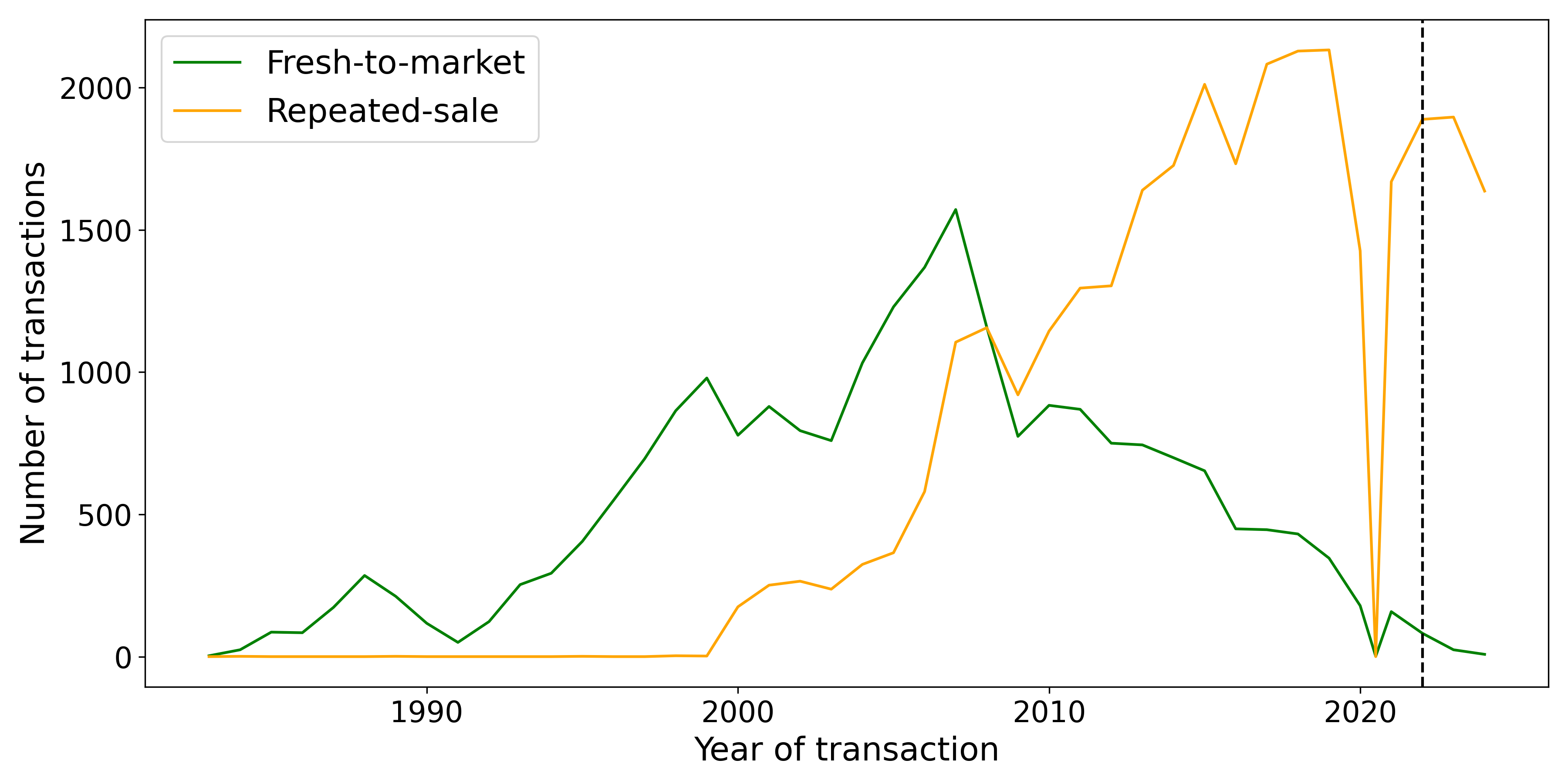}
    \caption{Number of transactions per transaction type}
    \label{fig:fresh_prev_dist}
\end{figure}

Figure \ref{fig:fresh_prev_dist} shows the number of fresh-to-market and repeat-sale transactions over time. Repeat sales become more prevalent in the later part of the sample, which matters for our empirical design: we use a temporal train–test split to mimic an out-of-sample forecasting exercise. As a result, the holdout period contains relatively more repeat-sale observations, while fresh-to-market works remain present but less frequent. We therefore report performance separately for fresh-to-market and previously auctioned works throughout the paper, and we place particular emphasis on fresh-to-market valuation where price history is unavailable and the potential contribution of visual features is greatest.

\subsection{Models}
\subsubsection{Neural Networks}
To effectively leverage both structured numerical data and unstructured image data for artwork price prediction, we design a multi-modal neural network model. This setup allows the model to learn complementary representations from different modalities, in our case tabular features (e.g., artist, size, medium) and image content, within a single model. 

\begin{figure}[H]
\centering
\scalebox{0.85}{%
\begin{tikzpicture}[
    inputnode/.style={circle, draw, fill=blue!20, minimum size=1cm, font=\small},
    imagenode/.style={circle, draw, fill=purple!20, minimum size=1cm, font=\small},
    layer/.style={rectangle, draw, rounded corners, minimum height=1cm, font=\small, align=center},
    tablayer/.style={layer, fill=green!20, minimum width=8.2cm},
    imglayer/.style={layer, fill=purple!10, minimum width=7.0cm},
    concatlayer/.style={layer, fill=orange!20, minimum width=7.4cm},
    fclayer/.style={layer, fill=red!20, minimum width=7.4cm},
    outlayer/.style={layer, fill=green!20, minimum width=7.4cm},
    arrow/.style={-{Latex[length=2mm]}, thick},
    node distance=1.2cm
]

% --- Projection blocks (no overlap) ---
\node[tablayer] (proj) at (0.0, 0.0)
{Feature Projection:\\
FC($n_{\text{feature}} \rightarrow d_{\text{tabular}}$) + ReLU};

\node[imglayer, anchor=west] (imgproj) at (4.9, 0.0)
{Image Projection:\\[1mm]
ResNet-50 $\rightarrow$ FC($\cdot \rightarrow d_{\text{image}}$)\\
or\\
ViT-Small $\rightarrow$ FC($\cdot \rightarrow d_{\text{image}}$)};

% --- Inputs (place them relative to the blocks, so arrows look clean) ---
\node[inputnode] (height) at ($(proj.north)+(-3.0,2.2)$) {Height};
\node[inputnode] (width)  at ($(proj.north)+(-1.5,2.2)$) {Width};
\node              (dots)  at ($(proj.north)+( 0.0,2.2)$) {$\cdots$};
\node[inputnode] (artist) at ($(proj.north)+( 1.5,2.2)$) {Artist};
\node[inputnode] (year)   at ($(proj.north)+( 3.0,2.2)$) {Year};

% Image node EXACTLY above imgproj -> vertical arrow
\node[imagenode] (image) at ($(imgproj.north)+(0,2.2)$) {Image};

% --- Concat node centered under both projections ---
\coordinate (midproj) at ($(proj.south)!0.5!(imgproj.south)$);
\node[concatlayer, below=1.4cm of midproj] (concat)
{Concat ($d_{\text{tabular}} + d_{\text{image}}$) + LayerNorm};

% --- FC stack ---
\node[fclayer, below=0.95cm of concat] (fc1)
{FC($d_{\text{tabular}} + d_{\text{image}} \rightarrow 128$) + BatchNorm + ReLU + Dropout};

\node[fclayer, below=0.70cm of fc1] (fc2)
{FC($128 \rightarrow 64$) + BatchNorm + ReLU};

\node[fclayer, below=0.70cm of fc2] (fc3)
{FC($64 \rightarrow 32$) + BatchNorm + ReLU};

\node[outlayer, below=0.75cm of fc3] (output)
{Output: FC($32 \rightarrow 1$)};

% --- Straight (diagonal) arrows: inputs -> feature projection ---
\draw[arrow] (height.south) -- ($(proj.north)+(-3.0,0)$);
\draw[arrow] (width.south)  -- ($(proj.north)+(-1.5,0)$);
\draw[arrow] (artist.south) -- ($(proj.north)+( 1.5,0)$);
\draw[arrow] (year.south)   -- ($(proj.north)+( 3.0,0)$);

% --- Straight vertical arrow: image -> image projection ---
\draw[arrow] (image.south) -- (imgproj.north);

% --- Projections -> concat (kept clean, no crossing boxes) ---
\draw[arrow] ($(proj.south)+( 2.0,0)$) -- ($(concat.north)+(-2.2,0)$);
\draw[arrow] ($(imgproj.south)+(-2.0,0)$) -- ($(concat.north)+( 2.2,0)$);

% --- Downstream arrows ---
\draw[arrow] (concat) -- (fc1);
\draw[arrow] (fc1) -- (fc2);
\draw[arrow] (fc2) -- (fc3);
\draw[arrow] (fc3) -- (output);

\end{tikzpicture}
}
\caption{Neural network architecture with structured metadata and image features}
\label{fig:art-model-combined}
\end{figure}
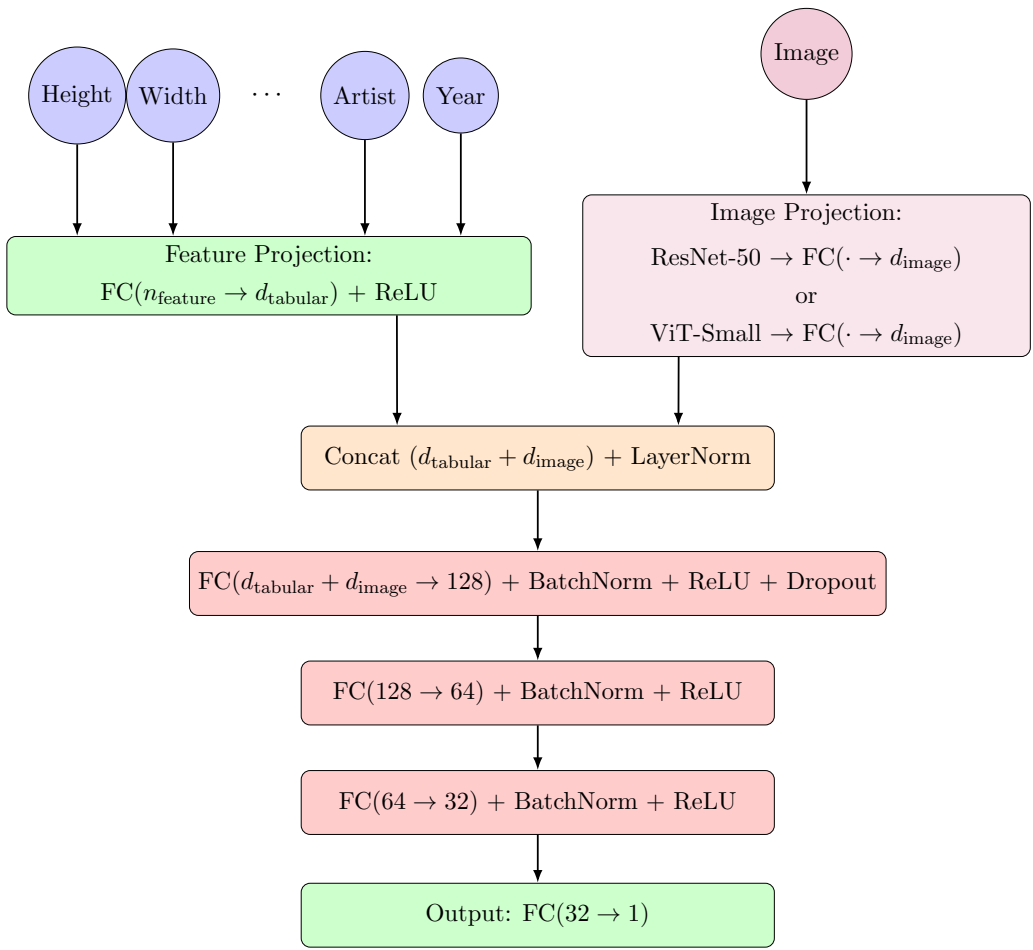

The neural network architecture is depicted in Figure~\ref{fig:art-model-combined}. The architecture has two parallel encoders—a tabular feature branch and an image branch—followed by a shared prediction head. This modular design lets us estimate (i) a tabular-only network (dropping the image branch) and (ii) the full multi-modal specification while keeping the downstream head fixed, so differences in performance can be attributed to the additional modality rather than to changes in the head architecture.

We use a Residual Network (ResNet-50) and a Vision Transformer (ViT-Small) as image encoders.\footnote{At the time of writing (December 2025), the foundational ResNet and ViT papers have accumulated roughly $300,000$ and $80,000$ citations on Google Scholar respectively. We use ResNet-50 and ViT-Small because both have mature implementations with widely available pretrained weights and similar orders of magnitude in capacity, facilitating a transparent CNN-versus-transformer comparison in our application.} Convolutional networks are naturally suited to extracting local visual cues (e.g., edges and texture), whereas transformers use self-attention to aggregate information across the image and represent longer-range relationships. We focus on these two backbones because they are widely used, computationally tractable, and broadly comparable in scale, which allows a clean comparison of how the inductive bias of the visual encoder affects performance in this setting.

The tabular branch maps the feature vector to a $d_{\text{tabular}}$-dimensional representation using a projection layer. Throughout, we set $d_{\text{tabular}}=100$; the number of input features $n_{\text{feature}}$ varies with the specification. The image branch uses a pretrained visual encoder (ResNet-50 or ViT-Small) to produce a $d_{\text{image}}$-dimensional embedding. We concatenate the two representations and apply LayerNorm before passing the fused vector through a fully connected network with ReLU activations; we use batch normalization and dropout for regularization. The network is trained end-to-end by minimizing mean squared error on log prices.

A key tuning parameter is $d_{\text{image}}$, which governs how much capacity is allocated to the visual channel. We vary $d_{\text{image}}$ across experiments to assess when, and for which subsamples, pixel information adds predictive value beyond tabular data.

\subsubsection{Linear Regression}
As a transparent benchmark, we estimate a standard hedonic pricing regression, which is extensively applied in artwork valuation (e.g., \cite{Renneboog}):
% As a benchmark model for our artwork price estimation, we employ a classical hedonic pricing model. These models have been extensively applied in real estate valuation and can be effectively extended to other asset classes, including artwork (e.g., \cite{Renneboog}). From a statistical standpoint, the following linear regression is estimated using a variety of characteristics believed to be correlated with asset prices:
\begin{equation}
    \log(p_{i}) =  x_{i}'\beta + \epsilon_{i}         ,
\end{equation}
where $p_i$ is the hammer price and $x_i$ collects observed characteristics. We work in logs to accommodate the wide price range. Hedonic models provide a useful baseline and an interpretable mapping from attributes to valuations (see, e.g., \citealp{Renneboog,valier2020performs}). At the same time, the pricing of artworks may exhibit nonlinearities and heterogeneous responses across the distribution \citep{pesando1993,malpezzi2003,scorcu2011right}, motivating the flexible machine-learning baseline we consider below.

\subsubsection{XGBoost}
We also use XGBoost \citep{chen2016xgboost} as a strong tabular baseline. Gradient-boosted trees handle mixed feature types and nonlinear interactions with minimal preprocessing and typically perform well in medium-sized tabular prediction problems \citep{friedman2001greedy,shwartz2022tabularsurvey,grinsztajn}. Unlike neural networks, XGBoost does not require feature scaling and is robust to complex interaction structures in high-dimensional one-hot encoded data.

\subsection{Estimation Procedure}

\paragraph{Train--test split and objective.}
We evaluate all models using a temporal split to approximate a realistic forecasting setting. The training sample covers 1970--2021 and the test sample covers 2022--2024. All price-prediction models are estimated by minimizing mean squared error on log prices.

\paragraph{Optimization and tuning.}
Neural networks are trained with the Adam optimizer \citep{kingma}. Hyperparameters are selected via a systematic search. For linear models and XGBoost, we use $k$-fold cross-validation on the training sample. For neural networks, tuning focuses on the learning rate, batch size, epochs, and regularization (dropout). We consider both standard dropout \citep{dropout} and an input-level dropout variant used in related work \citep{aubry}. For the image backbone, we either fine-tune end-to-end or freeze the pretrained encoder and train only the task-specific head; this provides a controlled trade-off between flexibility and overfitting risk. ViT-Small weights are taken from \cite{timm} and ResNet-50 weights from \cite{torchvision2016}.

For the linear hedonic benchmark, we also consider two standard extensions that add flexibility while preserving interpretability. First, we augment the feature set with second-order polynomial terms (including interactions) to capture simple nonlinearities. Second, we estimate a ridge (L2-regularized) version of the hedonic model, which stabilizes inference and prediction in the presence of many correlated covariates by shrinking coefficients on less informative predictors.

\section{Results}\label{sec:results}

\subsection{Price Prediction} 
This section evaluates out-of-sample performance for predicting log auction prices. We report test-set results separately for fresh-to-market works and repeated sales, since the information set differs sharply across the two groups.

A key design choice in the multimodal models is the dimension of the visual embedding, $d_\text{image}$, produced by the image encoder (ResNet-50 or ViT-Small). Increasing $d_\text{image}$ expands the capacity of the image branch and can improve fit when visual cues matter, but it also increases the risk of overfitting and computational cost. We therefore vary $d_\text{image}$ to gauge how much predictive value is contained in images and where returns to additional visual capacity flatten.

Across all models, performance is summarized using two complementary metrics: (i) $R^2$ in log-price space, which captures explained variance, and (ii) mean absolute percentage error (MAPE), which measures prediction accuracy in relative terms. We compare three model families: multimodal models (tabular + images), tabular-only machine learning baselines (neural network and XGBoost), and hedonic linear benchmarks, including the following variants:

\begin{itemize}
    \item \textbf{Hedonic:} unregularized linear benchmark.
    \item \textbf{Hedonic (regularized):} linear model with L2 regularization and polynomial/interaction terms up to degree two.
    \item \textbf{Hedonic with overpay:} adds an ``overpay'' measure (price relative to the previous estimate); repeated-sales only.
    \item \textbf{Hedonic with estimates:} full feature set plus auction house low/high estimates.
    \item \textbf{Estimates-only:} low/high estimates as the only regressors.
\end{itemize}

\subsubsection{Models with Previous Transaction Prices}\label{sec:withprev}
We first allow models to use the last observed transaction price when available, together with an indicator for whether a previous price exists. This specification is designed to exploit the strong persistence in repeated-sale prices and to capture potential anchoring effects. For fresh-to-market works, the previous-price feature is necessarily missing and is handled via a placeholder value together with the availability flag.

\begin{table}[H]
\centering
\scalebox{0.65}{
\begin{tabular}{lrcccccc}
\toprule
\textbf{Model} & \textbf{d\textsubscript{image}} & \textbf{MAPE  (\%)} & \textbf{MAPE (Previous) (\%)} & \textbf{MAPE (Fresh) (\%)} & \textbf{R\textsuperscript{2} } & \textbf{R\textsuperscript{2} (Previous)} & \textbf{R\textsuperscript{2} (Fresh)} \\
\midrule
\multicolumn{8}{l}{\textbf{Panel A: Multi-modal}} \\
\addlinespace

Multi-modal ResNet-50    & 10  & 4.9 & 4.8 & 6.6 & 0.79 & 0.79 &  0.51  \\
Multi-modal ResNet-50    & 100  & 4.9& 4.9& 5.9& 0.78 & 0.78 & 0.61   \\
Multi-modal ResNet-50    & 1000  & 5.0 & 5.0 & 5.7 & 0.78& 0.78& 0.64    \\
Multi-modal ResNet-50    & 10000  & 5.0 & 5.0 &5.7 & 0.78 & 0.78& 0.65    \\
Multi-modal ResNet-50    & 20000  & 5.1 & 5.1 & 6.0 & 0.78 & 0.78 & 0.62  \\ \addlinespace
Multi-modal ViT-Small    & 10 &  5.1 & 5.1 & 6.0 & 0.76 & 0.76 & 0.61 \\
Multi-modal ViT-Small    & 100 &   5.0&   5.0  & 6.0  & 0.78 & 0.78& 0.60\\
Multi-modal ViT-Small    & 1000 & 5.0  & 5.0  & 6.2  & 0.77 & 0.77  & 0.55 \\
Multi-modal ViT-Small    & 10000 & 5.1  & 5.1  & 5.9  & 0.77 & 0.78  & 0.61  \\
Multi-modal ViT-Small    & 20000 & 5.0  & 5.0  &  6.0  &  0.78 & 0.78  & 0.61   \\

\addlinespace
\multicolumn{8}{l}{\textbf{Panel B1: Tabular-only}} \\
\addlinespace
Neural Network           & —   &  5.2 & 5.2 & 6.3 &  0.77 &  0.77 & 0.54 \\
XGBoost                  & —   &  5.0 & 5.0  & 5.9 & 0.79 & 0.79 & 0.59\\

\addlinespace
\multicolumn{8}{l}{\textbf{Panel B2: Hedonic models}} \\
\addlinespace
Hedonic (regularized) & — & 5.9 & 5.9& 6.4 & 0.70 & 0.71 & 0.52 \\
Hedonic with overpay & — & — & 5.3 & — & — & 0.75 & — \\
Hedonic   & —   &  6.5 & 6.5 & 7.0 & 0.66 & 0.67 & 0.49 \\

Hedonic with estimates  & — & 2.8 & 2.8 & 3.3  & 0.93 & 0.93 & 0.90  \\
Estimates only  & —   & 2.9 & 2.9 & 3.4 & 0.93& 0.93 & 0.88  \\

\bottomrule
\end{tabular}}
\caption{Full model performance}
\label{tab:model-performance}
\end{table}

Table~\ref{tab:model-performance} summarizes the results. Panel A shows multi-modal models as a function of $d_{\text{image}}$. For {ResNet-50}, increasing $d_{\text{image}}$ primarily improves performance for \emph{fresh-to-market} works: $R^2$ rises from $0.51$ at $d_{\text{image}}=10$ to roughly $0.64$--$0.65$ around $d_{\text{image}}=1{,}000$--$10{,}000$, with little gain beyond that range. For \emph{previously auctioned} works, performance is largely flat in $d_{\text{image}}$ (around $R^2 \approx 0.78$--$0.79$), consistent with the idea that past transaction prices dominate the incremental information content of images once a sale history exists.

For {ViT-Small}, performance is comparatively stable across embedding sizes, with no clear monotone relationship between $d_{\text{image}}$ and either $R^2$ or MAPE. In this application, the transformer-based encoder appears less sensitive to the size of the exported embedding.

Comparing Panel A (multi-modal) to Panel B1 (tabular-only), the main pattern is that {tabular-only models perform well for repeated sales but degrade sharply for fresh-to-market lots}. XGBoost and the tabular neural network achieve $R^2 \approx 0.77$--$0.79$ for previously auctioned works, but only $R^2 \approx 0.54$--$0.59$ for fresh-to-market works. By contrast, the multi-modal ResNet-50 narrows this gap and reaches $R^2 \approx 0.61$--$0.65$ for fresh-to-market works when $d_{\text{image}}$ is sufficiently large, indicating that image features partially substitute for missing historical price anchors.

Panel~B2 shows that hedonic models without auction house estimates lag behind machine learning methods. Regularization improves fit relative to the plain hedonic benchmark, and the overpay variable is informative for repeated sales, but these models remain below tree-based and multi-modal approaches. The largest performance jump occurs when including auction house low/high estimates: both ``hedonic with estimates'' and ``estimates-only'' achieve $R^2 \approx 0.93$ and MAPE around $3\%$, and adding additional covariates provides only marginal gains. This highlights the predictive strength of expert pre-sale estimates and sets a high bar for purely data-driven models.

Overall, when past transaction prices are available, tabular features and price history already deliver strong performance; images matter most for fresh-to-market works, where history is absent and the model must rely on object-level characteristics captured visually.

\subsubsection{Models without Previous Prices}\label{withoutprev}
The previous-price feature is mechanically unavailable for fresh-to-market works and must be imputed in the specification above. A natural concern is that this imputation creates avoidable bias and complicates learning. We therefore re-estimate models without using the last transaction price. Instead, we retain only a binary indicator for whether a previous sale exists. This preserves the informational distinction between first-time and repeated-sale lots while avoiding any fictitious ``previous price'' signal.

Table~\ref{tab:model-performance:_without} reports the corresponding results (focusing on ResNet-50 for the multi-modal branch). Two patterns stand out. First, multi-modal performance for fresh-to-market works improves with embedding size up to an interior optimum: $d_{\text{image}}$ around $500$--$1{,}000$ yields the best test performance, while very large embeddings (e.g., $10{,}000$) slightly degrade accuracy, consistent with overfitting in high-dimensional representations. Second, removing explicit past prices reduces overall performance for previously auctioned works, yet image features become relatively more valuable in compensating for the missing price anchor.

\begin{table}[H]
\centering
\scalebox{0.6}{
\begin{tabular}{lrcccccc}
\toprule
\textbf{Model} & \textbf{d\textsubscript{image}} & \textbf{MAPE Total (\%)} & \textbf{MAPE (Previous) (\%)} & \textbf{MAPE (Fresh) (\%)} & \textbf{R\textsuperscript{2} Total} & \textbf{R\textsuperscript{2} (Previous)} & \textbf{R\textsuperscript{2} (Fresh)} \\
\midrule
\multicolumn{8}{l}{\textbf{Panel A: Multi-modal}} \\
\addlinespace

Multi-modal ResNet-50    & 10  & 5.2 & 5.2  & 5.7 & 0.76 & 0.76  & 0.64   \\

Multi-modal ResNet-50    & 100  & 5.2 & 5.1  & 6.1 & 0.77 & 0.77  &  0.57 \\
Multi-modal ResNet-50    & 500  & 5.0 & 5.0  & 5.7 & 0.78  &0.78   & 0.63  \\
Multi-modal ResNet-50    & 1000  & 5.0  & 5.0  & 5.3 & 0.78 & 0.78  & 0.67  \\

Multi-modal ResNet-50    & 10000  & 5.2 & 5.2  & 5.6 & 0.76 & 0.76  & 0.61   \\

\addlinespace
\multicolumn{8}{l}{\textbf{Panel B1: Tabular-only}} \\
\addlinespace
Neural Network           & —   &  6.0& 6.0    & 5.8 & 0.68   & 0.68   &  0.55  \\
XGBoost                  & —   & 5.6 & 5.6  & 5.7 & 0.74 & 0.74  & 0.63 \\

\addlinespace
\multicolumn{8}{l}{\textbf{Panel B2: Hedonic models}} \\
\addlinespace
Hedonic (regularized)  & —   & 6.8 & 6.9& 6.5  & 0.60& 0.60 & 0.51 \\
Hedonic   & —   & 7.4 & 7.4 & 6.7  & 0.56 & 0.56 & 0.51 \\

\bottomrule
\end{tabular}}
\caption{Model performance  without previous transaction price as feature}
\label{tab:model-performance:_without}
\end{table}

Relative to tabular-only baselines, the multi-modal ResNet-50 remains strongest for fresh-to-market works. XGBoost performs competitively overall, but the multi-modal model achieves higher $R^2$ for fresh-to-market samples (roughly $0.64$--$0.67$ at the best $d_{\text{image}}$) than tabular-only alternatives. Hedonic models perform the weakest, suggesting that linear structure is too restrictive once auction histories are incomplete and nonlinear interactions matter.

Given that this specification avoids imputation and performs best for fresh-to-market works, we use it as the baseline for subsequent analyses.

\subsubsection{Heterogeneity across Market Segments}

\begin{figure}[htbp]
\centering

% Panel A
\begin{minipage}{0.48\textwidth}
  \centering
  \includegraphics[width=\linewidth]{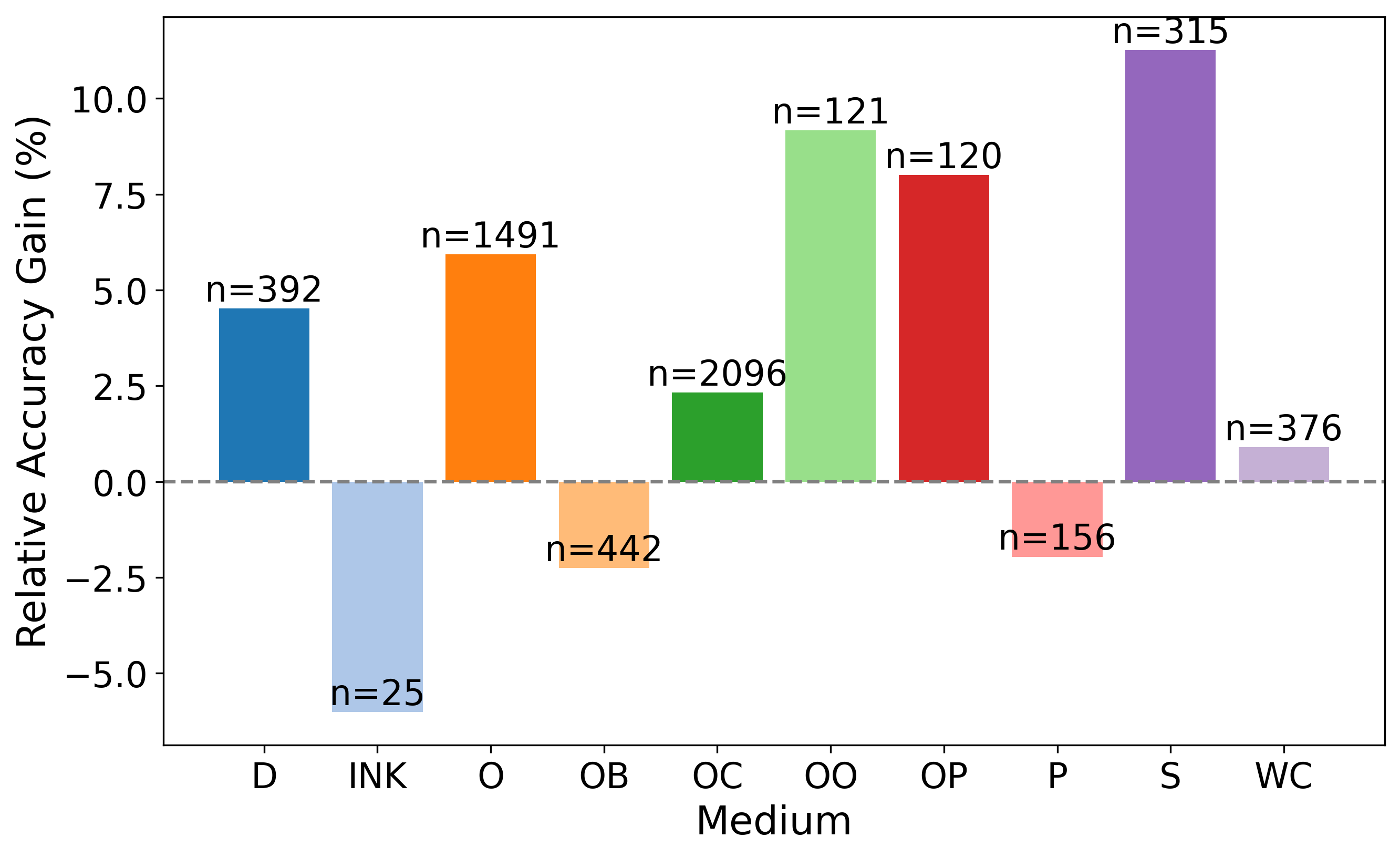}
  \subcaption{By medium}
  \label{fig:res_medium}
\end{minipage}%
\hfill
% Panel B
\begin{minipage}{0.48\textwidth}
  \centering
  \includegraphics[width=\linewidth]{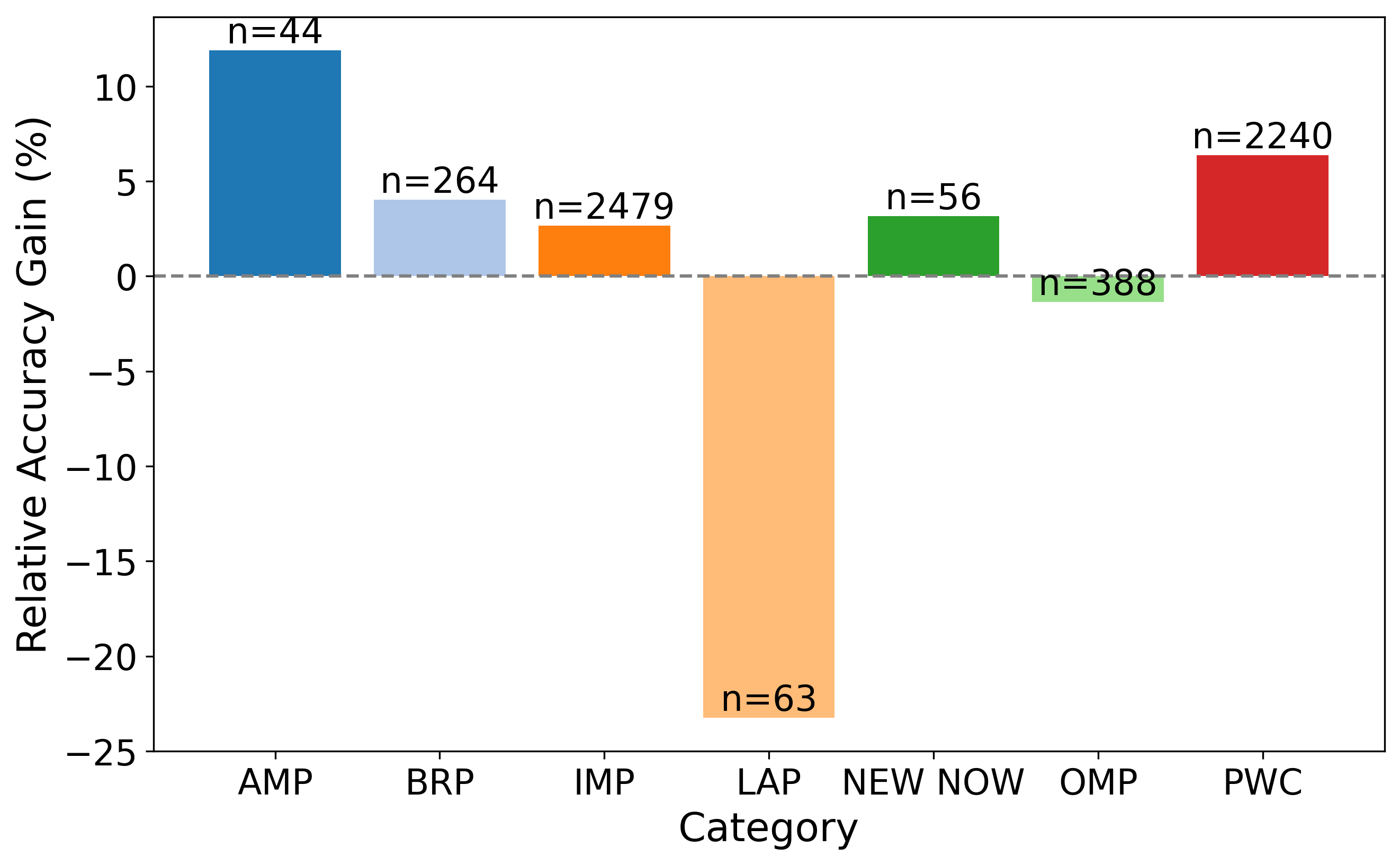}
  \subcaption{By category}
  \label{fig:res_category}
\end{minipage}

\caption{Relative accuracy gain from including images across subgroups}
\label{fig:subgroup_performance}
\end{figure}

Average performance can mask meaningful heterogeneity. We therefore quantify the incremental value of images within subgroups by computing the relative reduction in MAE when adding images to a tabular model:
\[
\frac{\text{MAE}_{\text{tab}} - \text{MAE}_{\text{img}}}{\text{MAE}_{\text{tab}}}.
\]
Figure~\ref{fig:subgroup_performance} reports this measure by medium and by category. The gains from images are largest in visually distinctive segments---such as American Paintings/Drawings and Sculptures, and Post-War and Contemporary---where style, composition, and surface properties plausibly carry valuation-relevant information not fully captured by metadata. Gains are also prominent for media where texture and material are central (e.g., oil-based works and sculpture). By contrast, some segments show little or even negative incremental gain, consistent with pricing being driven more by non-visual factors (e.g., provenance, institutional context, collector demand) than by pixel-level characteristics. As sample sizes vary across subgroups, these patterns should be interpreted cautiously.

\subsubsection{Predicting the Prediction Error}
The results above show that auction house estimates are extremely informative. This motivates a different question: can models explain when realized prices systematically deviate from the midpoint estimate? We therefore predict the auction house estimation error, defined as the log difference between realized prices and the midpoint estimate. Since the target is a log deviation, we evaluate models using MAE (in log units) alongside $R^2$, and we exclude previous transaction prices to keep the task comparable across fresh and repeated-sale works. Low and high estimates remain included as features by construction.

\begin{figure}[!ht]
    \centering
    \includegraphics[width=0.45\textwidth]{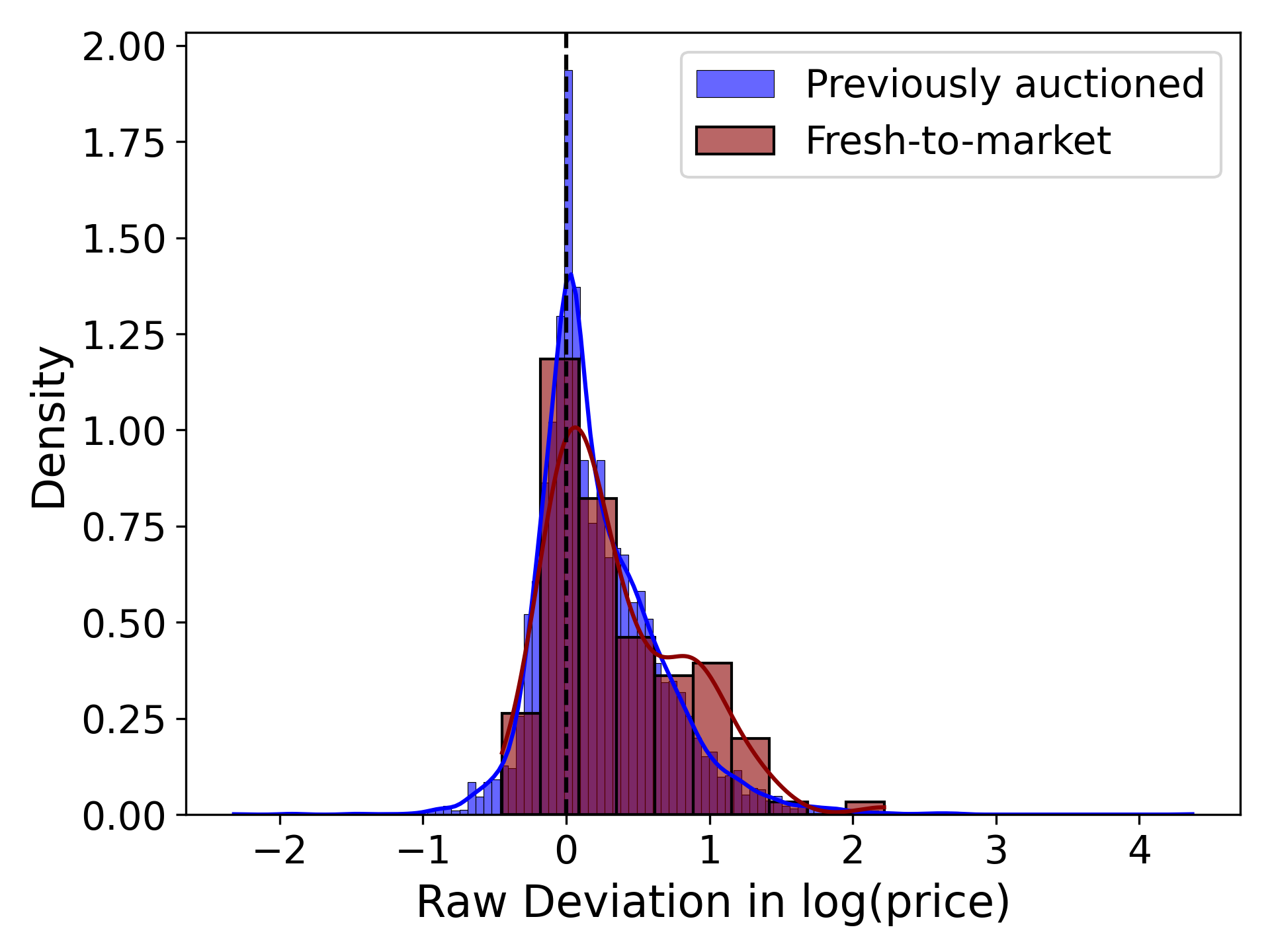}
    \caption{Auction house estimation error histogram}
    \label{fig:prediction_erors}
\end{figure}

\begin{figure}[!ht]
    \centering
    \includegraphics[width=0.5\textwidth]{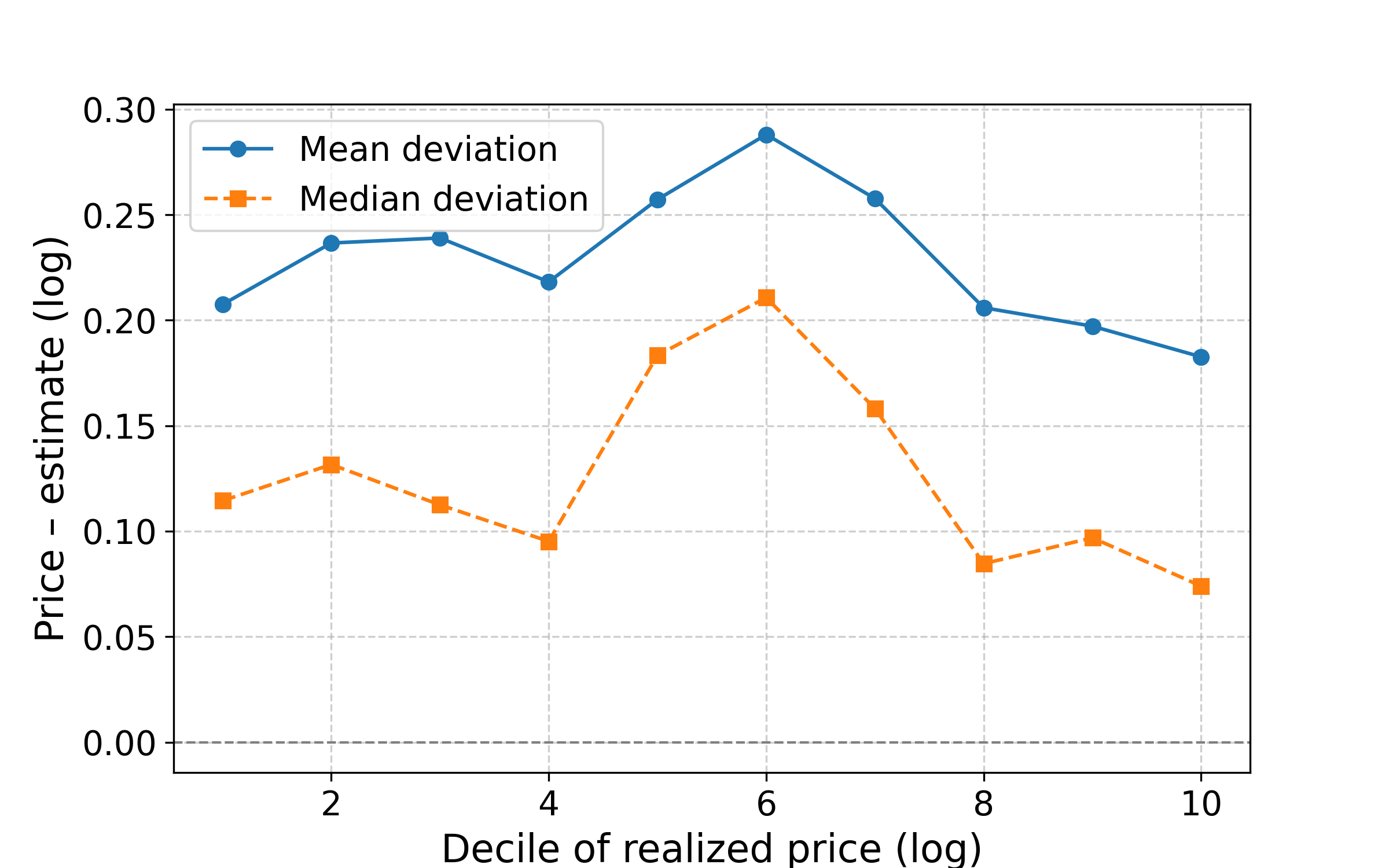}
    \caption{Estimation error per price decile}
    \label{fig:deciles_est_error}
\end{figure}

Figure~\ref{fig:prediction_erors} shows that estimation errors are tightly centered around zero, with fat tails and somewhat larger dispersion for fresh-to-market works. Figure~\ref{fig:deciles_est_error} further indicates a systematic pattern across the realized-price distribution: realized prices tend to exceed midpoint estimates across deciles, and the gap between mean and median deviations suggests right-skewness driven by a subset of strong overperformers.

Table~\ref{tab:error_prediction} reports predictive performance for this residual task. Overall, fit is modest: MAE is around $0.30$--$0.35$ log points and $R^2$ is generally below $0.15$. Multi-modal models provide at best small gains over tabular baselines. For fresh-to-market works, increasing $d_{\text{image}}$ tends to reduce stability and can worsen accuracy, consistent with the residual being noisy and only weakly predictable from observables. Taken together, these results suggest that most variation around auction house estimates is difficult to forecast, though predictability is somewhat higher for fresh-to-market works, where expert estimates are less tightly anchored to observable sale histories.

\begin{table}[H]
\centering
\scalebox{0.65}{
\begin{tabular}{lrcccccc}
\toprule
\textbf{Model} & \textbf{d\textsubscript{image}} & \textbf{MAE Total (\%)} & \textbf{MAE (Previous) (\%)} & \textbf{MAE (Fresh) (\%)} & \textbf{R\textsuperscript{2} Total} & \textbf{R\textsuperscript{2} (Previous)} & \textbf{R\textsuperscript{2} (Fresh)} \\
\midrule
\multicolumn{8}{l}{\textbf{Panel A: Multi-modal}} \\
\addlinespace

 Multi-modal ResNet-50   & 10  &  0.30 &  0.30 & 0.33 & 0.11   &  0.11 & 0.22  \\
 Multi-modal ResNet-50   & 100  &  0.31 & 0.31 & 0.34  & 0.11  & 0.11& 0.15  \\
 Multi-modal ResNet-50   & 1000  &  0.31 & 0.30 & 0.35 &  0.13 & 0.12 & 0.14   \\

 \addlinespace
 \multicolumn{8}{l}{\textbf{Panel B1: Tabular-only}} \\
\addlinespace
Neural network   & -  & 0.31  & 0.31  & 0.35  & 0.11  & 0.11  & 0.15  \\
XGBoost   & - & 0.33  & 0.33  & 0.36 &  0.08 & 0.08 & 0.18 \\
 \addlinespace
 \multicolumn{8}{l}{\textbf{Panel B2: Hedonic}} \\
\addlinespace
Hedonic (regularized)   & -  & 0.33  & 0.33 & 0.37 & 0.05   & 0.05  & 0.13 \\
Hedonic    & -  & 0.33  & 0.33 & 0.37 & 0.05   & 0.04  & 0.13 \\
\bottomrule
\end{tabular}}
\caption{Results for predicting estimation error}
\label{tab:error_prediction}
\end{table}

\subsection{Error Analysis}
Understanding where and why predictive models diverge from realized auction prices provides insights that complement performance metrics. In this section, we analyze errors both in comparison to auction house estimates and through a closer inspection of extreme model errors. Together, these perspectives highlight systematic biases and contextual factors that shape prediction accuracy.
\subsubsection{Predictions vs. Estimates}
In this analysis, we compare the predictions of an XGBoost model to the auction house's price estimates. Specifically, we investigate the difference between the model's predicted prices and the auction house average estimate. To understand which features explain this difference, we first use Lasso regression with cross-validation to select the most influential predictors. We then fit an Ordinary Least Squares (OLS) model on the top selected features to quantify their statistical significance (see \cite{belloni} for a detailed review of this procedure). Finally, we visualize residual distributions to assess the comparative accuracy of the model and the auction house estimates. For this analysis, we deliberately exclude auction house and artist identifiers as model features. These variables make up a large share of the available features and would dominate the feature selection process. More importantly, excluding them improves interpretability, as it allows us to highlight general drivers of differences between model predictions and auction house estimates rather than effects that are specific to individual artists or institutions.

Table \ref{tab:ols} shows the corresponding regression results. The OLS regression analyzes the difference between XGBoost model predictions and auction house price estimates  (in log terms) to identify systematic patterns in estimate deviations. While the model explains a very low  share of the variation (R² = 0.034), several predictors show statistically significant effects. For instance, the model tends to put a higher emphasis on the transaction year, implying a steady growth over time. Seasonal effects also emerge: months such as September and March are associated with larger estimates relative to model predictions, while July and December are linked to higher predictions. The coefficient of previous transaction prices suggests that when prior prices rise, auction houses are more reactive in their current estimates  than our model. Additionally, certain art market segments show consistent overestimation compared to model-based predictions. Missing location data  is also associated with lower model predictions. These findings point to potential biases in auction house pricing or model deficiencies that relate to time, context, and available historical data.

\begin{table}[!ht]
\centering

% First tabular
\resizebox{0.7\textwidth}{!}{%
\begin{tabular}{llll}
\toprule
\textbf{Dep. Variable:}           & $P_{\text{Model}} - P_{\text{Estimate}}$ & \textbf{R-squared:}      & 0.034   \\
\textbf{Model:}                   & OLS                                     & \textbf{Adj. R-squared:} & 0.032   \\
\textbf{Method:}                  & Least Squares                           & \textbf{F-statistic:}    & 19.43   \\
\textbf{No. Observations:}        & 5534                                    &                          &         \\
\bottomrule
\end{tabular}
}

\vspace{1em}

% Second tabular
\resizebox{0.7\textwidth}{!}{%
\begin{tabular}{lcccccc}
                                  & \textbf{coef} & \textbf{std err} & \textbf{t} & \textbf{P$>|$t$|$} & \textbf{[0.025} & \textbf{0.975]}  \\
\midrule
\textbf{const}                    & -0.3629  & 0.161     & -2.249  & 0.025  & -0.679 & -0.047 \\
\textbf{Transaction year}         &  0.5988  & 0.090     & 6.638   & 0.000  &  0.422 &  0.776 \\
\textbf{September}                & -0.2225  & 0.050     & -4.449  & 0.000  & -0.321 & -0.124 \\
\textbf{Previous price}           & -0.0238  & 0.004     & -5.401  & 0.000  & -0.032 & -0.015 \\
\textbf{December}                 &  0.1726  & 0.042     &  4.080  & 0.000  &  0.090 &  0.256 \\
\textbf{March}                    & -0.1159  & 0.027     & -4.267  & 0.000  & -0.169 & -0.063 \\
\textbf{Impressionism}            & -0.0768  & 0.020     & -3.922  & 0.000  & -0.115 & -0.038 \\
\textbf{July}                     &  0.1505  & 0.046     &  3.271  & 0.001  &  0.060 &  0.241 \\
\textbf{Unknown location}         & -0.3745  & 0.087     & -4.326  & 0.000  & -0.544 & -0.205 \\
\textbf{New Now}                  & -0.3016  & 0.095     & -3.167  & 0.002  & -0.488 & -0.115 \\
\textbf{Latin American Paintings} & -0.2652  & 0.091     & -2.920  & 0.004  & -0.443 & -0.087 \\
\bottomrule
\end{tabular}
}

\vspace{0.75em}

% Caption with controlled width
\captionsetup{width=0.7\textwidth}
\caption{OLS regression results for the difference between the auction house estimate and the XGBoost model prediction (in log space).}
\label{tab:ols}

\end{table}

\begin{figure}[H]
    \centering
    \includegraphics[width=0.6\textwidth]{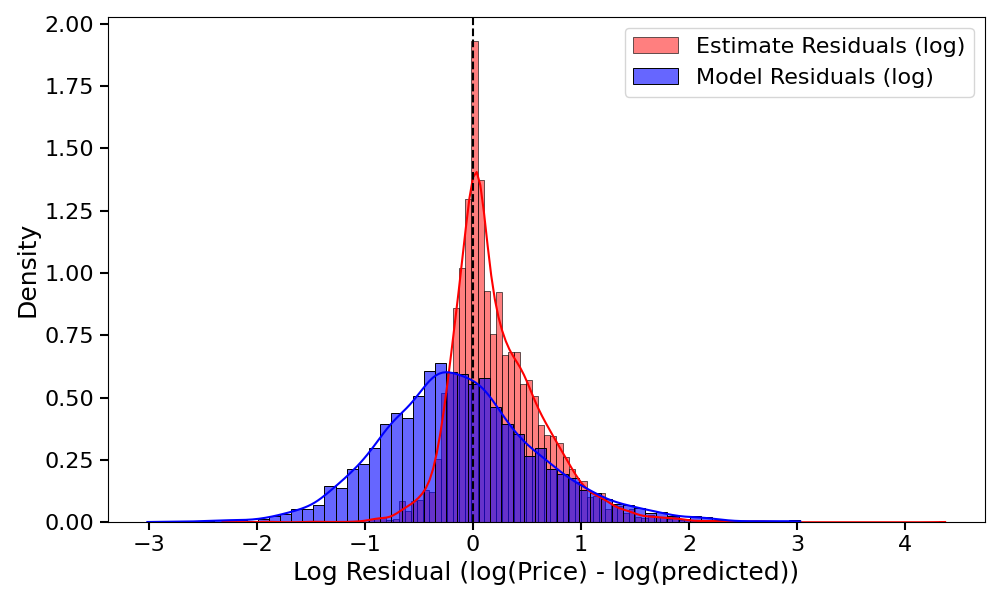}
    \caption{Histogram of XGBoost model and estimate residuals}
    \label{fig:dist_xgboost}
\end{figure}

The residual density plot in Figure \ref{fig:dist_xgboost} compares log-scale prediction errors from the auction house average estimates and the XGBoost model on the test set. The auction house residuals  are narrowly concentrated around zero, which reflects that realized prices tend to cluster closely around the published estimates. This likely illustrates the anchoring role of pre-sale guidance, which influences both bidder behavior and market expectations and  suggests that  auction houses provide estimates that are consistently close to outcomes. 
By contrast, XGBoost residuals are more dispersed. Importantly, the model’s distribution is also more symmetric, with both over- and underpredictions represented, whereas the auction house residuals show a sharper peak and less balance.

\subsubsection{Extreme Prediction Errors}
\begin{figure}[H]
    \centering
    \begin{subfigure}[b]{0.48\textwidth}
        \centering
        \includegraphics[width=\textwidth]{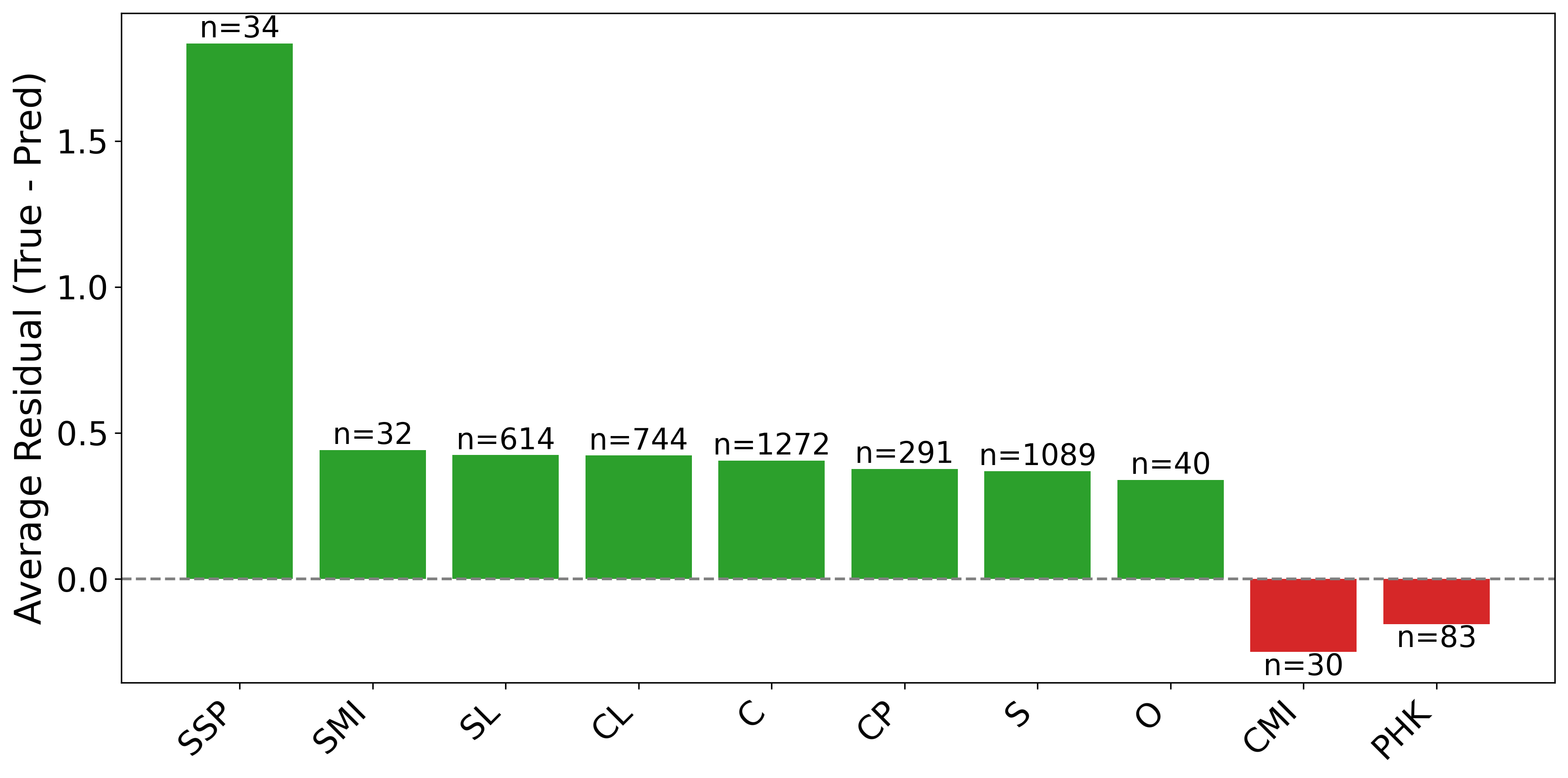}
        \caption{Auction house}
        \label{fig:transaction_house_extreme}
    \end{subfigure}
    \hfill
    \begin{subfigure}[b]{0.48\textwidth}
        \centering
        \includegraphics[width=\textwidth]{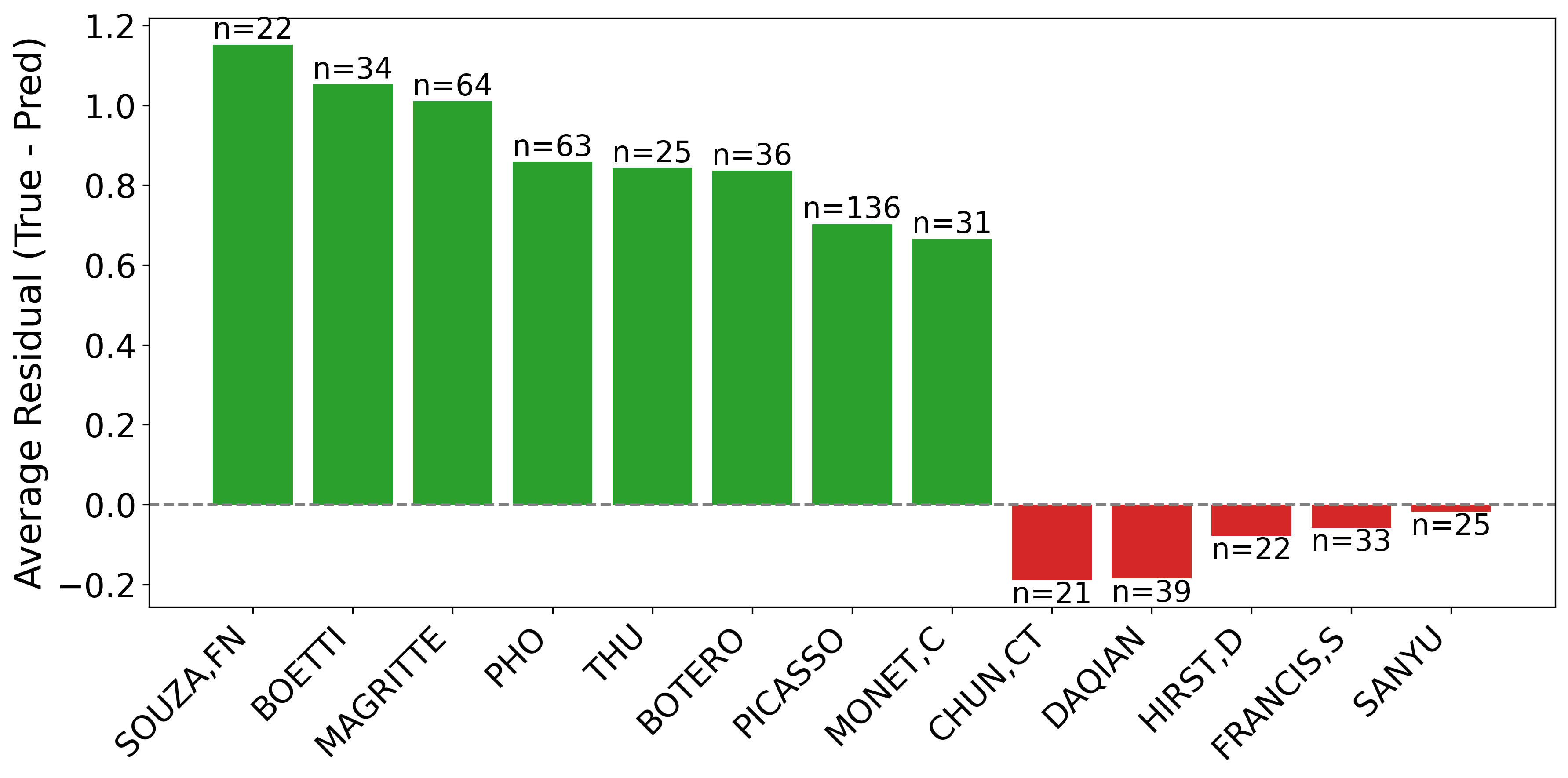}
        \caption{Artist}
        \label{fig:medium_extreme}
    \end{subfigure}
    
    \vspace{0.3cm}
    
    \begin{subfigure}[b]{0.48\textwidth}
        \centering
        \includegraphics[width=\textwidth]{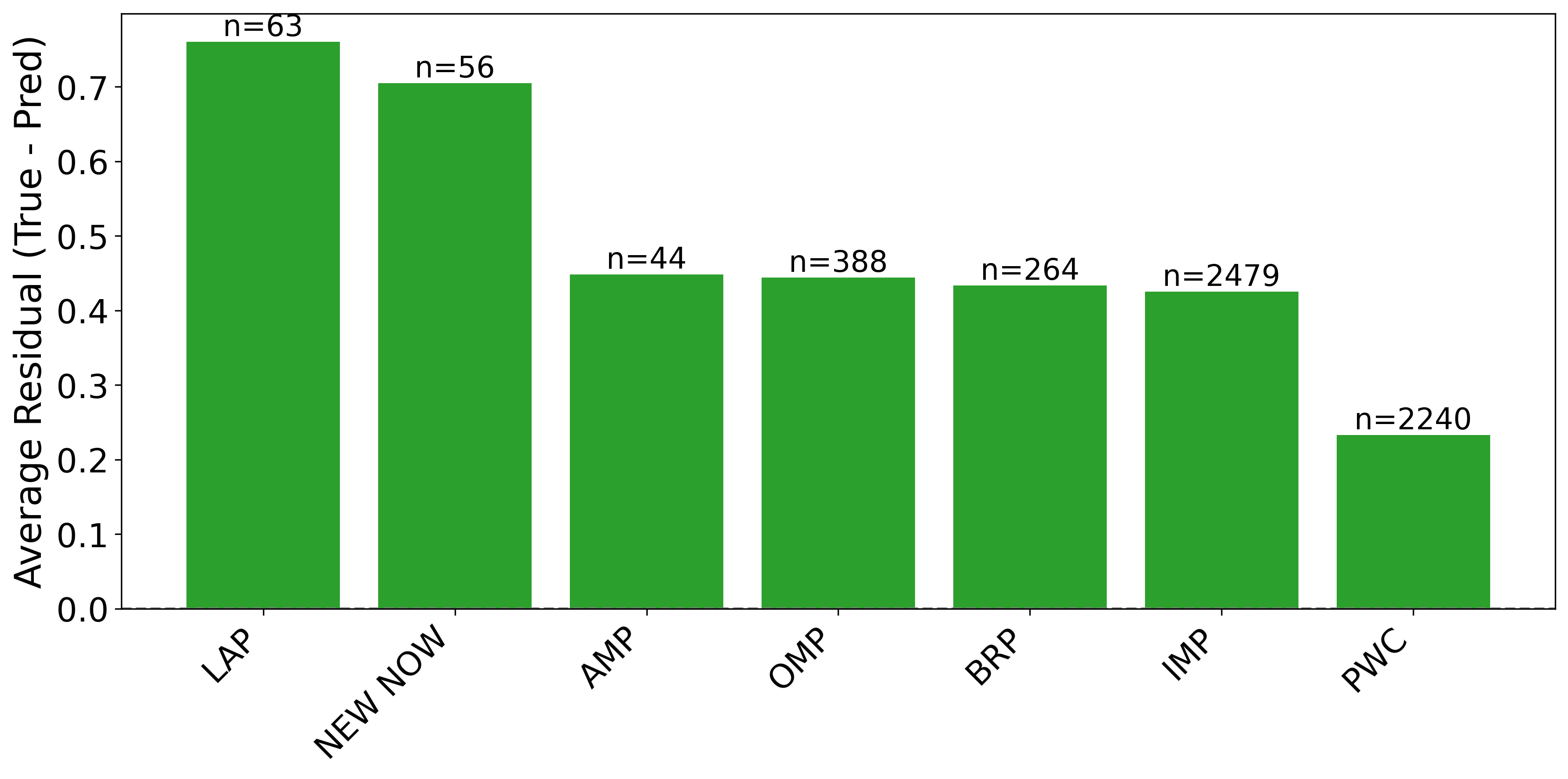}
        \caption{Category}
        \label{fig:category_extreme}
    \end{subfigure}
    \hfill
    \begin{subfigure}[b]{0.48\textwidth}
        \centering
        \includegraphics[width=\textwidth]{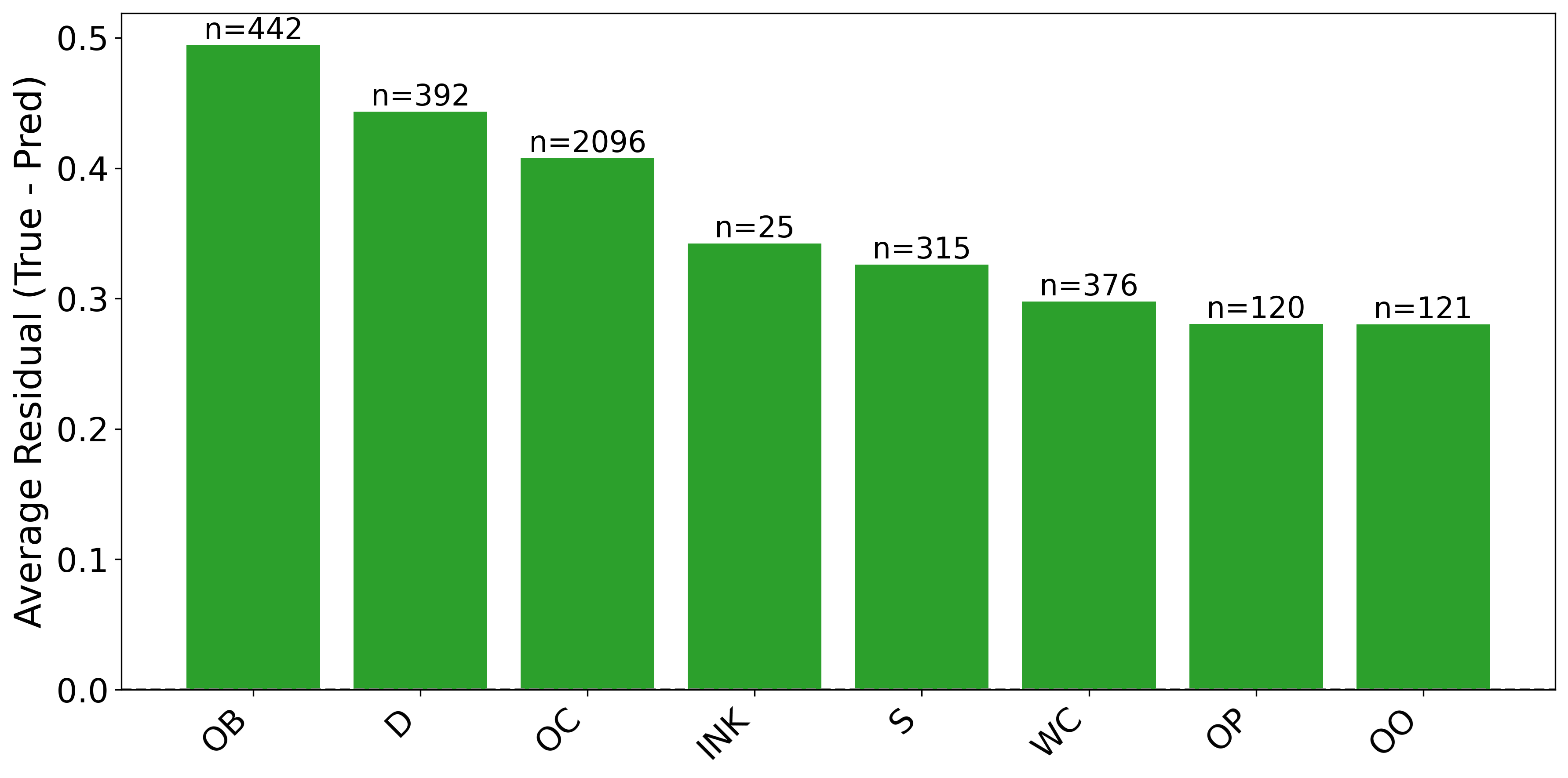}
        \caption{Medium}
        \label{fig:artist_extreme}
    \end{subfigure}
    
    \caption{Top under- and overpredicted groups by residual bias across different dimensions}
    \label{fig:extreme_groups}
\end{figure}

In this section, we examine systematic prediction errors by analyzing the average residuals across different grouping dimensions such as auction house, artist, category, and medium (Figure~\ref{fig:extreme_groups}). Instead of focusing on individual sales, this approach highlights where the model consistently under- or overestimates prices at the group level. By comparing average errors across these categories, we can better understand the boundaries of model performance and identify areas where predictions are shaped by market forces not fully captured in the training data. For this analysis, we rely on the multi-modal approach with a rich image embedding size of 1000. 

Residual bias differs substantially across dimensions. Artist-level errors show the greatest variance. This reflects the heterogeneity of reputations, collector bases, and market visibility. Transaction houses and media exhibit more moderate spreads, consistent with the idea that institutional and segment-level effects are relatively stable and easier to model. Frequent categories such as Post-War and Contemporary Art are fairly well calibrated, while those with fewer observations exhibit stronger bias.

Clear patterns also emerge \emph{within}  these dimensions. At the transaction house level, underpredictions are most pronounced for prestigious venues such as Sotheby’s and Christie’s, suggesting that competitive bidding and the signaling value of high-profile auctions systematically push realized prices above feature-based predictions. Among artists, the most underpredicted groups are often established, well-known names, which indicates that the model is unable to fully catch collector enthusiasm and trends in this segment. Conversely, overpredictions tend to occur for less prominent artists or segments with weaker demand signals.

Taken together, these results show that extreme errors are not random but clustered in predictable ways. Systematic underpredictions are associated with high-profile contexts where reputation plays a central role, while overpredictions appear in lower-demand segments. This highlights both the limits of structured and image features alone and the potential value of incorporating contextual or reputational signals—factors often better understood by human experts—to reduce systematic bias and improve predictive robustness.

\subsection{Feature Importance}
\subsubsection{Model Robustness}

The interpretation of feature impact remains one of the major shortcomings of modern machine learning approaches. Even though interpretation methods exist for a vast number of models, they are usually not straightforward.
Unlike classic linear models, where coefficients can be analyzed directly, weights in tree-based methods or deep learning models often carry little information. 
To uncover which factors most strongly influence our price predictions, we compare two complementary feature importance techniques with respect to our XGBoost model. The first makes use of XGBoost’s native gain metric, which measures the average reduction in variation by each feature's split across trees.  However, research shows that  this method tends to favor high‑cardinality or continuous features, since they provide more opportunities for splits, even if they aren’t truly informative \citep{zhang2023unbiasedgradientboostingdecision}.  In contrast, permutation importance (PFI) directly quantifies each feature’s practical predictive value by permuting its features in a test set and measuring the resulting performance decrease (as measured in MSE). This  approach captures both main effects and interactions but can underestimate importance for correlated features \citep{flora2022comparingexplanationmethodstraditional}. By applying both methods and comparing top-ranked features, we obtain a more robust and nuanced understanding: gain reveals structural significance within the model, while PFI validates which features drive prediction performance.

\begin{figure}[!ht]
\centering
\resizebox{0.64\textwidth}{!}{% scale entire figure environment
\begin{minipage}{\textwidth}
  \begin{subfigure}{0.48\textwidth}
    \centering
    \includegraphics[width=\linewidth]{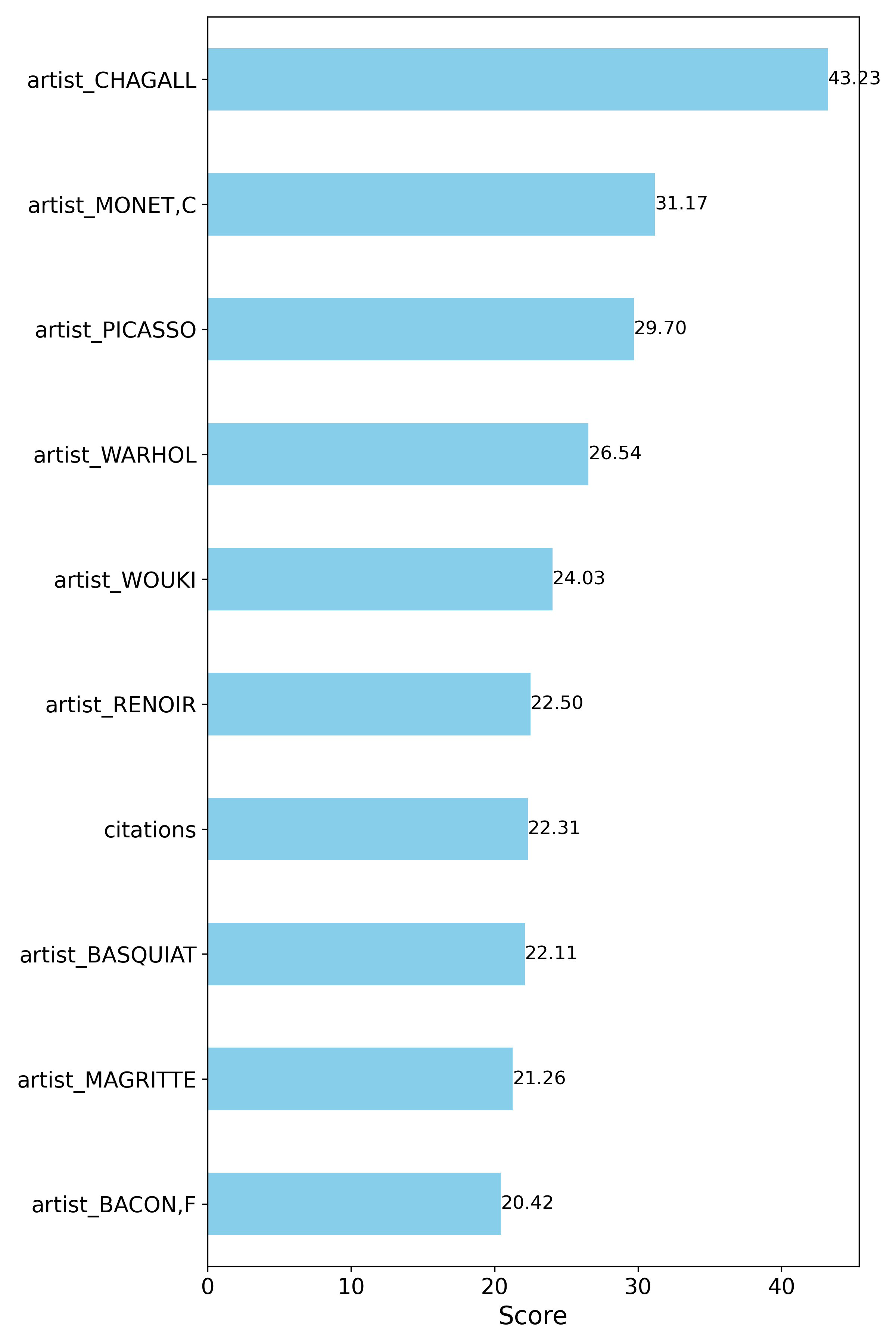}
    \caption{XGBoost gain importance}
    \label{fig:gain_imp}
  \end{subfigure}
  \hfill
  \begin{subfigure}{0.48\textwidth}
    \centering
    \includegraphics[width=\linewidth]{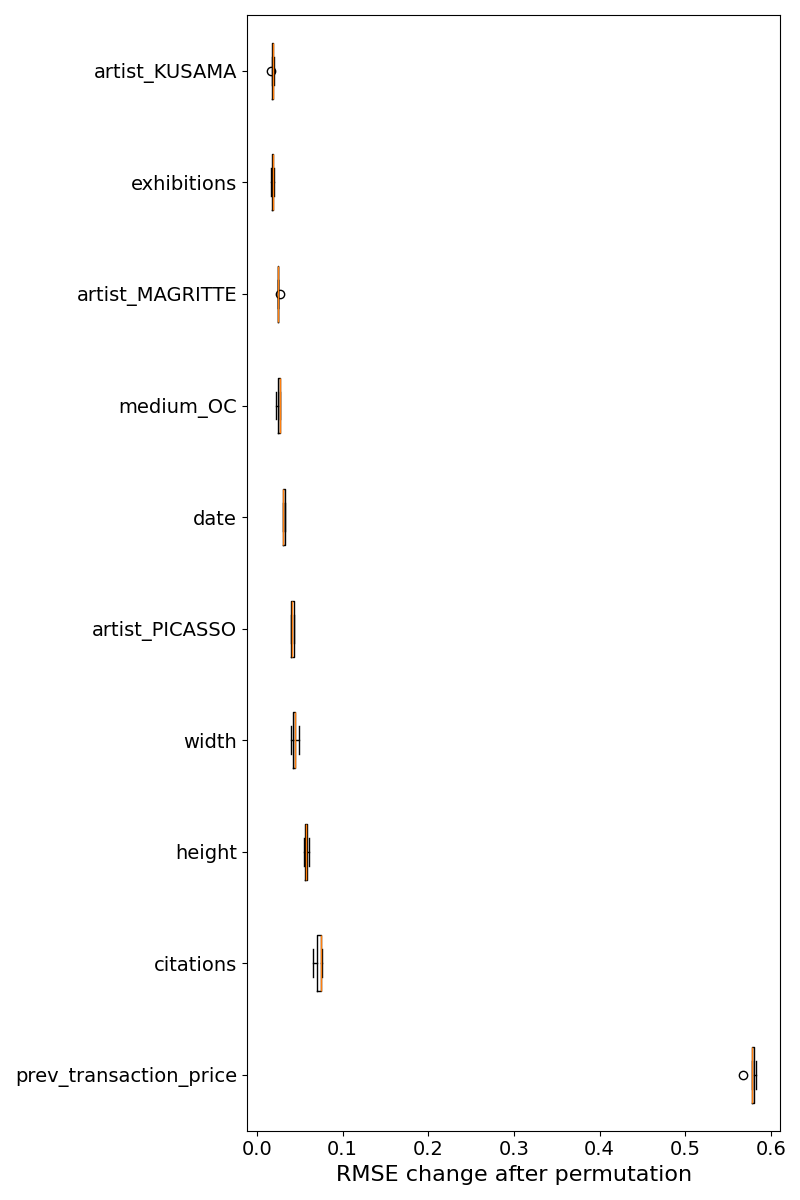}
    \caption{Permutation feature importance}
    \label{fig:perm_imp}
  \end{subfigure}
\end{minipage}
}% end resizebox
\caption{Comparison of feature importance methods.}
\label{fig:feature_importance_comparison}
\end{figure}

Figure \ref{fig:feature_importance_comparison} presents the  most influential predictors according to gain and permutation feature importance. The gain ranking highlights a concentration of artist dummies among the top variables, indicating that the model frequently uses artist identity to structure splits. In contrast, the permutation analysis demonstrates that the most substantial driver of predictive accuracy is the previous transaction price, which alone accounts for the largest deterioration in model performance when shuffled. Other factors, such as artwork dimensions, exhibitions, citations, and a small set of artist identifiers, exert only a moderate influence. Taken together, these results suggest that while artist-related categories shape the internal structure of the model, they add relatively little incremental predictive power. By far the strongest signal for pricing unseen artworks comes from historical market data, particularly previous sale prices, with physical and contextual characteristics playing a secondary role.

\subsubsection{Ablation Study}
To complement model training and feature importance analysis, we design a simple ablation study. While permutation feature importance (PFI) provides a granular ranking of individual predictors, it can be biased by correlations between features and does not directly show how groups of related variables contribute collectively. In contrast, ablation evaluates the impact of removing entire feature groups and comparing predictive performance to a full-feature baseline. We group variables by type, for instance, object-specific characteristics, artist- and provenance-related information, and drop them together. This approach provides an interpretable, group-level perspective on which dimensions of the data are most influential, which makes it a valuable analysis. We compare the following models:

\begin{enumerate}
  \item \textbf{Baseline model} – Train the model with the full feature set  and record performance metrics.  
  \item \textbf{Remove object-level attributes} – Drop height, width, shape, medium, citations, exhibitions, signed, dated, category   to test the importance of physical characteristics.  
  \item \textbf{Remove artist-level  features} – Drop artist  to assess the contribution of reputation.
  \item \textbf{Minimal transaction model} – Train on artist, transaction year, and previous transaction flag only, to evaluate how far predictive power can be sustained by a reduced set of variables.  
\end{enumerate}

This methodology follows the approach by \cite{aubry}. However, due to our repeated-sales data structure, we are able to assess the importance of these feature groups on both fresh-to-market and previously auctioned samples and are hence able to check whether their predictive power differs across these two segments. In particular, this allows us to examine whether object-level characteristics carry more weight when an artwork appears at auction for the first time, and whether artist reputation dominated in cases where an observable auction history exists. This extension provides a richer understanding of how the drivers of pricing vary depending on an artwork’s market trajectory. 
\begin{table}[H]
\centering
\scalebox{0.75}{
\begin{tabular}{lcccc}
\toprule
& \multicolumn{2}{c}{d\textsubscript{image} = 1000} & \multicolumn{2}{c}{d\textsubscript{image} = 10} \\
\cmidrule(lr){2-3} \cmidrule(lr){4-5}
\textbf{Model} & \textbf{R\textsuperscript{2} (Previous)} & \textbf{R\textsuperscript{2} (Fresh)} & \textbf{R\textsuperscript{2} (Previous)} & \textbf{R\textsuperscript{2} (Fresh)} \\
\midrule

Baseline (all features)                  &  0.78 & 0.67  &  0.76  & 0.64  \\
Remove object-level attributes           &  0.67 & 0.47  &  0.70 & 0.48   \\
Remove artist-level features             &  0.63 & 0.40  &  0.65 & 0.39   \\
Minimal model      & 0.68  & 0.36   & 0.70  & 0.46  \\

\bottomrule
\end{tabular}}
\caption{Ablation study: predictive performance (R\textsuperscript{2}) by feature group}
\label{tab:ablation_results}
\end{table}
Table \ref{tab:ablation_results} shows that dropping artist- or object-level features leads to sharp declines in explanatory power, especially for fresh-to-market works (e.g., R\textsuperscript{2} falls from 0.67 to 0.40 when artist information is removed with d\textsubscript{image}=1000). For repeated sales, predictive performance remains higher overall, but the minimal model relying only on artist, year, and a repeated-sale flag still achieves R\textsuperscript{2} $\approx$ 0.68-0.7. By contrast, the same model performs poorly for fresh-to-market works, which underlines the importance of artist reputation and object characteristics when no price anchor is available. The particularly strong drop in R\textsuperscript{2} for fresh-to-market works in the minimal model with d\textsubscript{image}=1000 suggests that high-dimensional image embeddings interact closely with artist- and object-level features. When these are removed, the model is left with little structured information and tends to overfit noise in the 1000-dimensional space, leading to poor generalization. In contrast, the lower-dimensional space (10-d) provides a coarser but more stable representation, which explains why the decline is less extreme in this setting.

Compared to \cite{aubry}, our richer image embedding space (1000 vs. 10 dimensions) improves absolute performance but confirms the same qualitative pattern: the drivers of pricing differ fundamentally depending on whether an artwork is new to the auction market.

\subsection{Visual Attribution to Pricing}

Prediction accuracy alone does not reveal which \emph{visual} cues the model uses. This is particularly important in a multi-modal setting, where rich metadata may substitute for, or interact with, image content. We therefore complement the performance results with two interpretability exercises. First, we use Grad-CAM to obtain \emph{local} explanations at the level of individual images, highlighting regions that are most influential for a given prediction. Second, we study the \emph{global} geometry of learned image representations by projecting image embeddings into two dimensions and examining whether the visual branch organizes artworks in economically meaningful ways.

\subsubsection{The Grad-CAM Method}
Deep neural networks do not yield coefficients that can be read like elasticities, and attribution is further complicated when tabular and image inputs enter jointly. We use Gradient-weighted Class Activation Mapping (Grad-CAM) \citep{Selvaraju_2017_ICCV} to visualize which parts of an artwork drive the prediction through the ResNet-50 image encoder. The key idea is to identify spatial regions in the final convolutional layer whose activations have the largest positive influence on the predicted log price.

Let $y$ denote the predicted log price and let $A^k \in \mathbb{R}^{H\times W}$ be the $k$-th feature map in the final convolutional layer. Grad-CAM proceeds as follows. First, compute the gradient of the output with respect to each spatial activation,
\[
\frac{\partial y}{\partial A^k_{ij}}.
\]
Second, aggregate these gradients across the spatial dimension to obtain a scalar importance weight for each channel,
\[
\alpha_k = \frac{1}{Z}\sum_{i}\sum_{j}\frac{\partial y}{\partial A^k_{ij}}, \qquad Z = H\times W.
\]
Third, form a weighted sum of feature maps and apply a rectifier,
\[
L_{\text{Grad-CAM}} = \text{ReLU}\!\left(\sum_k \alpha_k A^k\right),
\]
so that the heatmap highlights regions that contribute positively to the prediction. Finally, $L_{\text{Grad-CAM}}$ is upsampled to the input resolution and overlaid on the original artwork.

We implement Grad-CAM using the library of \cite{jacobgilpytorchcam} and report representative heatmaps for test-set images.\footnote{Additional Grad-CAM examples are reported in Appendix \ref{supp:visual}.} To isolate how structured features affect the reliance on images, we compare two model variants: an image-only baseline augmented with transaction year and a previous-sale availability flag, and the full multi-modal specification. The comparison is informative because it tests whether adding rich metadata reduces the need to extract price-relevant information from the image (a substitution margin), or instead sharpens attention to specific visual cues (a complementarity margin).

\begin{figure}[H]
    \centering
    \includegraphics[width=\textwidth, height= \textheight, keepaspectratio]{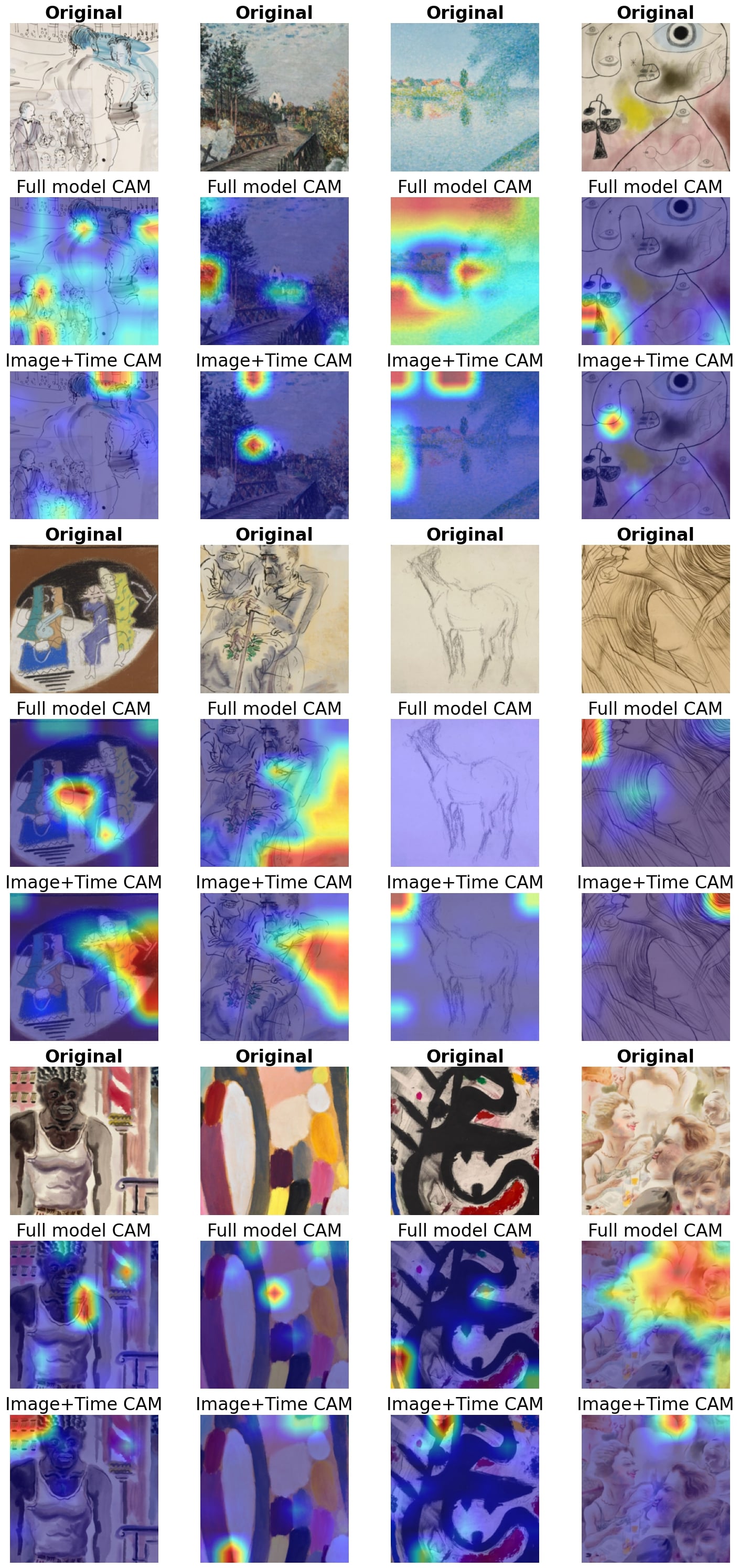}
    \caption{Grad-CAM visualizations for the small and feature-rich multi-modal models}
    \label{fig:gradcam}
\end{figure}

Figure~\ref{fig:gradcam} shows the resulting heatmaps. The patterns vary across works. In some cases, the full multi-modal model concentrates attention on visually salient regions (e.g., prominent subjects, high-contrast boundaries, or areas with distinctive texture), whereas the restricted model produces more scattered activations. In other cases, both variants attend to similar regions or generate diffuse maps, suggesting that attribution is sensitive to the particular work and that the role of the image branch is not uniform across the test set. A recurring caveat is that heatmaps sometimes emphasize visually simple regions (e.g., large monotone backgrounds), consistent with the possibility of shortcut learning: such regions may correlate with artist identity, medium, period, or photographic style rather than capturing economically meaningful aesthetics. Overall, Grad-CAM provides a useful diagnostic for whether the visual backbone relies on localized cues or broad, potentially spurious patterns, but it should be interpreted as suggestive rather than definitive evidence of causal visual drivers.

\subsubsection{Image Embeddings Interpretation}
Grad-CAM is inherently local: it explains one prediction at a time and does not directly reveal whether the model learns a coherent visual representation across the market. In order to understand the model's visual perception on a broader level (e.g., a style level such as Impressionism), we study the learned image embeddings produced by the ResNet-50 branch and ask whether they encode economically relevant similarity. Concretely, we take the $d_{\text{image}}=1000$ embedding vector for each artwork and project embeddings to two dimensions using Principal Component Analysis (PCA) \citep{jolliffe2016}. PCA identifies orthogonal directions that explain maximal variation, providing a low-dimensional view of how the model organizes images.

The purpose of this analysis is to test whether the multi-modal ResNet embeddings genuinely capture visual information from pixel data, or whether they merely reproduce categorical clusters already present in the structured features, essentially creating a “shortcut” without leveraging image content. By visualizing the embeddings, we can evaluate the extent to which the learned representation reflects meaningful visual similarity, art-historical distinctions, or market-relevant signals. We systematically compare two configurations of the multi-modal model (d\textsubscript{image}=1000)\footnote{In Appendix \ref{supp:embedding}, we provide the same visualization for ViT-Small. We observe similar patterns as with the ResNet model.}: 
\begin{itemize}
    \item \textbf{Feature-rich} Model with all relevant features
    \item \textbf{Minimal} Model with category, medium and transaction year as features
\end{itemize}

This setup allows us to further examine how the image model positions artworks in the embedding space under different amounts of contextual information. The comparison between the feature-rich and minimal models shows whether the inclusion of metadata shifts the embeddings toward groupings that align with market-relevant categories (such as artistic movements or media), or whether the visual model on its own emphasizes purely visual similarity. This provides a broader perspective on how visual features interact with non-visual attributes in shaping the model’s representation of artworks.

It is important to note that in a multi-modal setting, the resulting embedding space does not need to reproduce neat stylistic or material clusters. The image representations are optimized jointly with tabular features to predict auction prices, rather than to classify artworks by category or medium. As a result, the learned embeddings may reflect economic similarities or complementary price-relevant signals rather than strict art-historical boundaries. In this sense, the analysis of clusters serves primarily as an interpretability exercise.

\begin{figure}[ht!]
    \centering
    
    % --- Row 1: Category ---
    \begin{subfigure}[t]{0.48\textwidth}
        \centering
        \includegraphics[width=\linewidth]{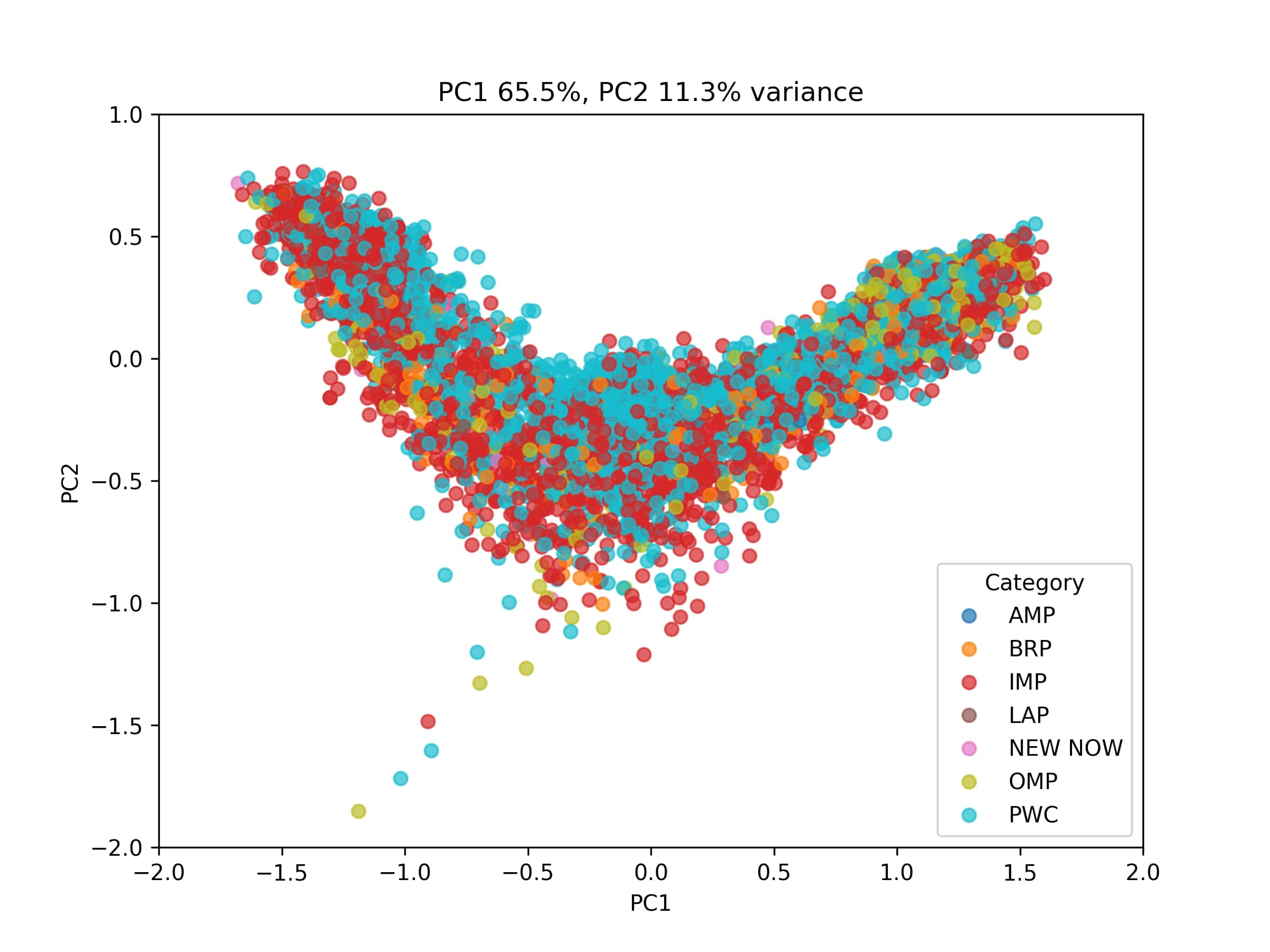}
        \caption{Feature-rich (category)}
        \label{fig:pca_cat_rich}
    \end{subfigure}
    \hfill
    \begin{subfigure}[t]{0.48\textwidth}
        \centering
        \includegraphics[width=\linewidth]{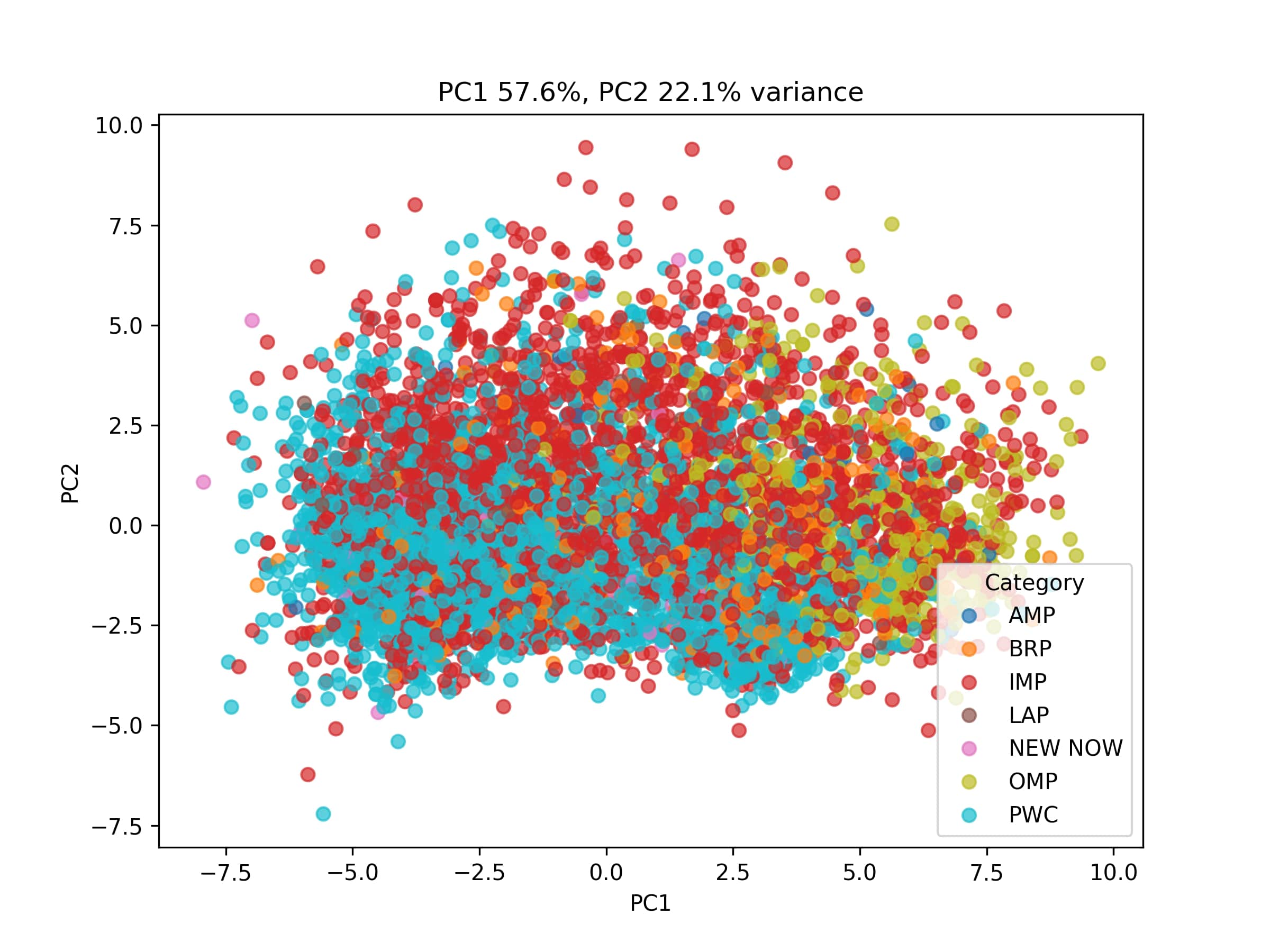}
        \caption{Minimal (category)}
        \label{fig:pca_cat_minimal}
    \end{subfigure}
    
    % --- Row 2: Medium ---
    \begin{subfigure}[t]{0.48\textwidth}
        \centering
        \includegraphics[width=\linewidth]{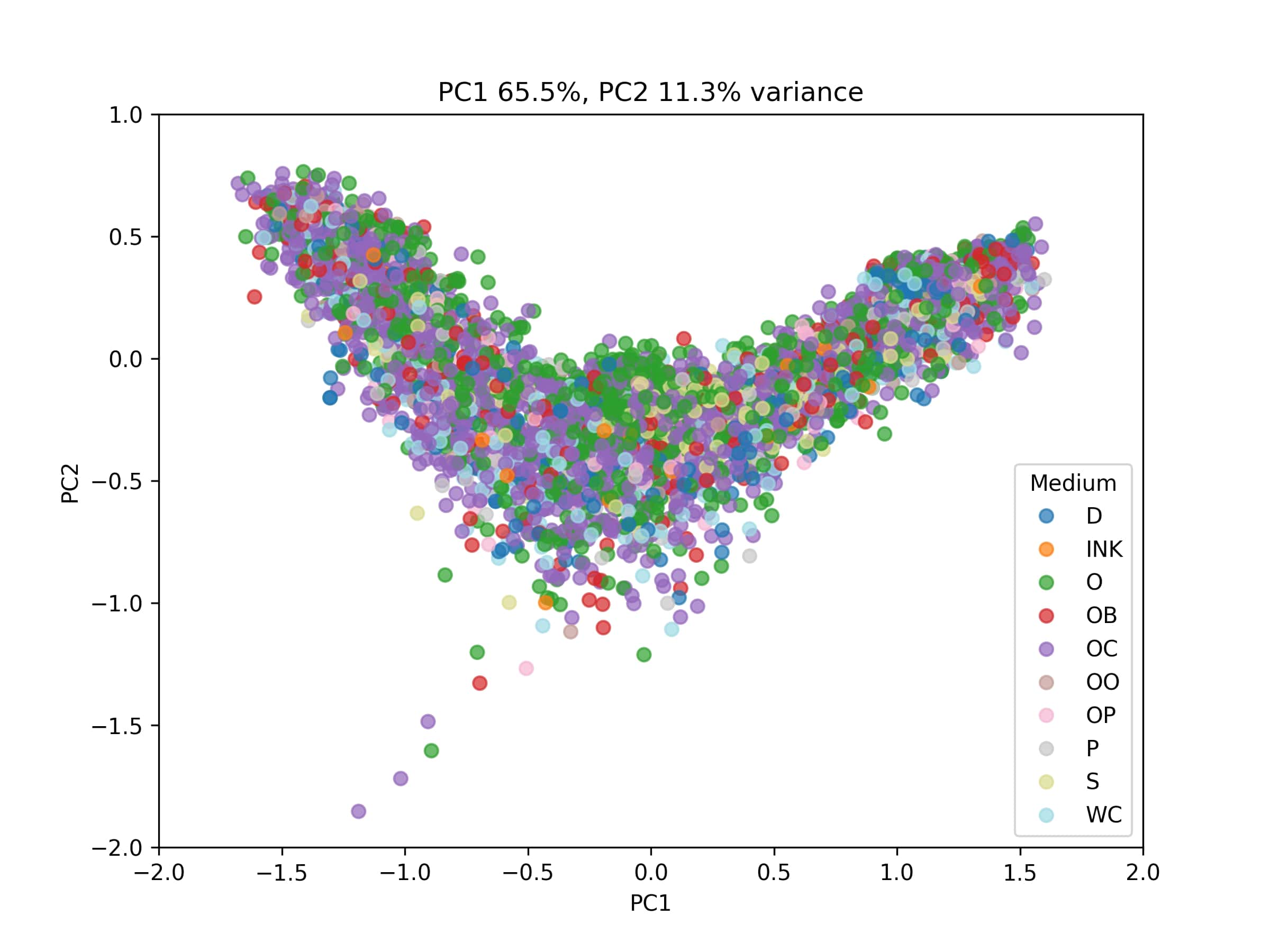}
        \caption{Feature-rich (medium)}
        \label{fig:pca_med_rich}
    \end{subfigure}
    \hfill
    \begin{subfigure}[t]{0.48\textwidth}
        \centering
        \includegraphics[width=\linewidth]{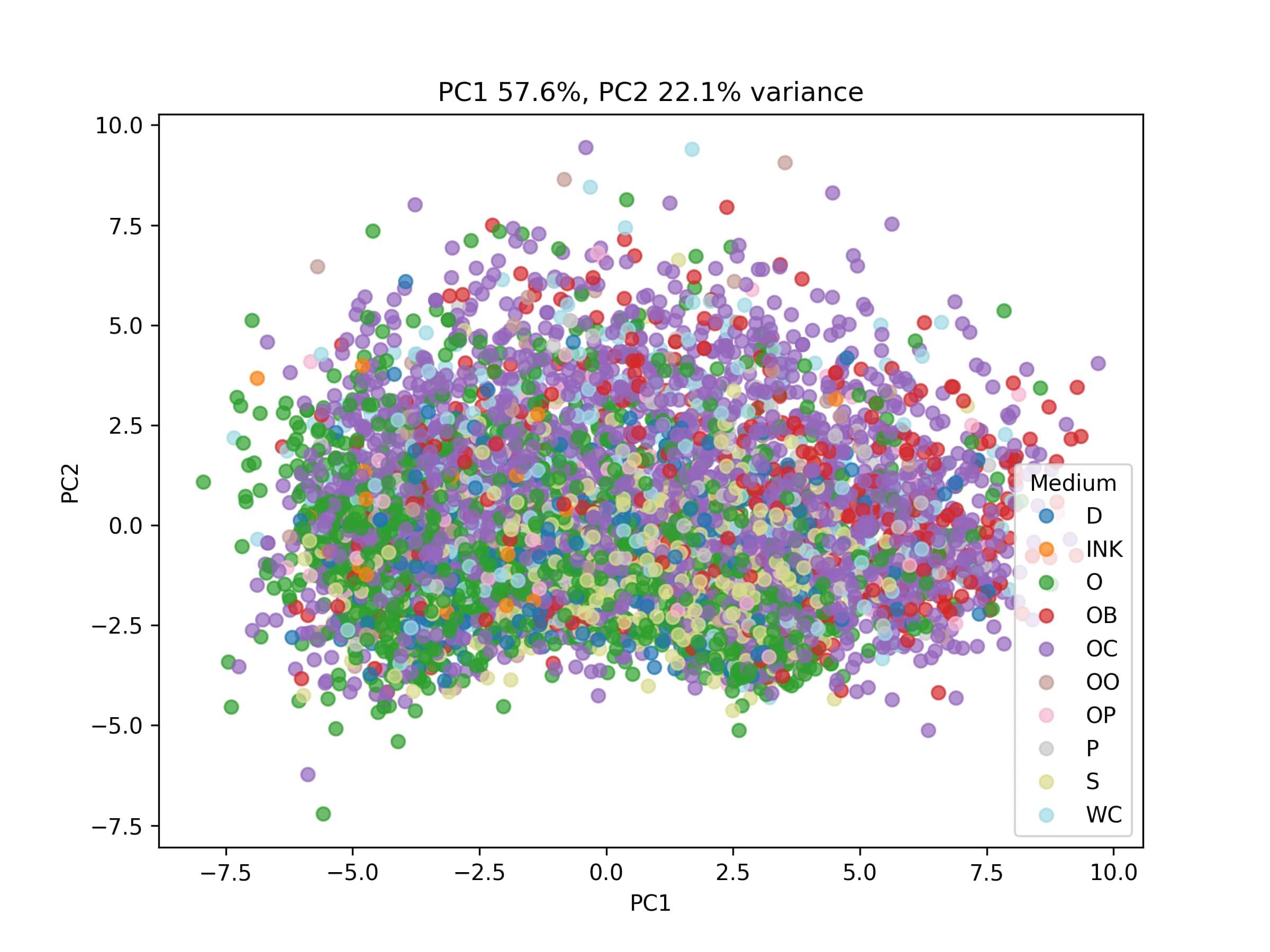}
        \caption{Minimal (medium)}
        \label{fig:pca_med_minimal}
    \end{subfigure}

    % --- Row 4: Predicted Price ---
    \begin{subfigure}[t]{0.48\textwidth}
        \centering
        \includegraphics[width=\linewidth]{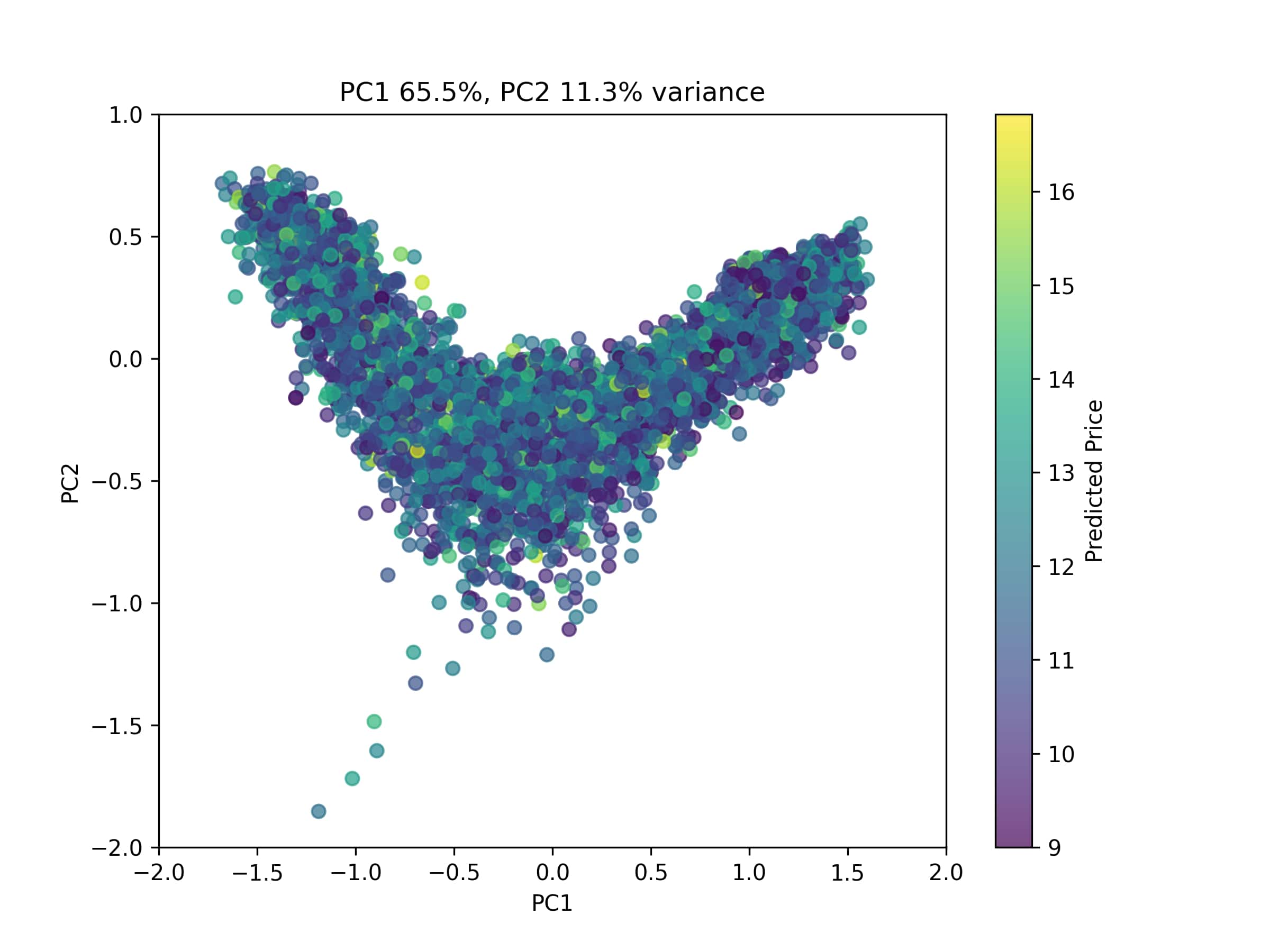}
        \caption{Feature-rich (predicted price)}
        \label{fig:pca_pred_rich}
    \end{subfigure}
    \hfill
    \begin{subfigure}[t]{0.48\textwidth}
        \centering
        \includegraphics[width=\linewidth]{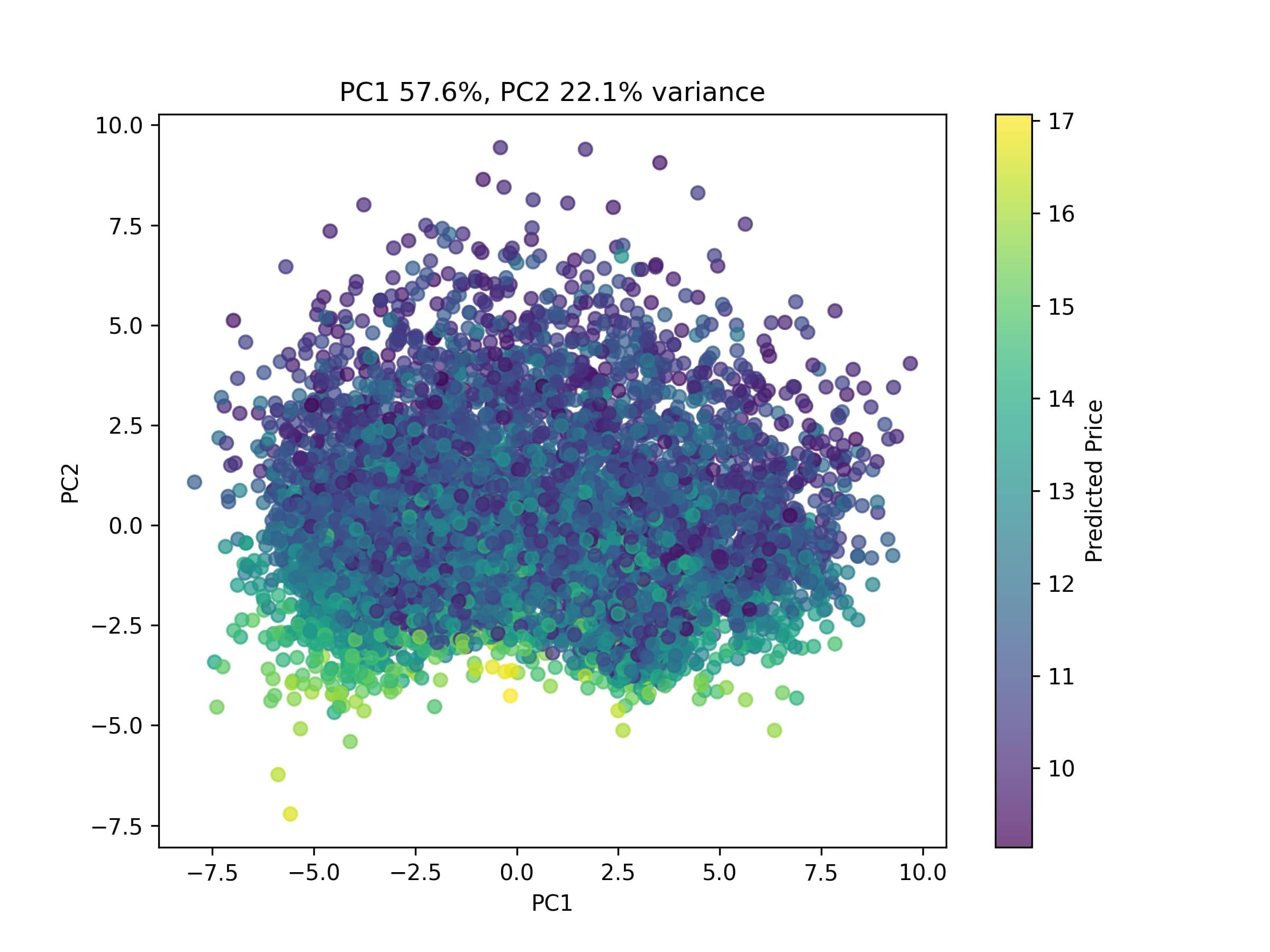}
        \caption{Minimal (predicted price)}
        \label{fig:pca_pred_minimal}
    \end{subfigure}
    
    \caption{PCA image embedding projections (ResNet-50 Model)}
    \label{fig:pca_comparison}
\end{figure}

The PCA projections of image embeddings in Figure \ref{fig:pca_comparison} highlight how the reliance on visual information shifts with model specification. In the feature-rich model, which combines over 700 structured predictors with image features and a corresponding 1000-dimensional output dimension, the embeddings form diffuse clouds with little visible separation by category or medium. The variance is highly concentrated in the first two components, with PC1 alone explaining 65.5\% and PC2 11.3\%. This indicates that most of the variation in the embedding space collapses into a single dominant direction, which reflects that the visual branch plays a relatively limited, complementary role once detailed tabular information is available.

By contrast, in the minimal model, where only category, medium, and transaction year are included as structured features, the image embeddings spread more broadly across the PCA space. Here, PC1 explains 57.6\% and PC2 22.1\%, a more balanced distribution that points to a stronger influence of the image branch. Nevertheless, the fact that two components already capture nearly 80\% of the variance suggests that the signal remains shallow: while the embeddings contain more visual information, they do not form sharply defined clusters.
Figure \ref{fig:pca_comparison} also plots the embedding spaces categorized by predicted prices. In the minimal model, more structure is visible, with higher- and lower-priced works spreading along clear gradients. In the feature-rich model, this separation is far less pronounced, as pricing information is largely absorbed by the structured predictors rather than the image embeddings.

The 2-dimensional projections themselves show some faint structure. For instance, media such as sculpture or watercolor occasionally appear more distinct, and categories like Old Masters separate partially at the margins, yet the overall embedding space shows substantial overlap. Oil-based media in particular are almost entirely blended, and categories such as Post-War and Impressionist intermingle without clear boundaries. This suggests that the embeddings capture subtle cues that support pricing, rather than reproducing clean categorical distinctions.

Taken together, these findings highlight that visual embeddings provide weak but complementary signals, becoming more influential when structured predictors are scarce, while serving mainly as supportive information in richer model configurations. It also underlines that models do not merely generate embeddings that serve as shortcuts for categorical features.

% \iffalse
\subsection{Combining Auction House Estimates with Machine Learning}

Auction house estimates are highly informative benchmarks, and our earlier results show that they outperform machine learning models trained only on structured and image data. This raises an important question: can machine learning contribute anything beyond what expert forecasts already capture? To address this, we construct a two-stage ensemble framework that explicitly integrates both sources of information.

In the first stage, we train a baseline XGBoost model using all features except the auction house’s low and high estimates. Predictions are generated via K-fold cross-validation, yielding out-of-fold estimates for the training set and stable averages for the test set. These predictions capture purely data-driven pricing signals based on metadata (artist, medium, size, date, etc.). In the second stage, we combine the first-stage predictions with the low and high estimates as inputs to a second XGBoost model. This stacked architecture allows us to evaluate whether machine learning provides incremental value once institutional forecasts are included.

\subsubsection{Predictive Accuracy}

On headline metrics, differences between the ensemble and estimates-only models are modest. For works with previous sales, the ensemble achieves an R\textsuperscript{2}  of 0.938 and MAPE of 2.74\%, compared to 0.936 and 2.77\% for estimates-only. For fresh works, the ensemble records an R\textsuperscript{2} of 0.912 and MAPE of 2.84\%, while estimates-only yields 0.917 and 2.82\%. Mean absolute errors are essentially identical across both setups. At face value, these results suggest that machine learning adds little to expert estimates. However, such aggregate metrics can mask systematic biases in the error distribution, where the ensemble shows clearer advantages.

\subsubsection{Error Distribution and Bias Correction}

\begin{figure}[htbp]
    \centering
    \includegraphics[width=0.5\textwidth]{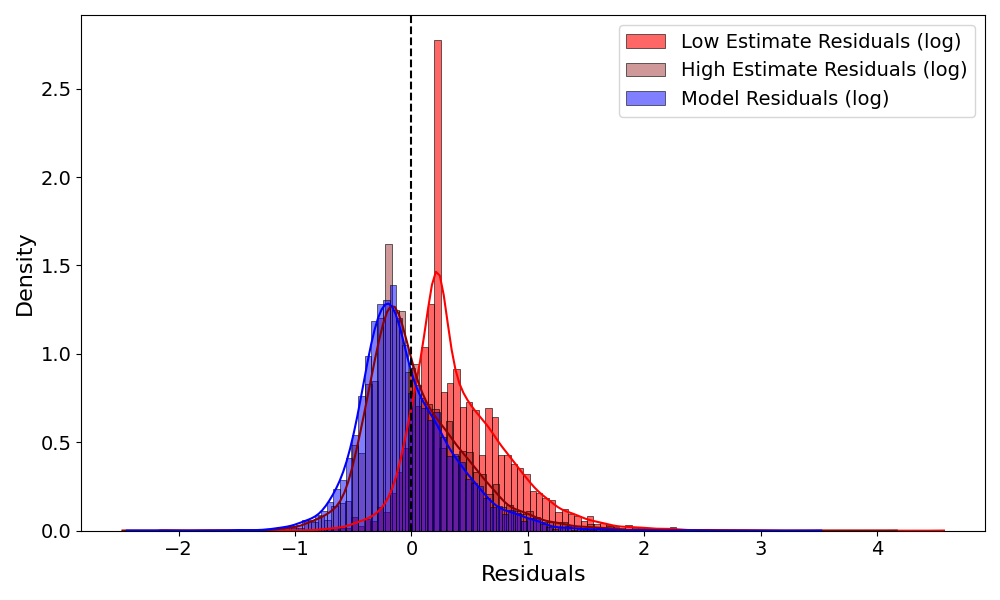}
    \caption{Residual distributions of auction house estimates and the ensemble model}
    \label{fig:resid_dist}
\end{figure}

Residual analysis reveals important differences. As expected, auction house low estimates are systematically right-skewed, tending to underpredict high-value works, while high estimates are closer but still conservative. The ensemble reduces this skew and produces residuals that are more symmetrically distributed and slightly more concentrated around zero (Figure~\ref{fig:resid_dist}).

\begin{figure}[H]
    \centering
    \includegraphics[width=0.5\textwidth]{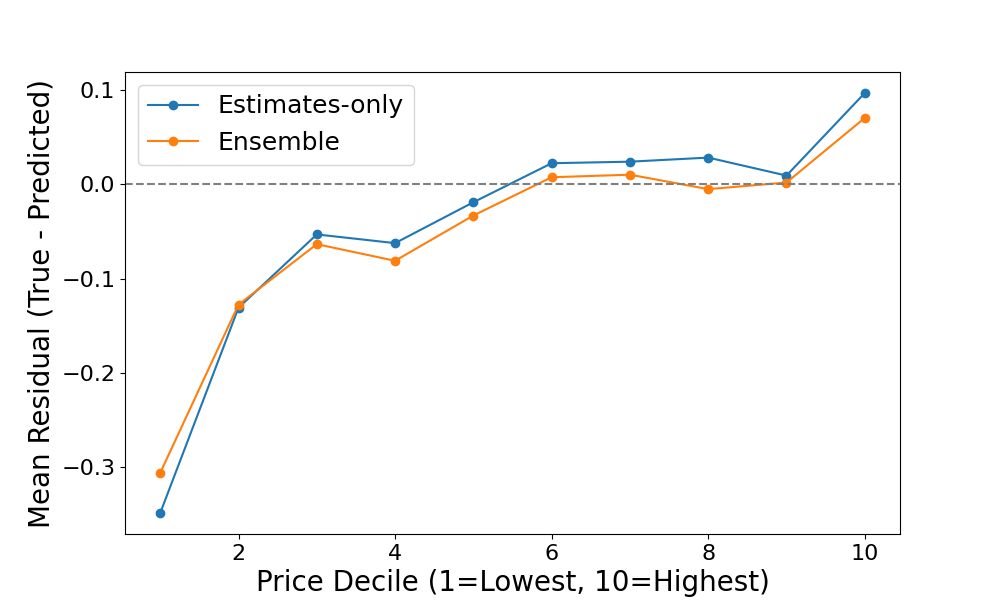}
    \caption{Mean residual  by realized-price decile}
    \label{fig:resid_deciles}
\end{figure}

A decile-level analysis based on our ensemble model and an XGBoost model with low and high estimates as only features confirms this pattern (Figure~\ref{fig:resid_deciles}). We sort artworks into deciles based on realized prices and plot the corresponding average residuals. For the lowest-priced works, both models tend to overpredict (negative residuals), but the ensemble moderates this bias. At the top end of the price distribution, both models tend to underpredict, yet again the ensemble is closer to zero. In other words, the ensemble does not alter the overall shape of the error curve, but it compresses the extremes and reduces systematic misses at the tails. 

This correction is especially relevant in practice. Overprediction in the low-value segment risks deterring buyers, while underprediction of high-profile works can damage institutional credibility. By narrowing these biases, the ensemble contributes to fairer and more balanced valuations, even when overall accuracy remains similar.

\subsubsection{Interpreting the Ensemble}

\begin{figure}[!ht]
    \centering
    \includegraphics[width=0.8\textwidth]{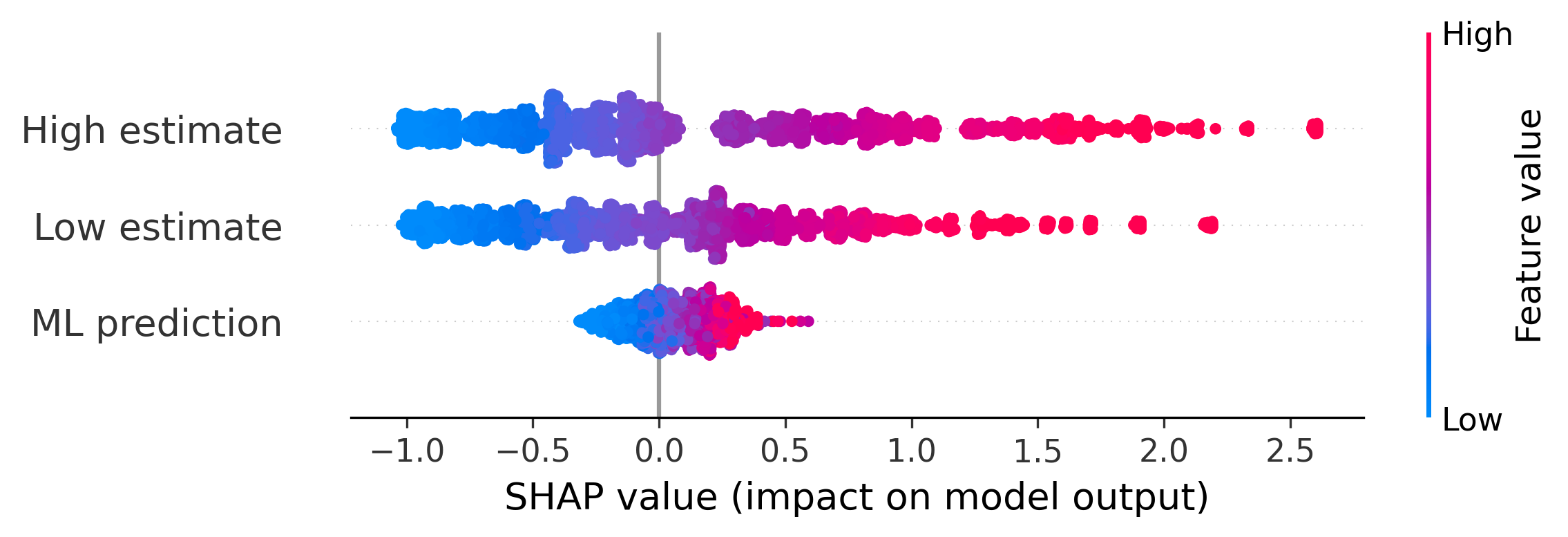}
    \caption{Overall SHAP summary plot for the ensemble model}
    \label{fig:shap_beeswarm}
\end{figure}

To understand how the ensemble integrates auction house forecasts with data-driven signals, we apply SHAP (SHapley Additive exPlanations) values \citep{shap}, a game-theoretic method that attributes each feature’s contribution to a model prediction relative to a baseline expectation, defined as the model’s average prediction across all samples. Unlike permutation importance, which measures the drop in accuracy when a feature is shuffled, SHAP provides signed contributions for each observation, showing whether a variable pushes the prediction above or below the model’s baseline expectation. As Figure~\ref{fig:shap_beeswarm} shows, the high estimate generally shifts predictions upward, though with some variation across cases. The low estimate tends to pull predictions downward, but again with a spread of effects depending on the observation. The machine learning prediction displays the most balanced distribution around zero, sometimes amplifying the auction house forecasts and sometimes counteracting them.

\begin{figure}[!ht]
    \centering
    \includegraphics[width=0.75\textwidth]{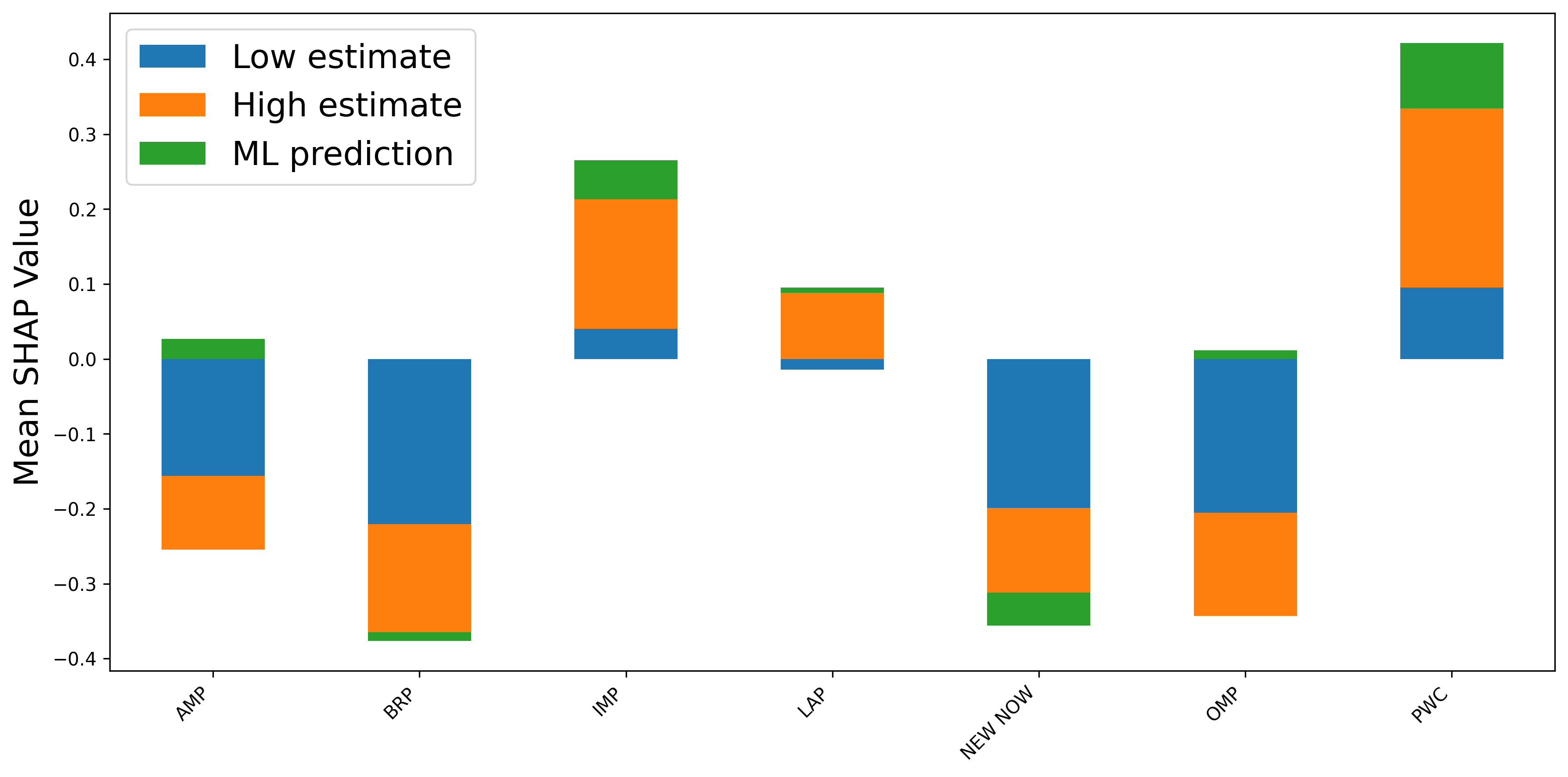}
    \caption{Average SHAP contributions by artwork category}
    \label{fig:shap_category}
\end{figure}

Category-level analysis in Figure~\ref{fig:shap_category}. reveals systematic differences. In segments such as Old Masters, both low and high estimates act as anchors that the ensemble largely respects, while the ML signal has little influence. In more volatile and speculative categories such as Post-War and Contemporary, the ensemble leans more heavily on the high estimate, with the ML signal providing additional upward adjustment where auction houses appear systematically cautious. Importantly, the direction of mean SHAP values per category is mechanically tied to the baseline expectation: if a category’s average price level is above the overall mean prediction, contributions are predominantly positive, whereas categories below the baseline generally show negative mean SHAP values.

Combining auction house estimates with machine learning yields only marginal improvements in headline accuracy, but it meaningfully reshapes the error distribution. The ensemble reduces systematic underprediction of high-value works and moderates overprediction of lower-value segments, which provides better calibration across the market. SHAP analysis further shows that the ensemble does not simply average its inputs but considers inputs depending on context. In short, the value of the ensemble lies less in squeezing out higher R\textsuperscript{2} and more in bias correction and calibration, which  are critical in high-stakes valuation settings such as the art market.

% \fi

\subsection{Predicting the Transaction Probability}
In the art market, understanding which artworks are likely to be sold and at what price is essential for collectors, sellers, auction houses, and investors. Traditionally, auction houses provide high and low price estimates that serve as reference points or anchors during the bidding process. While these estimates are intended to guide buyers and sellers, they can inadvertently introduce bias. This phenomenon, known as anchoring, causes bidders and sellers to fixate on the given price range, often leading to inflated reservation prices set by sellers. Consequently, artworks priced too high relative to actual demand may remain unsold, which has a distortional effect on markets. 

If successful, obtaining transaction probabilities through a model offers practical implications for auction houses and sellers: by quantifying the transaction probability while considering anchoring bias, pricing strategies can be adjusted to reduce overestimation, which potentially lowers reservation prices and increases sales success rates. 

\subsubsection{Data and Methodology}
Our dataset includes 4180 unsold samples (after filtering). To balance the classes in our classification problem, we randomly sample an equal number of observations from our previously used dataset of sold artworks.

We apply similar preprocessing steps as before. However, due to data scarcity, a temporal split is no longer feasible. Instead, we ensure that the ratio of sold to unsold artworks is balanced in both the training and test sets. Additionally, due to our objective, we no longer impose a minimum transaction price. 
Subsequently, we focus on neural networks as our classification model, predicting the likelihood of an artwork being sold at auction based on its characteristics. The objective is to identify the key drivers of auction success and to examine how predictive features differ between sold and unsold works. In contrast to the deeper multi-modal architecture used for price prediction (Figure~\ref{fig:art-model-combined}), the classification network is deliberately kept smaller, with fewer layers. This reflects both the lower complexity of the binary classification task and the more limited sample size available for training, where a shallower network helps reduce the risk of overfitting. We train the model using Binary Cross-Entropy (BCE) as the optimization objective and evaluate performance through ROC curves and confusion matrices. To complement the neural network results, we also employ XGBoost as a classifier and assess feature importance using feature permutation, which helps to interpret the signals that distinguish sold from unsold lots.

\subsubsection{Prediction Results}

\begin{figure}[!ht]
    \centering
    % First row: two side-by-side
    \begin{subfigure}{0.45\textwidth}
        \centering
        \includegraphics[width=\linewidth]{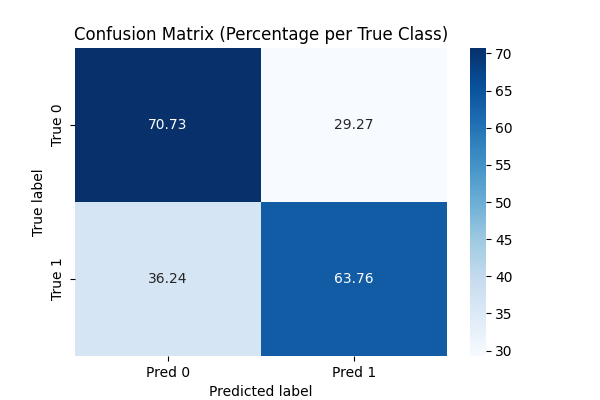}
        \caption{Multi-modal (d\textsubscript{image} = 10)}
    \end{subfigure}
    \hfill
    \begin{subfigure}{0.45\textwidth}
        \centering
        \includegraphics[width=\linewidth]{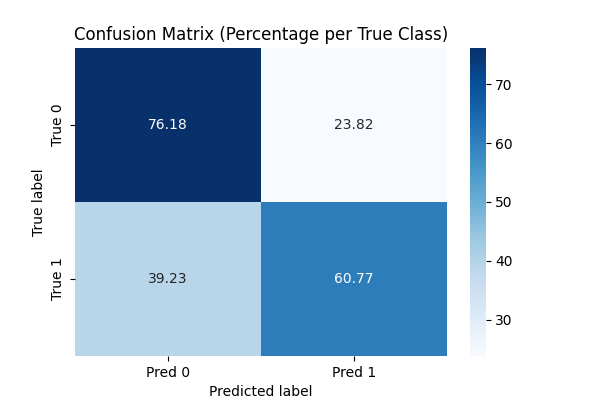}
        \caption{Multi-modal (d\textsubscript{image} = 1000)}
    \end{subfigure}

    % Second row: one centered
    \vskip\baselineskip
    \begin{subfigure}{0.5\textwidth}
        \centering
        \includegraphics[width=\linewidth]{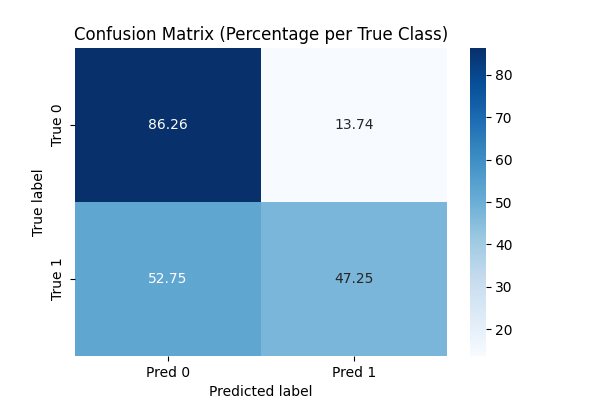}
        \caption{Tabular-only}
    \end{subfigure}

    \caption{Confusion matrices for classification models}
    \label{fig:conf_matrices}
\end{figure}

The comparative evaluation of the three models in Figure \ref{fig:conf_matrices} reveals modest but noteworthy differences in predictive capacity. The tabular-only model demonstrates strong recall for the unsold class (0.86) but substantially weaker recall for the sold class (0.47), resulting in an overall accuracy of 0.67. This reflects a conservative bias toward predicting works as unsold. The multi-modal model with small image embeddings (10 dimensions) provides a more balanced classification, achieving recall values of 0.71 (unsold) and 0.64 (sold), with an overall accuracy of 0.67. This suggests that incorporating compact image features improves sensitivity to sales while maintaining reasonable performance for unsold predictions. The multi-modal model with large image embeddings (1000 dimensions) achieved intermediate results, with recall values of 0.77 (unsold) and 0.61 (sold) and an overall accuracy of 0.69, indicating that the high-dimensional image representation  only yields a minor performance advantage.

\begin{figure}[!ht]
    \centering
    \includegraphics[width=0.4\linewidth]{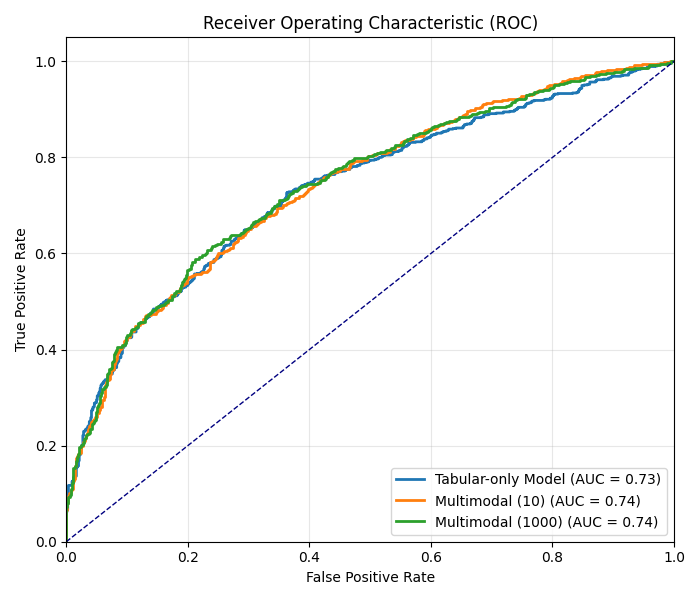}
    \caption{Receiver Operating Characteristic (ROC) curves 
    }
    \label{fig:roc_curves}
\end{figure}

The ROC curves further reinforce these observations (Figure \ref{fig:roc_curves}). Both multi-modal approaches obtain an AUC of 0.74, while the tabular-only model achieves 0.73. Although the multi-modal approaches capture additional signals, the improvements are small, and all models remain clustered around AUC values, which indicates limited discriminative power.

Taken together, the results highlight that while the models capture some meaningful structure in the data, their overall accuracy (0.67–0.69) and AUC values (0.73-0.74) suggest only moderate predictive utility. These limitations likely stem from the inherent complexity of the art market, the overlap between sold and unsold features, and the limited signal extracted from image data. From a practical perspective, however, the small multi-modal embedding provides a balanced trade-off between specificity and sensitivity, which suggests that compact image features can complement tabular information without over-complicating the model.

\newlength{\subcolumnwidth}
\newenvironment{subcolumns}[1][0.45\columnwidth]
 {\valign\bgroup\hsize=#1\setlength{\subcolumnwidth}{\hsize}\vfil##\vfil\cr}
 {\crcr\egroup}
\newcommand{\nextsubcolumn}[1][]{%
  \cr\noalign{\hfill}
  \if\relax\detokenize{#1}\relax\else\hsize=#1\setlength{\subcolumnwidth}{\hsize}\fi
}
\newcommand{\nextsubfigure}{\vfill}

\subsubsection{Feature Distribution and Importance}
\begin{figure}[!ht]
\begin{subcolumns}
  \subfloat[Feature permutation\label{permutation_class}]{%
    \includegraphics[width=\subcolumnwidth]{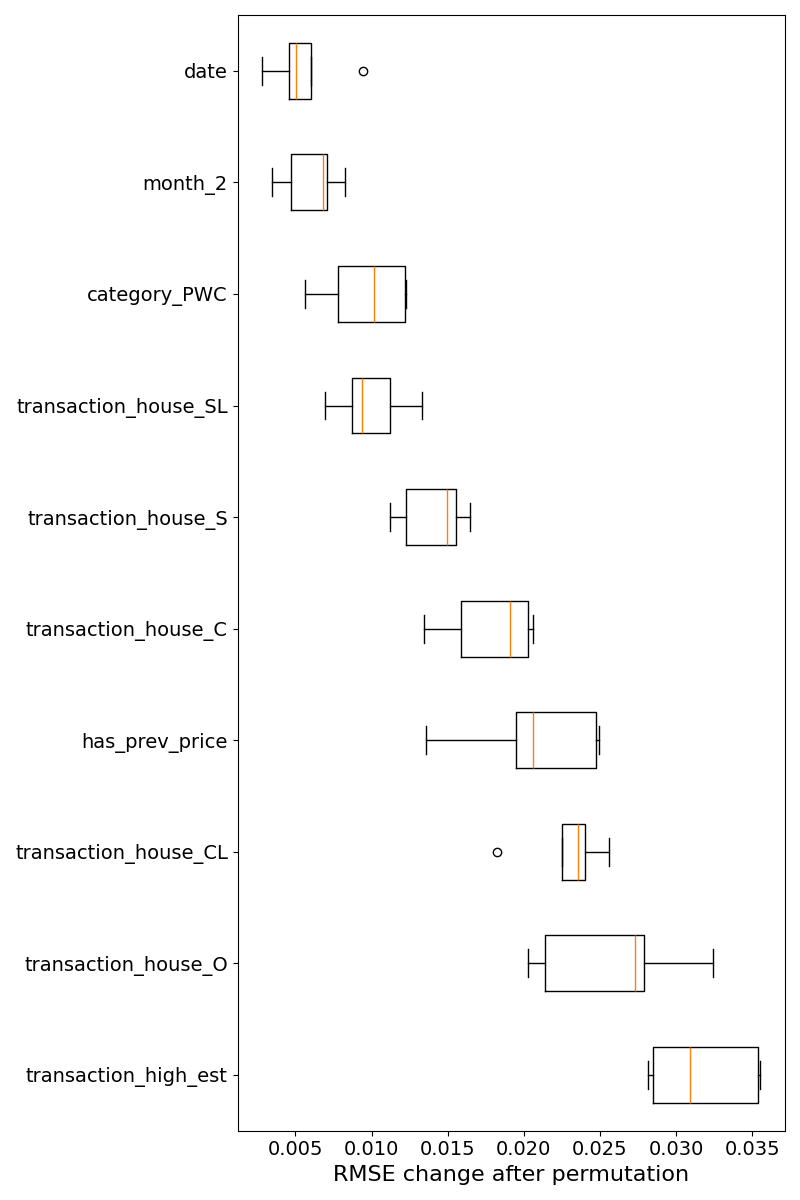}
  }
\nextsubcolumn
  \subfloat[Distribution of low estimates for error types\label{fig:feature_dist_low}]{%
    \includegraphics[width=\subcolumnwidth]{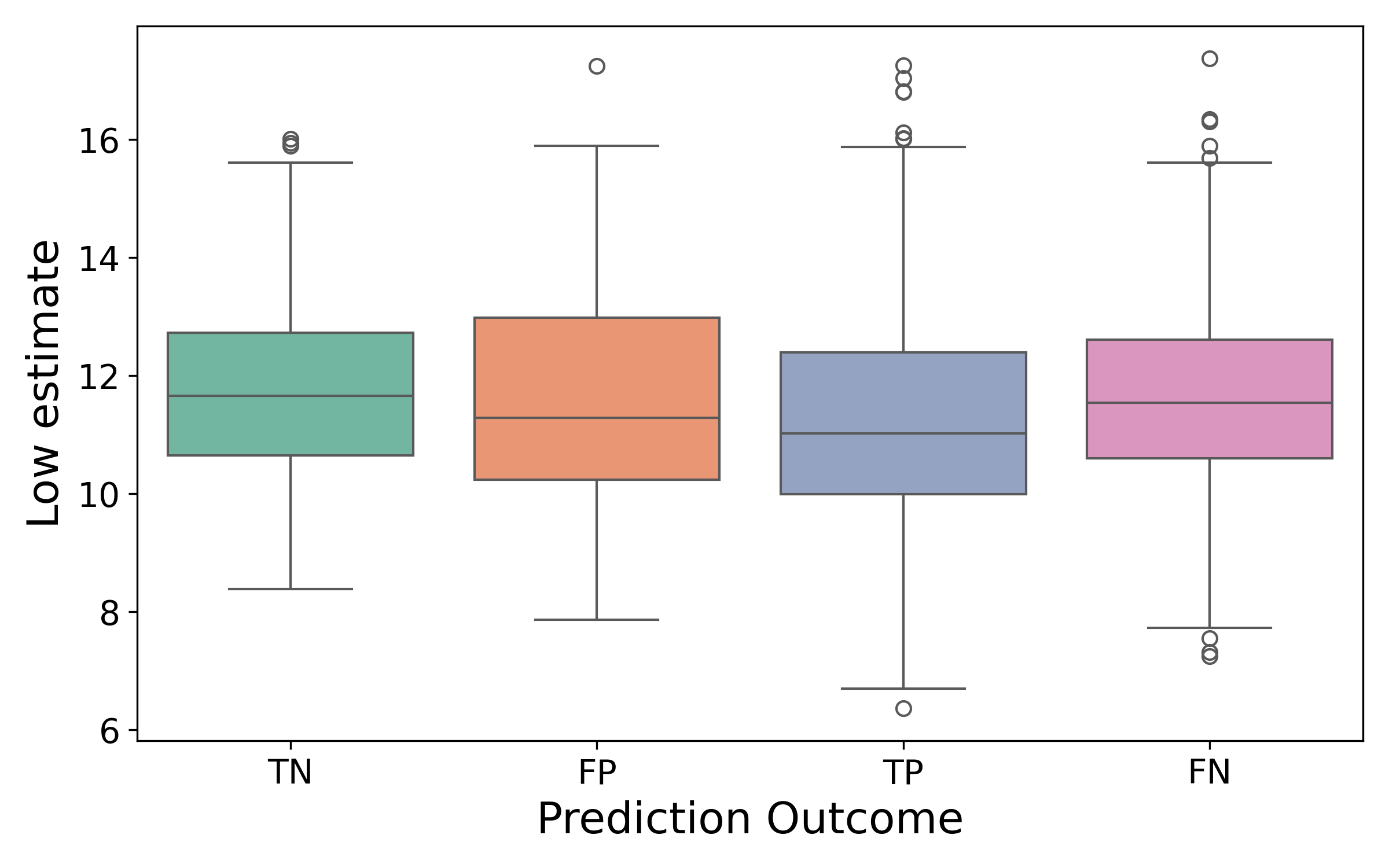}
  }
\nextsubfigure
  \subfloat[Distribution of high estimates for error types\label{fig:feature_dist_high}]{%
    \includegraphics[width=\subcolumnwidth]{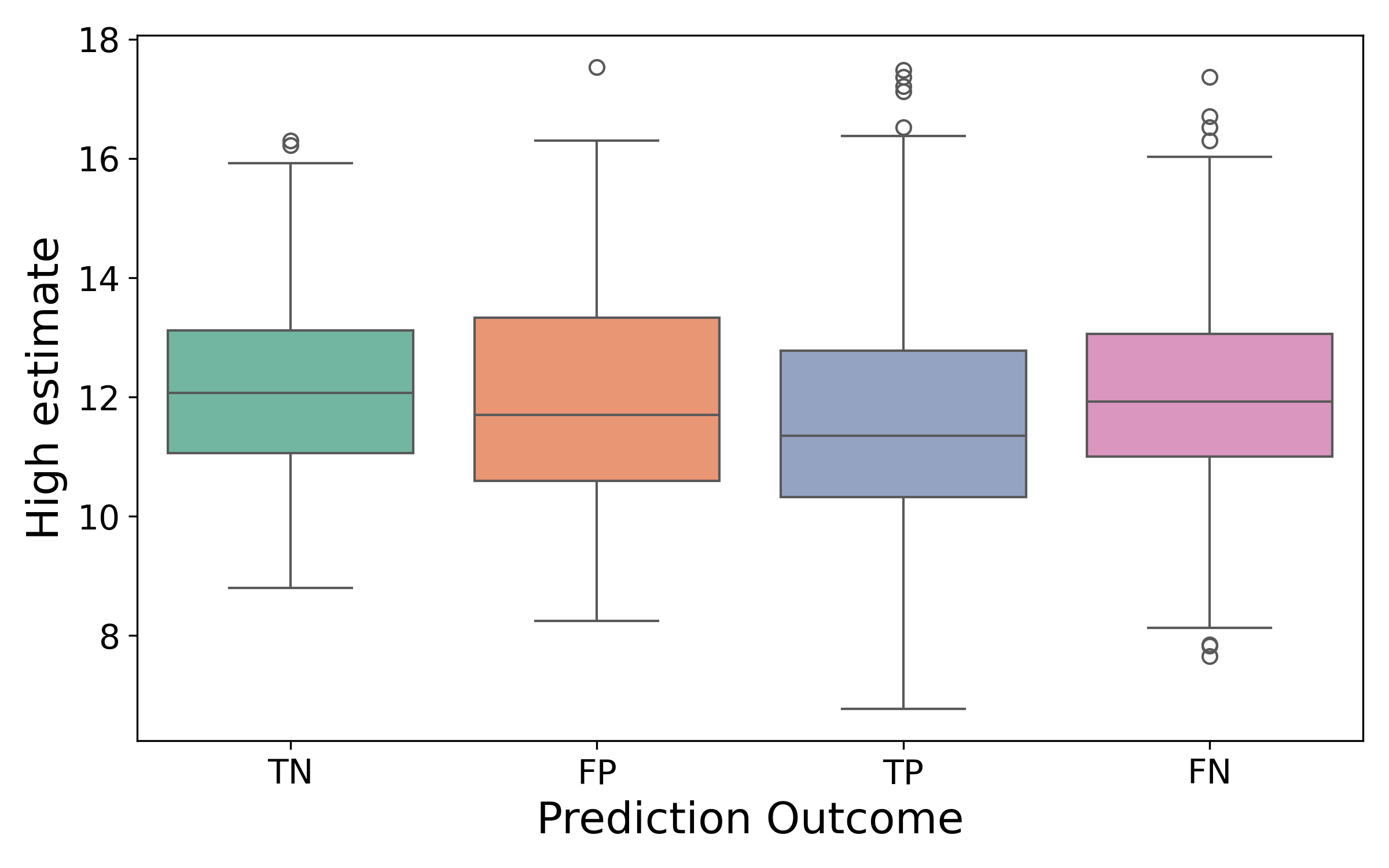}
  }
\end{subcolumns}
\caption{Feature importance in classification}
\end{figure}

For feature importance and feature distribution analysis, we rely on a benchmark XGBoost classification model, which again only uses tabular data. Figure \ref{permutation_class} displays the corresponding ten most important features detected by feature permutation. It indicates that the high auction house estimate, the availability of a transaction history, and auction house identity are the most influential predictors. While low estimates are not part of this table, they might still be crucial but are shaded by their strong correlation with high estimates. 

These results suggest that the model captures meaningful market signals rather than spurious temporal patterns.

Additionally, an analysis of feature distributions reveals that true positives (correctly predicted sales) tend to have slightly lower average low and high estimates compared to other prediction error types (see Figure \ref{fig:feature_dist_low} and \ref{fig:feature_dist_high}). This suggests that artworks with more modest price expectations are somewhat easier for the model to correctly classify as sold, potentially reflecting a pattern where lower-valued items are more likely to meet market demand. The contrary seems to apply for true negatives. These findings underline the importance of price estimates on auction outcomes in general. For auction houses, this implies that conservative pricing (particularly in the form of realistic high estimates) can meaningfully influence the likelihood of a successful sale. Setting estimates too aggressively may reduce the probability of sale, while aligning expectations with market conditions appears to improve outcomes.

At the same time, these insights should be interpreted with caution: the classification task relies on a substantially smaller sample size than the price prediction models, which limits the stability of the estimates and increases the risk of overfitting. The observed relationships are therefore best understood as indicative patterns rather than definitive causal mechanisms.

\section{Conclusion}
This paper evaluates whether computer vision adds economically meaningful information for art valuation once standard structured predictors are available. Using a repeated-sales auction panel matched to artwork images, we benchmark classical hedonic specifications and strong tabular machine-learning methods against multi-modal architectures that fuse image embeddings with metadata. The central finding is state-dependent: when an artwork has an auction history, price anchors and reputation-related covariates dominate and images add little beyond strong tabular baselines; when an artwork is fresh to the auction market, visual embeddings provide incremental predictive content and materially improve valuation accuracy relative to tabular-only models. This pattern supports a simple organizing principle: unstructured visual information is most valuable precisely when structured history is thin.

Two additional results help interpret these gains. First, across model classes, performance is sensitive to the way visual information is compressed into the joint representation: moderate-dimensional embeddings perform best, consistent with a bias--variance trade-off in data-constrained settings. Second, interpretability diagnostics suggest that the visual branch often attends to broad compositional cues and surface patterns and, at times, to superficial correlations. These findings indicate that image models should be viewed as complements to structured signals and expert judgment, rather than stand-alone valuation tools.

We also assess performance relative to the market’s operative benchmark: auction-house presale estimates. Presale estimates remain the single most informative predictor of realized prices, consistent with experts’ access to information that is only partially observed in public data (e.g., condition and provenance) and with the anchoring role of estimates in the auction process. Nevertheless, combining estimates with machine-learning signals yields a practical improvement in calibration: the ensemble modestly reduces systematic overprediction in lower-value segments and underprediction at the top end, even when headline accuracy changes little. In this sense, the contribution of machine learning lies less in replacing expert forecasts than in sharpening them—particularly in the tails and in settings where historical anchors are absent.

Several caveats qualify the interpretation. Our identification of “fresh-to-market” valuation is constrained by the limited number of first-time sales in the out-of-sample period and by the fact that auction records omit or imperfectly measure key drivers such as condition, provenance, and private-sale information. These limitations point naturally to future work. Richer data—especially text from catalogues and condition reports, provenance histories, and broader coverage beyond frequently traded artists and major houses—would allow a cleaner assessment of what visual models learn versus what they proxy for. Methodologically, extending the framework to incorporate dynamics (e.g., time-varying tastes and market states) and alternative calibration objectives would further clarify when multi-modal learning can add value in illiquid, image-native asset markets.

\newpage
\appendix
% \begingroup
% \makeatletter
% \let\ps@plain\ps@empty
% \appendixpage
% \makeatother
% \endgroup
% \noappendicestocpagenum
% \addappheadtotoc

\section*{Appendix}
\section{Artistic Abbreviations}\label{app:abbr}
In Table \ref{tab:abbreviations}, we provide a list of abbreviations of categorical variables.
\begin{table}[ht!]
\captionsetup{list=no} % <--- prevents entry in List of Tables
\centering
\begin{tabular}{ll}
\toprule
\textbf{Abbreviation} & \textbf{Meaning} \\
\midrule
OO & Oil on other \\
OB & Oil on board, panel, wood \\
OC & Oil on canvas \\
OP & Oil on paper \\
OM & Oil on metal \\
WC & Watercolor \\
S  & Sculpture \\
O  & Other \\
D  & Drawing \\
P  & Pastel \\
INK & Ink \\
\midrule
IMP & Impressionist and Modern Art \\
PWC & Post-War and Contemporary Art \\
BRP & British Art \\
AMP & American Paintings/Drawings and Sculptures \\
LAP & Latin American Art \\
OMP & Old Master and 19th Century Art \\
\midrule
SSP & Sotheby's Paris \\
SMI & Sotheby's Milano \\
SL  & Sotheby's London \\
CL  & Christie's London \\
C   & Christie's New York \\
CP  & Christie's Paris \\
S   & Sotheby's New York \\
O   & Other \\
CM  & Christie's Monaco \\
PMK & Phillips Hong Kong \\
\bottomrule
\end{tabular}
\caption{List of abbreviations}
\label{tab:abbreviations}
\end{table}

\section{Additional Grad-CAM Examples}\label{supp:visual}
\begin{figure}[H]
    \captionsetup{list=no} % <--- prevents entry in List of Tables
    \centering
    \includegraphics[height= 0.85\textheight, keepaspectratio]{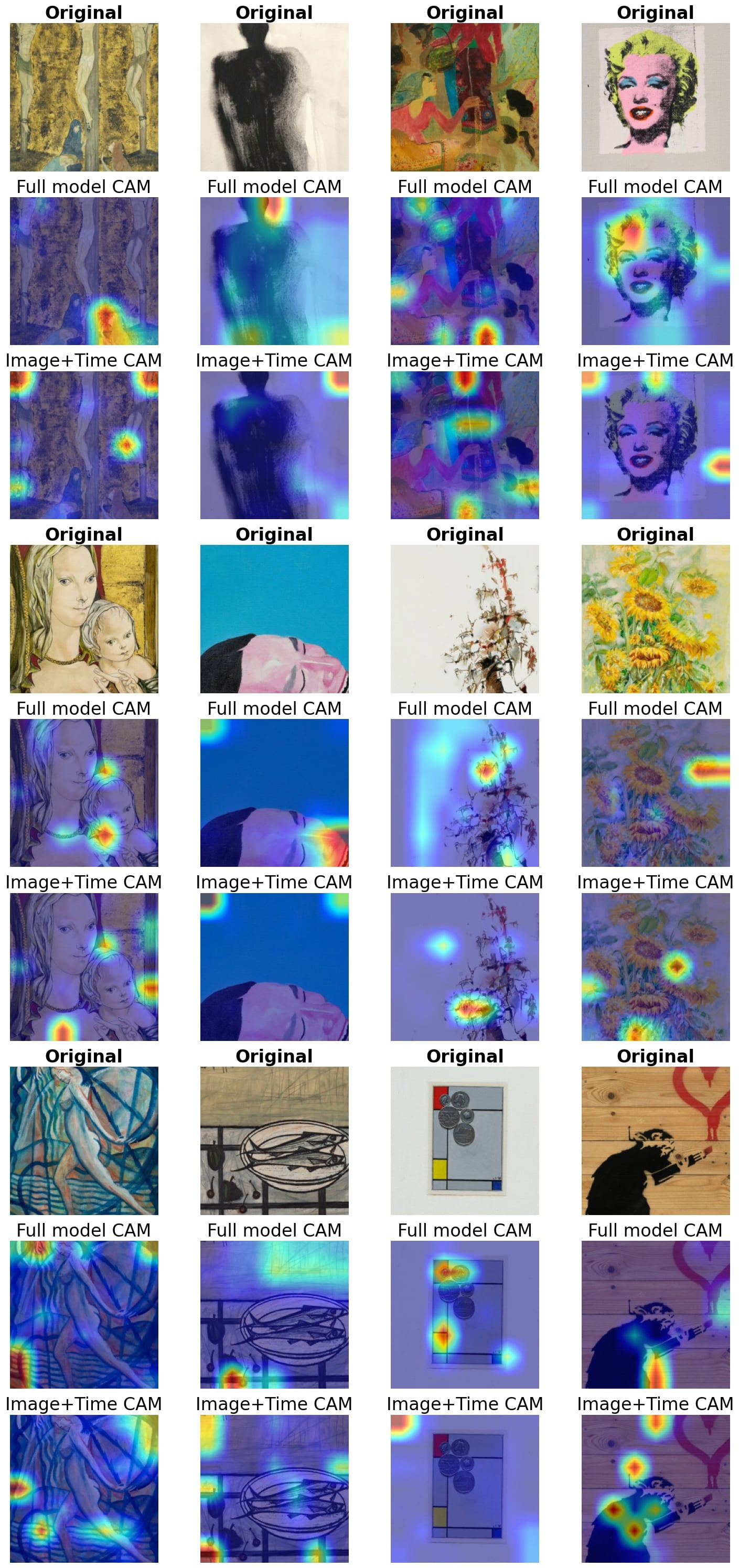}
    \caption{Comparison of Grad-CAM visualizations for the small and feature-rich multi-modal models (further examples)}    
\end{figure}

\begin{figure}[ht!]
    \captionsetup{list=no} % <--- prevents entry in List of Tables
    \centering
    \includegraphics[height= 0.9\textheight, keepaspectratio]{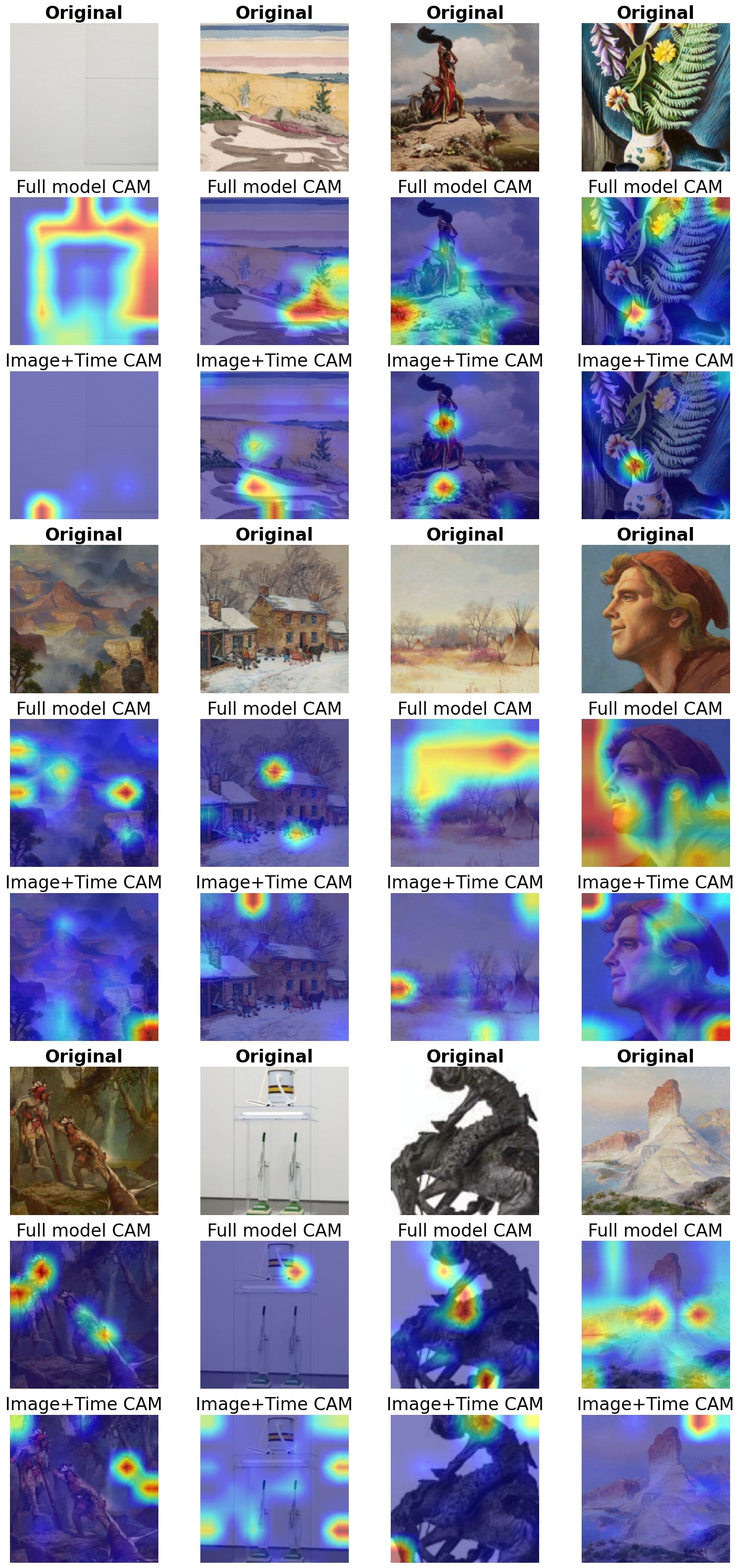}
    \caption{Comparison of Grad-CAM visualizations for the small and feature-rich multi-modal models (further examples)}
    
\end{figure}

\section{ViT Embedding Visualizations}\label{supp:embedding}

\begin{figure}[H]
    \centering
        \captionsetup{list=no} % <--- prevents entry in List of Tables
    
    % --- Row 1: Category ---
    \begin{subfigure}[t]{0.48\textwidth}
        \centering
        \includegraphics[width=\linewidth]{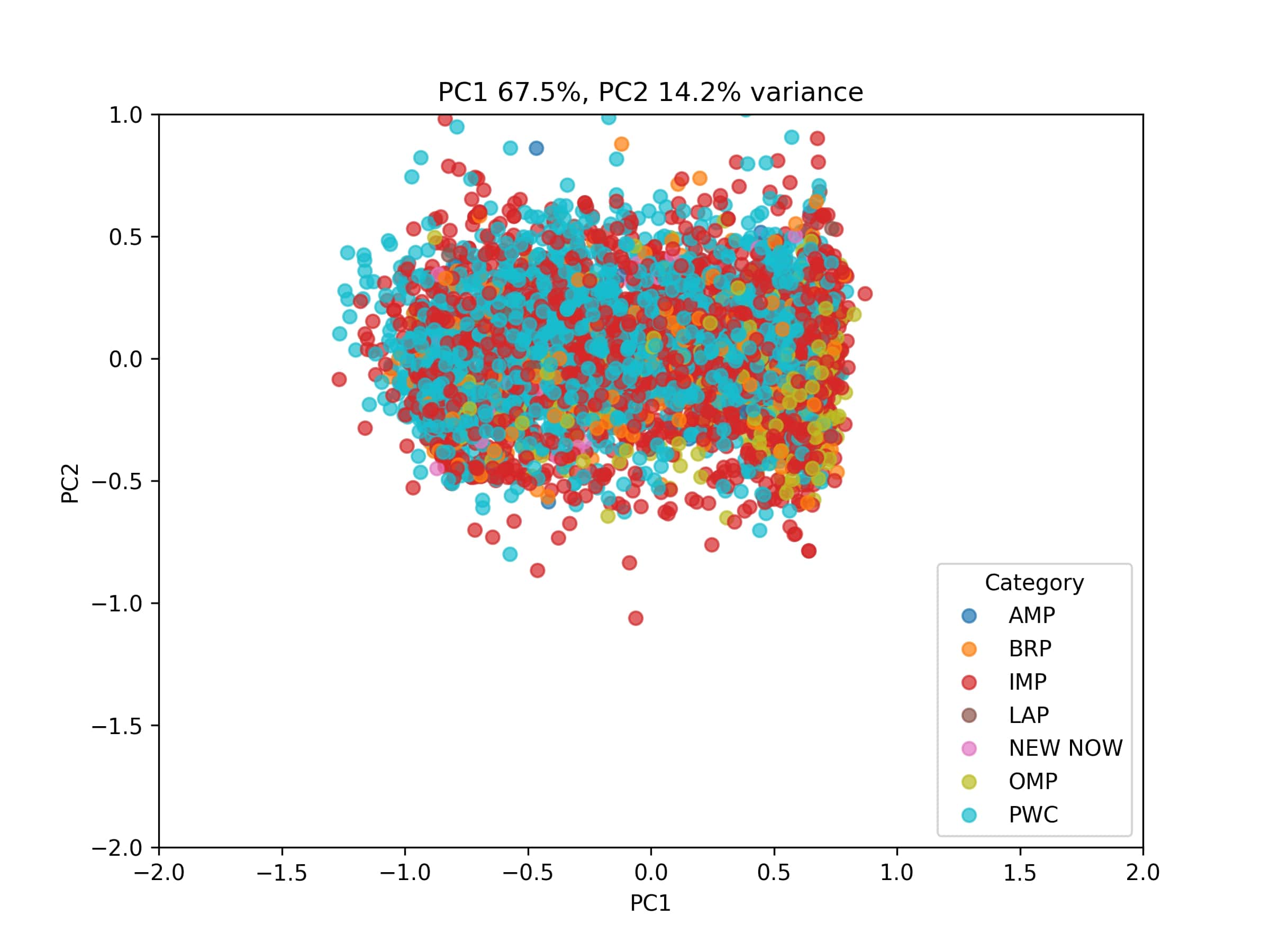}
        \caption{Feature-rich (category)}
        
    \end{subfigure}
    \hfill
    \begin{subfigure}[t]{0.48\textwidth}
        \centering
        \includegraphics[width=\linewidth]{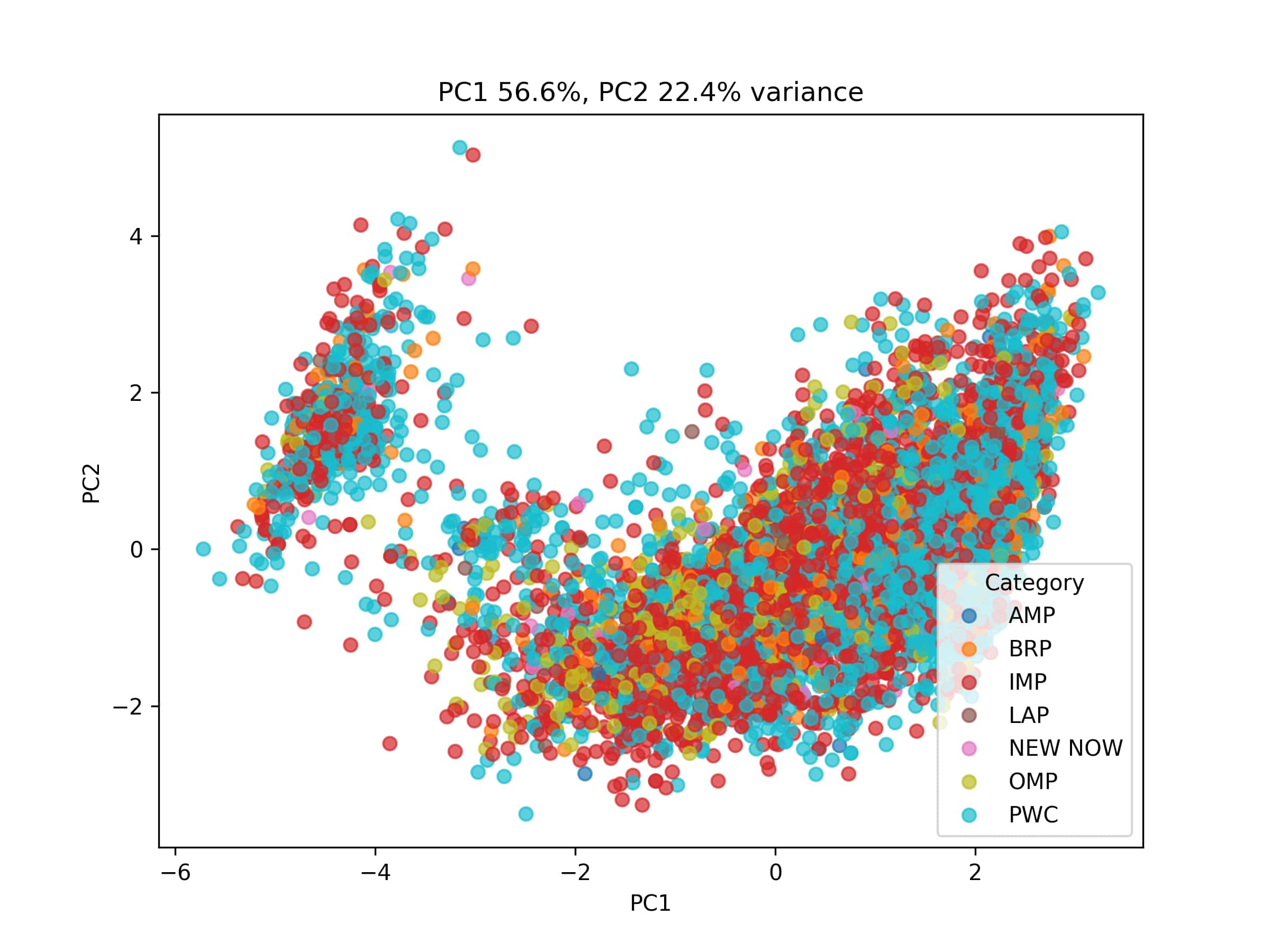}
        \caption{Minimal (category)}
        
    \end{subfigure}
    
    % --- Row 2: Medium ---
    \begin{subfigure}[t]{0.48\textwidth}
        \centering
        \includegraphics[width=\linewidth]{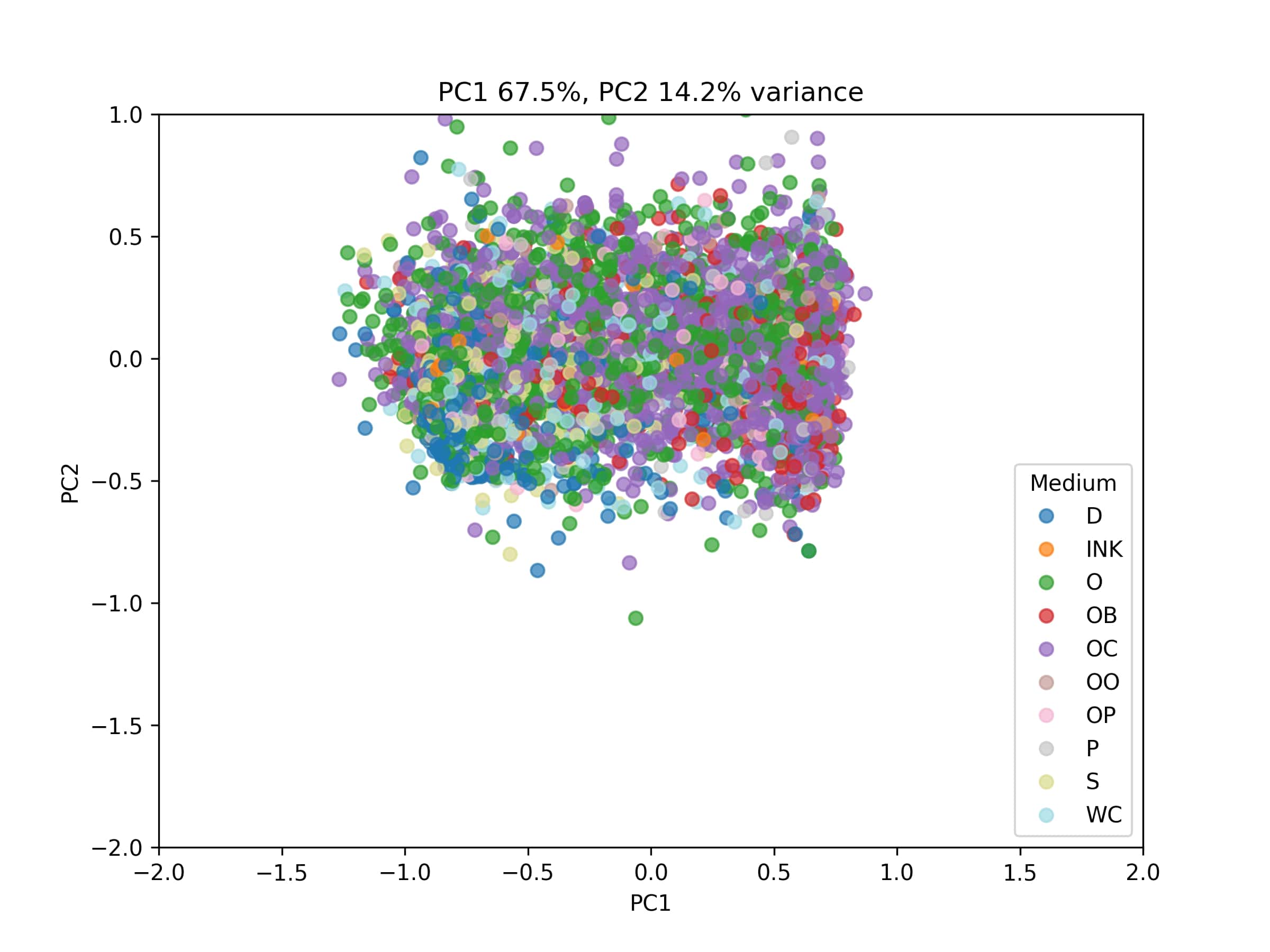}
        \caption{Feature-rich (medium)}
       
    \end{subfigure}
    \hfill
    \begin{subfigure}[t]{0.48\textwidth}
        \centering
        \includegraphics[width=\linewidth]{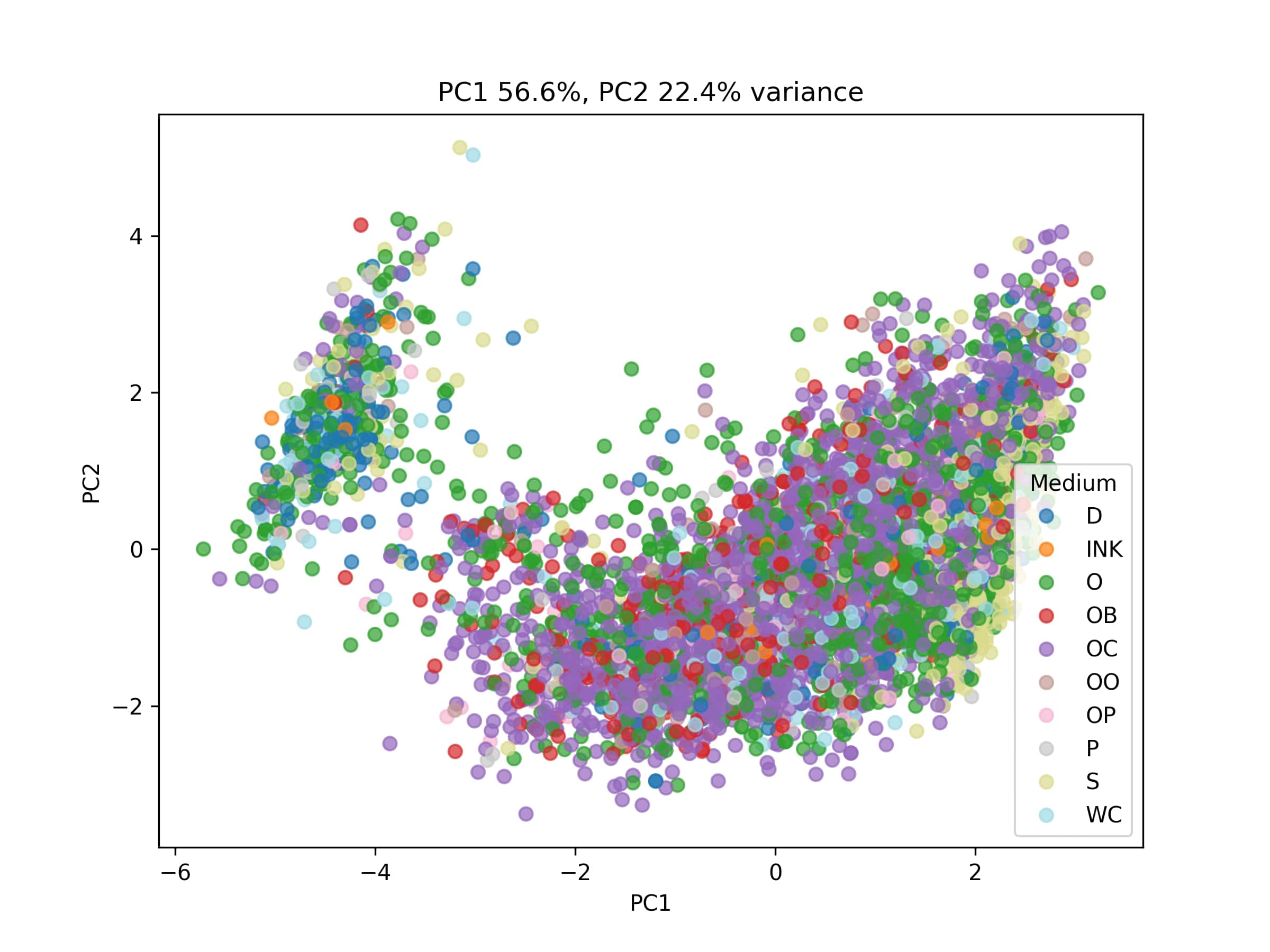}
        \caption{Minimal (medium)}
       
    \end{subfigure}

    % --- Row 4: Predicted Price ---
    \begin{subfigure}[t]{0.48\textwidth}
        \centering
        \includegraphics[width=\linewidth]{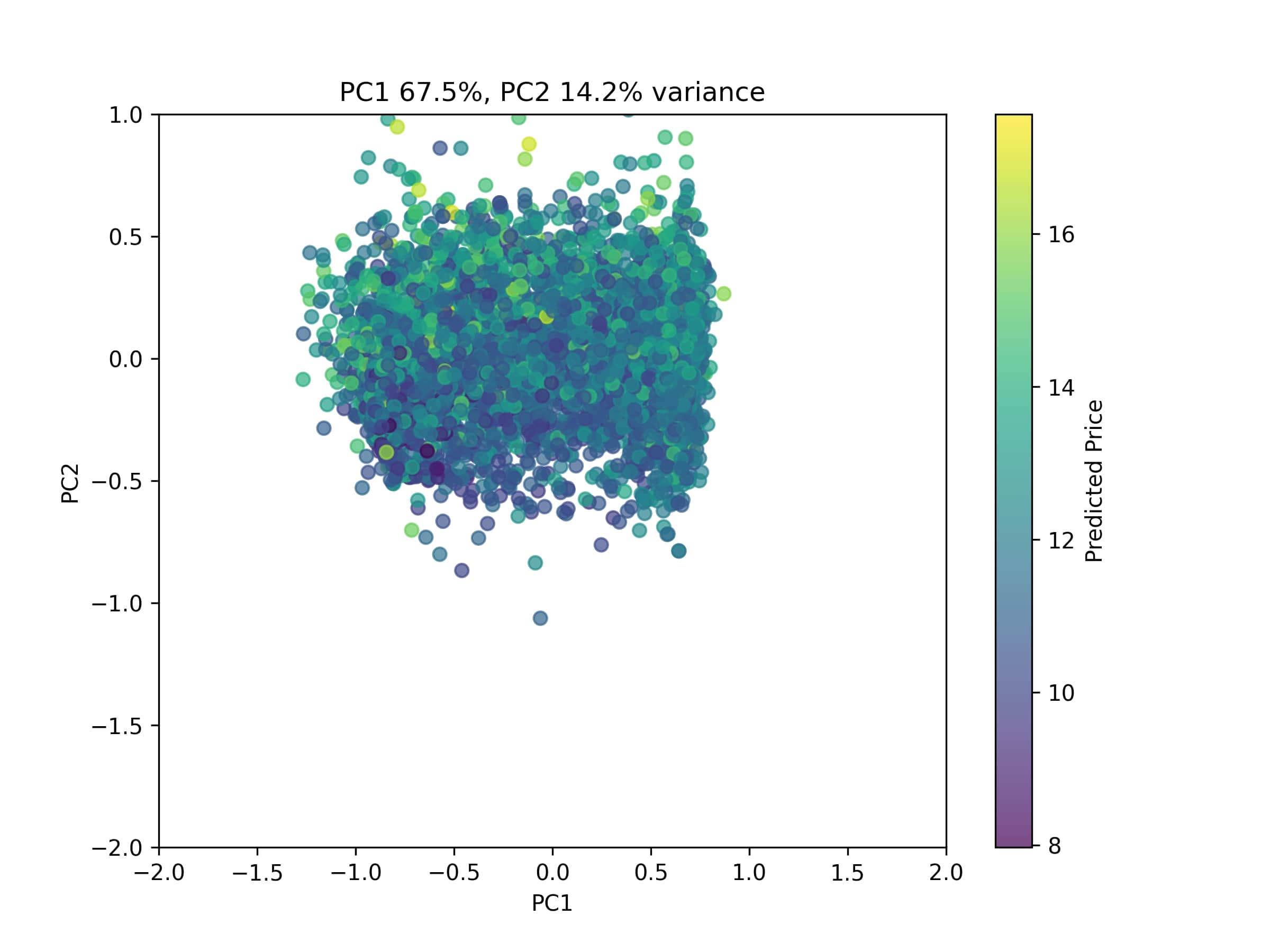}
        \caption{Feature-rich (predicted price)}
        
    \end{subfigure}
    \hfill
    \begin{subfigure}[t]{0.48\textwidth}
        \centering
        \includegraphics[width=\linewidth]{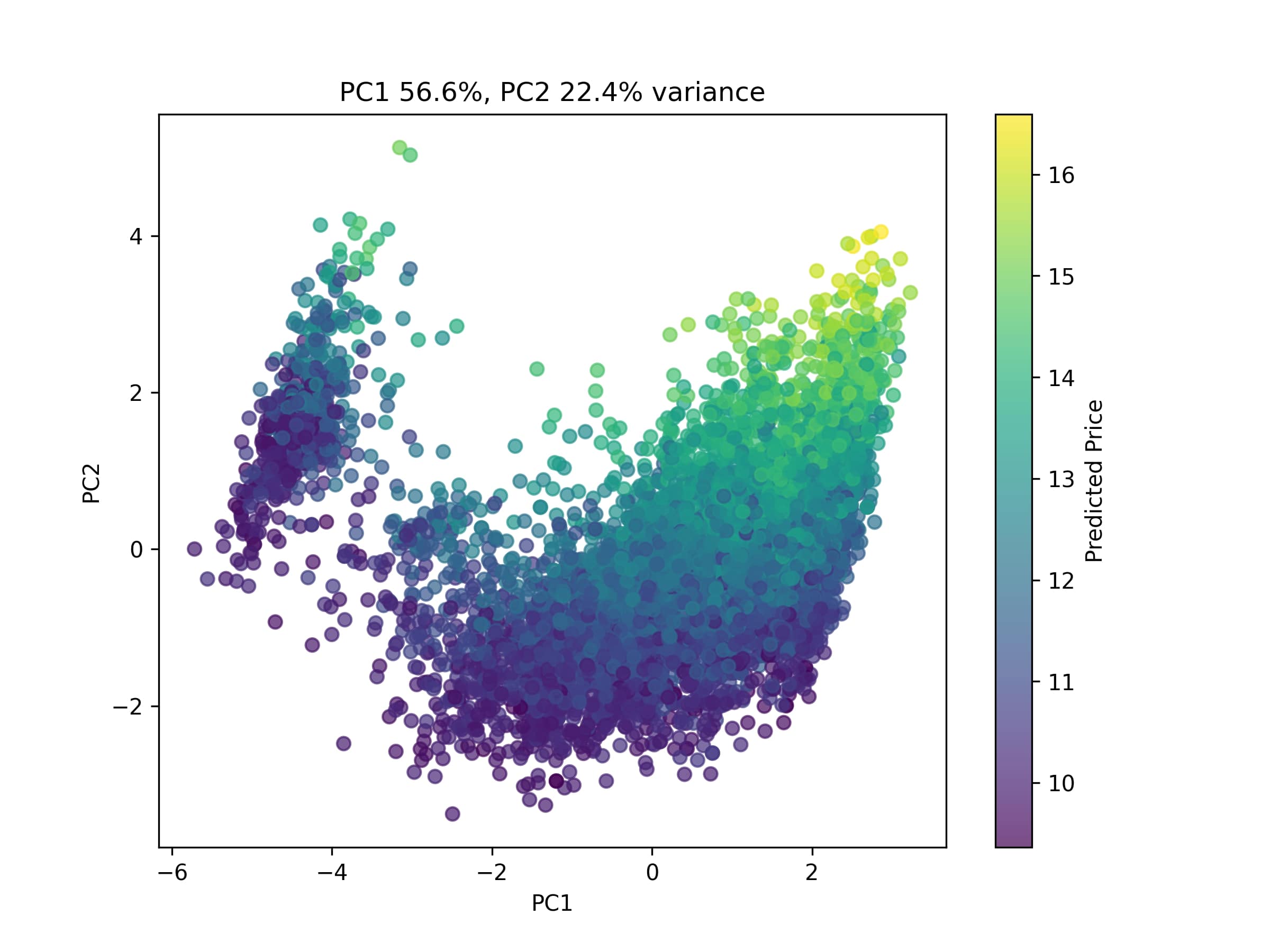}
        \caption{Minimal (predicted price)}
     
    \end{subfigure}
    
    \caption{PCA image embedding projections (ViT-Small model)}
   
\end{figure}

\singlespacing
\setlength\bibsep{0pt}
\bibliographystyle{aer}
\bibliography{references}

\end{document}